%% file: main.tex
\documentclass[sigconf]{acmart}
\AtBeginDocument{%
  }
\usepackage{float}
\usepackage{graphicx}
\usepackage{textcomp}
\usepackage{colortbl}
\usepackage{adjustbox}
\usepackage{caption}
\usepackage{paralist}
\usepackage{tabularx}
\usepackage{import}
\usepackage{algpseudocode,algorithm}
\usepackage{listings}
\usepackage{multirow}
\usepackage{caption}
\usepackage{subcaption}
\usepackage{hyperref}
\usepackage{circuitikz}
\usepackage{pgf}
\usepackage{pgfplots}
\DeclareUnicodeCharacter{2212}{−}
\usepgfplotslibrary{groupplots,dateplot}
\usepackage{xcolor}
\usepackage{bm}
\usepackage{pgfplotstable}
\usepackage[many]{tcolorbox}
\pgfplotsset{compat=1.8}
\usepackage{adjustbox}
\usepackage{changepage}
\usepackage{enumitem}
\usepackage{dblfloatfix}
\usepackage{minted}
\usepackage{svg}
\usepackage{xpatch}

\usepackage[utf8]{inputenc}

\usepackage{amsmath}
\usepackage{amssymb}

\usepackage{listings}
\usepackage{xcolor}

\tcbset{
  highlightbox/.style={
    colback=gray!10,              
    colframe=gray!50,             
    boxrule=0.4pt,                
    arc=5pt,                      
    left=5pt,right=5pt,top=3pt,bottom=3pt, 
    enhanced,                     
  }
}

\definecolor{vscKeyword}{HTML}{0000FF}   
\definecolor{vscType}{HTML}{267F99}      
\definecolor{vscString}{HTML}{A31515}    
\definecolor{vscComment}{HTML}{008000}   
\definecolor{vscPreproc}{HTML}{AF00DB}   
\definecolor{vscFunc}{HTML}{795E26}      

\lstdefinestyle{vscodeLightC}{
  language=C,
  basicstyle=\ttfamily,
  numbers=none, numberstyle=\tiny\color{gray},
  frame=none, breaklines=true, columns=fullflexible, keepspaces=true,
  showstringspaces=false,
  commentstyle=\color{vscComment}\ttfamily,
  stringstyle=\color{vscString}\ttfamily,
  keywordstyle=[1]\color{vscKeyword}\bfseries,   
  keywordstyle=[2]\color{vscType}\bfseries,      
  keywordstyle=[3]\color{vscPreproc}\bfseries,   
  morekeywords=[1]{if,else,for,while,return,break,continue,sizeof},
  morekeywords=[2]{int,uint32_t,uint64_t,size_t,void,char,unsigned,signed,const,static,struct,union,enum,volatile},
  otherkeywords={\#define,\#include,\#if,\#ifdef,\#ifndef,\#endif,\#pragma},
  keywordstyle=\color{vscKeyword}\bfseries,
  emph={memset,p256_double,p256_add_mixed,CCOPY,NEQ,EQ},
  emphstyle=\color{vscFunc}
}

\setcopyright{none}
\renewcommand\footnotetextcopyrightpermission[1]{}

\begin{document}

\title{PowerFuzz: Power–Based Black-Box Firmware Fuzzing}

\author{Dakshina Tharindu, Sahan Sanjaya, Philip Baptist, Prabhat Mishra}
\affiliation{%
  \institution{University of Florida}
  \city{Gainesville}
  \state{Florida}
  \country{USA}}




\begin{abstract}
Fuzzing is widely used for software and hardware verification, offering an effective alternative to random testing. While gray-box fuzzers benefit from full visibility into the system under test and can leverage execution feedback such as branch coverage, these approaches are not applicable when verifying systems whose firmware or binaries are not publicly available. In such scenarios, obtaining coverage information for guiding the fuzzer becomes infeasible. In this paper, we introduce PowerFuzz, a statistical black-box fuzzing framework that leverages power side-channel measurements as a substitute for binary instrumentation, requiring no internal visibility into the target firmware. A central challenge in black-box firmware fuzzing is determining the executed branches during test execution. To address this challenge, we use power traces to identify branches utilizing a sliding window followed by a growing window full-trace correlation method. This approach also enables the construction of a high-level control-flow graph of the black-box firmware, which we utilize to drive the fuzzer to unexplored execution paths. Extensive evaluation using three embedded hardware platforms and ten firmware benchmarks demonstrates that PowerFuzz can provide branch coverage comparable (within 13.5\%) to gray-box fuzzers while significantly outperforming (up to 22\%) state-of-the-art black-box fuzzers.

\end{abstract}




\maketitle

\input{sections/1_intro}

\input{sections/2_related}
\input{sections/3_methodology}

\input{sections/4_experiments}
\input{sections/5_conclusion}


\bibliographystyle{ACM-Reference-Format}
\bibliography{refs}

\end{document}

%% file: sections/1_intro.tex
\section{Introduction}\label{sec:introduction}


The rapid proliferation of embedded devices has transformed modern computing infrastructure. From industrial control systems and medical implants to automotive controllers and smart home appliances, embedded systems now constitute the backbone of critical infrastructure worldwide. As of 2024, over 19.8 billion Internet of Things (IoT) devices are deployed globally, with projections exceeding 30 billion by 2030~\cite{statista_iot}. Many embedded systems operate in safety-critical and security-sensitive environments where failures can have catastrophic consequences~\cite{igure2006security}. Despite their pervasive role, the systematic verification and testing of embedded firmware remains significantly understudied relative to traditional software, largely due to the difficulty of observing firmware execution behavior without direct access to the binary or hardware debugging infrastructure. Ensuring the correctness of embedded firmware through rigorous automated testing is therefore not merely a technical challenge, but an urgent practical necessity for the reliable operation of the systems that underpin modern infrastructure.


\begin{figure}[htp]
  \centering
  \begin{subfigure}{0.45\linewidth}
    \centering
    \input{figs/sota_overview}
    \caption{State-of-the-art graybox fuzzing using branch coverage}
    \label{fig:sota_overview}
  \end{subfigure}
  \hfill
  \begin{subfigure}{0.45\linewidth}
    \centering
    \input{figs/powerfuzz_overview}
    \caption{Black-box firmware fuzzing using power traces}
    \label{fig:powerfuzz_overview}
  \end{subfigure}
  \caption{Architectural comparison of gray-box and power-based black-box firmware fuzzing frameworks. (a) State-of-the-art gray-box fuzzing leverages binary instrumentation to extract branch coverage feedback from a firmware image with full internal visibility. (b) PowerFuzz enables the same coverage-guided fuzzing loop in a fully black-box setting by substituting binary-level branch coverage with power-based branch coverage derived from power side-channel measurements, requiring no access to the firmware binary.}
\label{fig:overview}
\end{figure}
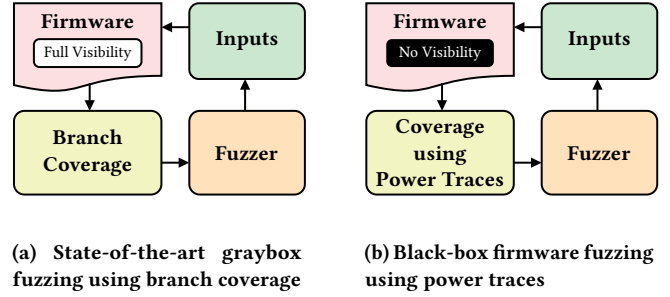

Fuzzing has emerged as one of the most effective techniques for vulnerability discovery in software systems. It was originally introduced as a random input generation method~\cite{miller1990empirical} and then it has evolved over the past three decades into a sophisticated, feedback-driven discipline. As shown in Figure~\ref{fig:sota_overview}, modern coverage-guided fuzzers such as AFL~\cite{zalewski_afl_whitepaper}, libFuzzer~\cite{crump2023libafl}, and HonggFuzz~\cite{honggfuzz} leverage compile-time or binary-level instrumentation to collect branch coverage feedback, enabling the fuzzer to systematically explore deeper and more diverse execution paths. This paradigm of coverage-guided fuzzing has been extended from traditional software to firmware testing. Systems such as Firmadyne~\cite{chen2016towards}, Avatar~\cite{zaddach2014avatar}, and Fuzzware~\cite{scharnowski2022fuzzware} attempt to rehost or emulate firmware in a controlled environment, enabling instrumentation-based feedback collection. The central insight underlying all effective fuzzing is that a meaningful feedback loop, one that informs the fuzzer whether a new input has triggered previously unseen program behavior, is indispensable for efficient exploration of a firmware's state space. Without such a feedback loop, the fuzzer operates blindly, unable to distinguish a productive mutation from a redundant one.

However, rehosting and instrumentation-based approaches require full visibility of the firmware binary. Therefore, they have limited applicability since embedded firmware is frequently proprietary and shipped in encrypted or obfuscated form. For example, manufacturers of commercial-off-the-shelf (COTS) devices such as routers, PLCs, and medical devices rarely disclose firmware binaries~\cite{costin2014large}. The absence of JTAG debug ports, secure boot enforcement, and locked fuse bits further prevent binary extraction on modern microcontrollers. This creates a fundamental challenge: while traditional high-level validation techniques and gray-box fuzzing rely on internal branch coverage to evaluate performance, such metrics are inaccessible when the internal design, source code, or memory interfaces are proprietary. Furthermore, the inherent constraints of embedded systems, such as limited I/O bandwidth and restricted computational power severely limit the runtime feedback available during fuzzing. Therefore, the absence of effective fuzzing techniques for fully black-box firmware, constitutes a critical and largely unaddressed problem in embedded systems security.

To design effective fuzzing in a black-box setting, two fundamental challenges must be overcome. First, the feedback loop that drives coverage-guided mutation must be reconstructed without any visibility into the firmware binary. Second, the fuzzer must develop behavioral insight into the firmware's internal execution to generate semantically meaningful inputs that exercise diverse code paths rather than relying on purely random mutation, which is well-known to be inefficient for state-dependent firmware~\cite{natella2022stateafl}. These requirements are in direct conflict with the black-box constraint: a black-box tester lacks meaningful feedback and behavioral insight when no binary access or debugging infrastructure is available. Any viable solution must therefore derive equivalent information from external observables of the device under test, without modifying or instrumenting the firmware itself.

In this paper, we address this challenge through a novel insight. The power consumption of a microcontroller unit (MCU) is an externally observable side-channel that encodes fine-grained information about its internal execution. Power traces vary measurably as a function of the instructions executed, data operated upon, and branches taken within the firmware. This physical phenomenon, long exploited in cryptographic side-channel attacks~\cite{kocher1999differential}, has not previously been leveraged as a fuzzing feedback mechanism. We present \textbf{PowerFuzz}, the first power-based black-box firmware fuzzing framework. Figure~\ref{fig:powerfuzz_overview} shows the PowerFuzz framework, which operates under the same fuzzing paradigm but in a fully black-box setting replacing binary-level branch coverage with our power-based branch coverage derived from physical side-channel measurements. PowerFuzz integrates an AFL-based input generation engine with a real-time power trace acquisition pipeline. When the fuzzer dispatches an input to the target device, a synchronized acquisition system captures the resulting power trace as a time-series signal. We introduce a trace comparison algorithm that identifies execution divergences between power traces, enabling us to dynamically construct a novel tree structure, referred to as \textit{Trace-guided Control Flow Graph} (TCFG). We make an explicit distinction between CFG and TCFG since we cannot guarantee that TCFG would be isomorphic to CFG unless the underlying fuzzer is able to cover all the branches. TCFG is a hierarchical representation of the firmware's execution behavior built entirely from power traces based on side-channel measurements, without any binary access. Unlike a conventional Control Flow Graph (CFG) derived from static binary analysis, TCFG is constructed incrementally at runtime: each node stores a power trace segment corresponding to an identified basic block, and each edge represents a branch transition observed through trace divergence.

TCFG is iteratively refined during the fuzzing process, and serves as a structural proxy for branch coverage, feeding back into the fuzzer's mutation engine to prioritize inputs that explore new branches. This architectural correspondence demonstrates that PowerFuzz is an adaptation of coverage-guided fuzzing to address the firmware visibility constraints in embedded systems.

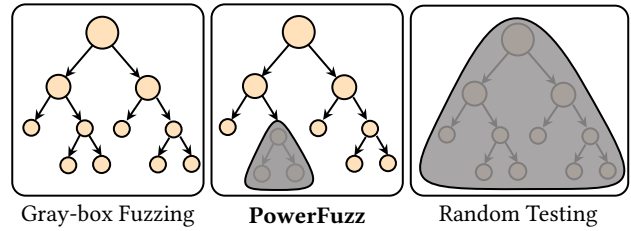
\begin{figure}[htp]
    \centering
    \vspace{-0.2in}
    \input{figs/cfg-overview}
    \vspace{-0.2in}
    \caption{Conceptual comparison of control flow awareness across firmware testing approaches.} 
    \label{fig:cfg-overview}
\end{figure}

Figure~\ref{fig:cfg-overview} contextualizes PowerFuzz within the broader landscape of firmware testing approaches by visualizing the degree of control flow awareness available to each method. In gray-box fuzzing (Figure~\ref{fig:cfg-overview}, left), binary instrumentation provides the fuzzer with complete structural knowledge of the firmware's control flow. The fuzzer can observe every branch taken, map the full control of the firmware, and exploit this precise structural knowledge to generate inputs that systematically drive execution into unexplored paths. At the opposite extreme, random testing (Figure~\ref{fig:cfg-overview}, right) operates without any knowledge  of the firmware's internal structure. The control flow is entirely opaque in case of random testing: inputs are generated without any structural guidance, and there is no mechanism to distinguish a mutation that reaches a new branch from one that redundantly re-executes an already-covered path. PowerFuzz occupies a middle ground between these two extremes (Figure~\ref{fig:cfg-overview}, center). Without access to the firmware binary, PowerFuzz cannot directly observe the control flow. However, PowerFuzz progressively gathers structural knowledge of the firmware as the fuzzing advances by dynamically constructing a power-trace-driven TCFG from power side-channel measurements. This partial-but-growing visibility is the defining characteristic of PowerFuzz.

In summary, this paper makes the following contributions:

\begin{itemize}
    \item We present the first power side-channel-based black-box firmware fuzzing framework. 
    \item We implement a branch identification method using captured power traces by incorporating a sliding-window analysis followed by a growing-window full-trace correlation technique along with dynamic time warping technique.
    \item We enable dynamic generation of trace-guided control flow graph without internal firmware visibility by accumulating and comparing power traces across the fuzzing iterations.
    \item Experimental evaluation using three embedded platforms and ten firmware benchmarks demonstrates that PowerFuzz can achieve fuzzing performance comparable to gray-box fuzzers and significantly outperform state-of-the-art black-box fuzzers based on electromagnetic (EM) emanations.
\end{itemize}

The remainder of this paper is organized as follows. Section~\ref{sec:background_related} provides relevant background and surveys related efforts. Section~\ref{sec:methodology} describes PowerFuzz methodology. Section~\ref{sec:experiments} presents experimental results. Finally, Section~\ref{sec:conclusion} concludes the paper.

%% file: figs/sota_overview.tex
\tikzset{every picture/.style={line width=0.75pt}} 

\begin{tikzpicture}[x=0.75pt,y=0.75pt,yscale=-1,xscale=1]

\draw  [color={rgb, 255:red, 0; green, 0; blue, 0 }  ,draw opacity=1 ][fill={rgb, 255:red, 252; green, 224; blue, 225 }  ,fill opacity=1 ] (62,76) -- (136,76) -- (136,112.42) .. controls (89.75,112.42) and (99,125.55) .. (62,117.05) -- cycle ;
\draw  [color={rgb, 255:red, 0; green, 0; blue, 0 }  ,draw opacity=1 ][fill={rgb, 255:red, 255; green, 255; blue, 255 }  ,fill opacity=1 ] (72,96.8) .. controls (72,95.25) and (73.25,94) .. (74.8,94) -- (123.2,94) .. controls (124.75,94) and (126,95.25) .. (126,96.8) -- (126,105.2) .. controls (126,106.75) and (124.75,108) .. (123.2,108) -- (74.8,108) .. controls (73.25,108) and (72,106.75) .. (72,105.2) -- cycle ;
\draw  [color={rgb, 255:red, 0; green, 0; blue, 0 }  ,draw opacity=1 ][fill={rgb, 255:red, 204; green, 231; blue, 207 }  ,fill opacity=1 ] (150,80.75) .. controls (150,78.13) and (152.13,76) .. (154.75,76) -- (203.25,76) .. controls (205.87,76) and (208,78.13) .. (208,80.75) -- (208,109.25) .. controls (208,111.87) and (205.87,114) .. (203.25,114) -- (154.75,114) .. controls (152.13,114) and (150,111.87) .. (150,109.25) -- cycle ;
\draw  [color={rgb, 255:red, 0; green, 0; blue, 0 }  ,draw opacity=1 ][fill={rgb, 255:red, 242; green, 244; blue, 193 }  ,fill opacity=1 ] (62,135.25) .. controls (62,132.35) and (64.35,130) .. (67.25,130) -- (130.75,130) .. controls (133.65,130) and (136,132.35) .. (136,135.25) -- (136,166.75) .. controls (136,169.65) and (133.65,172) .. (130.75,172) -- (67.25,172) .. controls (64.35,172) and (62,169.65) .. (62,166.75) -- cycle ;
\draw  [color={rgb, 255:red, 0; green, 0; blue, 0 }  ,draw opacity=1 ][fill={rgb, 255:red, 255; green, 226; blue, 187 }  ,fill opacity=1 ] (150,135.25) .. controls (150,132.35) and (152.35,130) .. (155.25,130) -- (202.75,130) .. controls (205.65,130) and (208,132.35) .. (208,135.25) -- (208,166.75) .. controls (208,169.65) and (205.65,172) .. (202.75,172) -- (155.25,172) .. controls (152.35,172) and (150,169.65) .. (150,166.75) -- cycle ;
\draw    (178,130) -- (178,117) ;
\draw [shift={(178,114)}, rotate = 90] [fill={rgb, 255:red, 0; green, 0; blue, 0 }  ][line width=0.08]  [draw opacity=0] (5.36,-2.57) -- (0,0) -- (5.36,2.57) -- cycle    ;
\draw    (150,88) -- (139,88) ;
\draw [shift={(136,88)}, rotate = 360] [fill={rgb, 255:red, 0; green, 0; blue, 0 }  ][line width=0.08]  [draw opacity=0] (5.36,-2.57) -- (0,0) -- (5.36,2.57) -- cycle    ;
\draw    (136,156) -- (147,156) ;
\draw [shift={(150,156)}, rotate = 180] [fill={rgb, 255:red, 0; green, 0; blue, 0 }  ][line width=0.08]  [draw opacity=0] (5.36,-2.57) -- (0,0) -- (5.36,2.57) -- cycle    ;
\draw    (100,116) -- (100,127) ;
\draw [shift={(100,130)}, rotate = 270] [fill={rgb, 255:red, 0; green, 0; blue, 0 }  ][line width=0.08]  [draw opacity=0] (5.36,-2.57) -- (0,0) -- (5.36,2.57) -- cycle    ;

\draw (99,85) node  [font=\small] [align=left] {\begin{minipage}[lt]{50.32pt}\setlength\topsep{0pt}
\begin{center}
\textbf{Firmware}
\end{center}

\end{minipage}};
\draw (99,101) node   [align=left] {\begin{minipage}[lt]{36.72pt}\setlength\topsep{0pt}
\begin{center}
{\scriptsize Full Visibility}
\end{center}

\end{minipage}};
\draw (179,95) node  [font=\small] [align=left] {\begin{minipage}[lt]{39.44pt}\setlength\topsep{0pt}
\begin{center}
\textbf{Inputs}
\end{center}

\end{minipage}};
\draw (99,151) node  [font=\small] [align=left] {\begin{minipage}[lt]{50.24pt}\setlength\topsep{0pt}
\begin{center}
\textbf{Branch}\\\textbf{Coverage}
\end{center}

\end{minipage}};
\draw (178.63,151.06) node  [font=\small] [align=left] {\begin{minipage}[lt]{39.95pt}\setlength\topsep{0pt}
\begin{center}
\textbf{Fuzzer}
\end{center}

\end{minipage}};

\end{tikzpicture}

%% file: figs/powerfuzz_overview.tex
\tikzset{every picture/.style={line width=0.75pt}} 

\begin{tikzpicture}[x=0.75pt,y=0.75pt,yscale=-1,xscale=1]

\draw  [color={rgb, 255:red, 0; green, 0; blue, 0 }  ,draw opacity=1 ][fill={rgb, 255:red, 252; green, 224; blue, 225 }  ,fill opacity=1 ] (82,96) -- (156,96) -- (156,132.42) .. controls (109.75,132.42) and (119,145.55) .. (82,137.05) -- cycle ;
\draw  [draw opacity=0][fill={rgb, 255:red, 0; green, 0; blue, 0 }  ,fill opacity=1 ] (92,116.8) .. controls (92,115.25) and (93.25,114) .. (94.8,114) -- (143.2,114) .. controls (144.75,114) and (146,115.25) .. (146,116.8) -- (146,125.2) .. controls (146,126.75) and (144.75,128) .. (143.2,128) -- (94.8,128) .. controls (93.25,128) and (92,126.75) .. (92,125.2) -- cycle ;
\draw  [color={rgb, 255:red, 0; green, 0; blue, 0 }  ,draw opacity=1 ][fill={rgb, 255:red, 204; green, 231; blue, 207 }  ,fill opacity=1 ] (170,100.75) .. controls (170,98.13) and (172.13,96) .. (174.75,96) -- (223.25,96) .. controls (225.87,96) and (228,98.13) .. (228,100.75) -- (228,129.25) .. controls (228,131.87) and (225.87,134) .. (223.25,134) -- (174.75,134) .. controls (172.13,134) and (170,131.87) .. (170,129.25) -- cycle ;
\draw  [color={rgb, 255:red, 0; green, 0; blue, 0 }  ,draw opacity=1 ][fill={rgb, 255:red, 242; green, 244; blue, 193 }  ,fill opacity=1 ] (82,155.25) .. controls (82,152.35) and (84.35,150) .. (87.25,150) -- (150.75,150) .. controls (153.65,150) and (156,152.35) .. (156,155.25) -- (156,186.75) .. controls (156,189.65) and (153.65,192) .. (150.75,192) -- (87.25,192) .. controls (84.35,192) and (82,189.65) .. (82,186.75) -- cycle ;
\draw  [color={rgb, 255:red, 0; green, 0; blue, 0 }  ,draw opacity=1 ][fill={rgb, 255:red, 255; green, 226; blue, 187 }  ,fill opacity=1 ] (170,155.25) .. controls (170,152.35) and (172.35,150) .. (175.25,150) -- (222.75,150) .. controls (225.65,150) and (228,152.35) .. (228,155.25) -- (228,186.75) .. controls (228,189.65) and (225.65,192) .. (222.75,192) -- (175.25,192) .. controls (172.35,192) and (170,189.65) .. (170,186.75) -- cycle ;
\draw    (198,150) -- (198,137) ;
\draw [shift={(198,134)}, rotate = 90] [fill={rgb, 255:red, 0; green, 0; blue, 0 }  ][line width=0.08]  [draw opacity=0] (5.36,-2.57) -- (0,0) -- (5.36,2.57) -- cycle    ;
\draw    (170,108) -- (159,108) ;
\draw [shift={(156,108)}, rotate = 360] [fill={rgb, 255:red, 0; green, 0; blue, 0 }  ][line width=0.08]  [draw opacity=0] (5.36,-2.57) -- (0,0) -- (5.36,2.57) -- cycle    ;
\draw    (156,176) -- (167,176) ;
\draw [shift={(170,176)}, rotate = 180] [fill={rgb, 255:red, 0; green, 0; blue, 0 }  ][line width=0.08]  [draw opacity=0] (5.36,-2.57) -- (0,0) -- (5.36,2.57) -- cycle    ;
\draw    (120,136) -- (120,147) ;
\draw [shift={(120,150)}, rotate = 270] [fill={rgb, 255:red, 0; green, 0; blue, 0 }  ][line width=0.08]  [draw opacity=0] (5.36,-2.57) -- (0,0) -- (5.36,2.57) -- cycle    ;

\draw (119,105) node  [font=\small] [align=left] {\begin{minipage}[lt]{50.32pt}\setlength\topsep{0pt}
\begin{center}
\textbf{Firmware}
\end{center}

\end{minipage}};
\draw (119,121) node   [align=left] {\begin{minipage}[lt]{36.72pt}\setlength\topsep{0pt}
\begin{center}
{\scriptsize \textcolor[rgb]{1,1,1}{No Visibility}}
\end{center}

\end{minipage}};
\draw (199,115) node  [font=\small] [align=left] {\begin{minipage}[lt]{39.44pt}\setlength\topsep{0pt}
\begin{center}
\textbf{Inputs}
\end{center}

\end{minipage}};
\draw (119,171) node  [font=\small] [align=left] {\begin{minipage}[lt]{50.24pt}\setlength\topsep{0pt}
\begin{center}
\textbf{Coverage using}\\\textbf{Power Traces}
\end{center}

\end{minipage}};
\draw (198.63,171.06) node  [font=\small] [align=left] {\begin{minipage}[lt]{39.95pt}\setlength\topsep{0pt}
\begin{center}
\textbf{Fuzzer}
\end{center}

\end{minipage}};

\end{tikzpicture}

%% file: figs/cfg-overview.tex
\tikzset{every picture/.style={line width=0.75pt}} 

\begin{tikzpicture}[x=0.75pt,y=0.75pt,yscale=-1,xscale=1]

\draw  [fill={rgb, 255:red, 255; green, 226; blue, 187 }  ,fill opacity=1 ] (201.57,134.43) .. controls (201.57,132.22) and (203.36,130.43) .. (205.57,130.43) .. controls (207.78,130.43) and (209.57,132.22) .. (209.57,134.43) .. controls (209.57,136.64) and (207.78,138.43) .. (205.57,138.43) .. controls (203.36,138.43) and (201.57,136.64) .. (201.57,134.43) -- cycle ;
\draw  [fill={rgb, 255:red, 255; green, 226; blue, 187 }  ,fill opacity=1 ] (193.29,153) .. controls (193.29,150.79) and (195.08,149) .. (197.29,149) .. controls (199.49,149) and (201.29,150.79) .. (201.29,153) .. controls (201.29,155.21) and (199.49,157) .. (197.29,157) .. controls (195.08,157) and (193.29,155.21) .. (193.29,153) -- cycle ;
\draw  [fill={rgb, 255:red, 255; green, 226; blue, 187 }  ,fill opacity=1 ] (209.71,152.57) .. controls (209.71,150.2) and (211.63,148.29) .. (214,148.29) .. controls (216.37,148.29) and (218.29,150.2) .. (218.29,152.57) .. controls (218.29,154.94) and (216.37,156.86) .. (214,156.86) .. controls (211.63,156.86) and (209.71,154.94) .. (209.71,152.57) -- cycle ;
\draw    (207.71,138) -- (212.44,145.73) ;
\draw [shift={(214,148.29)}, rotate = 238.57] [fill={rgb, 255:red, 0; green, 0; blue, 0 }  ][line width=0.08]  [draw opacity=0] (5.36,-2.57) -- (0,0) -- (5.36,2.57) -- (3.56,0) -- cycle    ;
\draw    (203.14,137.71) -- (198.67,146.34) ;
\draw [shift={(197.29,149)}, rotate = 297.43] [fill={rgb, 255:red, 0; green, 0; blue, 0 }  ][line width=0.08]  [draw opacity=0] (5.36,-2.57) -- (0,0) -- (5.36,2.57) -- (3.56,0) -- cycle    ;
\draw  [fill={rgb, 255:red, 255; green, 226; blue, 187 }  ,fill opacity=1 ] (207.79,81.86) .. controls (207.79,77.44) and (211.37,73.86) .. (215.79,73.86) .. controls (220.2,73.86) and (223.79,77.44) .. (223.79,81.86) .. controls (223.79,86.28) and (220.2,89.86) .. (215.79,89.86) .. controls (211.37,89.86) and (207.79,86.28) .. (207.79,81.86) -- cycle ;
\draw  [fill={rgb, 255:red, 255; green, 226; blue, 187 }  ,fill opacity=1 ] (187.79,109) .. controls (187.79,105.69) and (190.47,103) .. (193.79,103) .. controls (197.1,103) and (199.79,105.69) .. (199.79,109) .. controls (199.79,112.31) and (197.1,115) .. (193.79,115) .. controls (190.47,115) and (187.79,112.31) .. (187.79,109) -- cycle ;
\draw  [fill={rgb, 255:red, 255; green, 226; blue, 187 }  ,fill opacity=1 ] (232.64,109.57) .. controls (232.64,106.26) and (235.33,103.57) .. (238.64,103.57) .. controls (241.96,103.57) and (244.64,106.26) .. (244.64,109.57) .. controls (244.64,112.89) and (241.96,115.57) .. (238.64,115.57) .. controls (235.33,115.57) and (232.64,112.89) .. (232.64,109.57) -- cycle ;
\draw    (211.14,88) -- (199.1,102.01) ;
\draw [shift={(197.14,104.29)}, rotate = 310.68] [fill={rgb, 255:red, 0; green, 0; blue, 0 }  ][line width=0.08]  [draw opacity=0] (5.36,-2.57) -- (0,0) -- (5.36,2.57) -- (3.56,0) -- cycle    ;
\draw    (220.71,88.43) -- (232.76,102.44) ;
\draw [shift={(234.71,104.71)}, rotate = 229.32] [fill={rgb, 255:red, 0; green, 0; blue, 0 }  ][line width=0.08]  [draw opacity=0] (5.36,-2.57) -- (0,0) -- (5.36,2.57) -- (3.56,0) -- cycle    ;
\draw  [fill={rgb, 255:red, 255; green, 226; blue, 187 }  ,fill opacity=1 ] (176.14,129.57) .. controls (176.14,127.36) and (177.93,125.57) .. (180.14,125.57) .. controls (182.35,125.57) and (184.14,127.36) .. (184.14,129.57) .. controls (184.14,131.78) and (182.35,133.57) .. (180.14,133.57) .. controls (177.93,133.57) and (176.14,131.78) .. (176.14,129.57) -- cycle ;
\draw  [fill={rgb, 255:red, 255; green, 226; blue, 187 }  ,fill opacity=1 ] (221.57,130.14) .. controls (221.57,127.93) and (223.36,126.14) .. (225.57,126.14) .. controls (227.78,126.14) and (229.57,127.93) .. (229.57,130.14) .. controls (229.57,132.35) and (227.78,134.14) .. (225.57,134.14) .. controls (223.36,134.14) and (221.57,132.35) .. (221.57,130.14) -- cycle ;
\draw  [fill={rgb, 255:red, 255; green, 226; blue, 187 }  ,fill opacity=1 ] (247.57,130.43) .. controls (247.57,128.22) and (249.36,126.43) .. (251.57,126.43) .. controls (253.78,126.43) and (255.57,128.22) .. (255.57,130.43) .. controls (255.57,132.64) and (253.78,134.43) .. (251.57,134.43) .. controls (249.36,134.43) and (247.57,132.64) .. (247.57,130.43) -- cycle ;
\draw    (190,113.14) -- (183.58,123.45) ;
\draw [shift={(182,126)}, rotate = 301.89] [fill={rgb, 255:red, 0; green, 0; blue, 0 }  ][line width=0.08]  [draw opacity=0] (5.36,-2.57) -- (0,0) -- (5.36,2.57) -- (3.56,0) -- cycle    ;
\draw    (197.43,113.71) -- (203.71,123.88) ;
\draw [shift={(205.29,126.43)}, rotate = 238.28] [fill={rgb, 255:red, 0; green, 0; blue, 0 }  ][line width=0.08]  [draw opacity=0] (5.36,-2.57) -- (0,0) -- (5.36,2.57) -- (3.56,0) -- cycle    ;
\draw    (235.43,115.14) -- (229.69,123.52) ;
\draw [shift={(228,126)}, rotate = 304.38] [fill={rgb, 255:red, 0; green, 0; blue, 0 }  ][line width=0.08]  [draw opacity=0] (5.36,-2.57) -- (0,0) -- (5.36,2.57) -- (3.56,0) -- cycle    ;
\draw    (243.14,114.57) -- (248.46,123.43) ;
\draw [shift={(250,126)}, rotate = 239.04] [fill={rgb, 255:red, 0; green, 0; blue, 0 }  ][line width=0.08]  [draw opacity=0] (5.36,-2.57) -- (0,0) -- (5.36,2.57) -- (3.56,0) -- cycle    ;
\draw  [fill={rgb, 255:red, 255; green, 226; blue, 187 }  ,fill opacity=1 ] (239.71,148) .. controls (239.71,145.79) and (241.51,144) .. (243.71,144) .. controls (245.92,144) and (247.71,145.79) .. (247.71,148) .. controls (247.71,150.21) and (245.92,152) .. (243.71,152) .. controls (241.51,152) and (239.71,150.21) .. (239.71,148) -- cycle ;
\draw  [fill={rgb, 255:red, 255; green, 226; blue, 187 }  ,fill opacity=1 ] (256.57,147.71) .. controls (256.57,145.51) and (258.36,143.71) .. (260.57,143.71) .. controls (262.78,143.71) and (264.57,145.51) .. (264.57,147.71) .. controls (264.57,149.92) and (262.78,151.71) .. (260.57,151.71) .. controls (258.36,151.71) and (256.57,149.92) .. (256.57,147.71) -- cycle ;
\draw    (249.43,134.57) -- (245.27,141.43) ;
\draw [shift={(243.71,144)}, rotate = 301.22] [fill={rgb, 255:red, 0; green, 0; blue, 0 }  ][line width=0.08]  [draw opacity=0] (5.36,-2.57) -- (0,0) -- (5.36,2.57) -- (3.56,0) -- cycle    ;
\draw    (254.86,134) -- (259.05,141.13) ;
\draw [shift={(260.57,143.71)}, rotate = 239.53] [fill={rgb, 255:red, 0; green, 0; blue, 0 }  ][line width=0.08]  [draw opacity=0] (5.36,-2.57) -- (0,0) -- (5.36,2.57) -- (3.56,0) -- cycle    ;
\draw   (172,74.74) .. controls (172,71.02) and (175.02,68) .. (178.74,68) -- (261.26,68) .. controls (264.98,68) and (268,71.02) .. (268,74.74) -- (268,157.26) .. controls (268,160.98) and (264.98,164) .. (261.26,164) -- (178.74,164) .. controls (175.02,164) and (172,160.98) .. (172,157.26) -- cycle ;
\draw  [fill={rgb, 255:red, 255; green, 226; blue, 187 }  ,fill opacity=1 ] (109.2,82.2) .. controls (109.2,77.78) and (112.78,74.2) .. (117.2,74.2) .. controls (121.62,74.2) and (125.2,77.78) .. (125.2,82.2) .. controls (125.2,86.62) and (121.62,90.2) .. (117.2,90.2) .. controls (112.78,90.2) and (109.2,86.62) .. (109.2,82.2) -- cycle ;
\draw  [fill={rgb, 255:red, 255; green, 226; blue, 187 }  ,fill opacity=1 ] (89.2,109.35) .. controls (89.2,106.03) and (91.89,103.35) .. (95.2,103.35) .. controls (98.52,103.35) and (101.2,106.03) .. (101.2,109.35) .. controls (101.2,112.66) and (98.52,115.35) .. (95.2,115.35) .. controls (91.89,115.35) and (89.2,112.66) .. (89.2,109.35) -- cycle ;
\draw  [fill={rgb, 255:red, 255; green, 226; blue, 187 }  ,fill opacity=1 ] (134.06,109.92) .. controls (134.06,106.6) and (136.75,103.92) .. (140.06,103.92) .. controls (143.37,103.92) and (146.06,106.6) .. (146.06,109.92) .. controls (146.06,113.23) and (143.37,115.92) .. (140.06,115.92) .. controls (136.75,115.92) and (134.06,113.23) .. (134.06,109.92) -- cycle ;
\draw    (112.56,88.35) -- (100.52,102.36) ;
\draw [shift={(98.56,104.63)}, rotate = 310.68] [fill={rgb, 255:red, 0; green, 0; blue, 0 }  ][line width=0.08]  [draw opacity=0] (5.36,-2.57) -- (0,0) -- (5.36,2.57) -- (3.56,0) -- cycle    ;
\draw    (122.13,88.77) -- (134.18,102.78) ;
\draw [shift={(136.13,105.06)}, rotate = 229.32] [fill={rgb, 255:red, 0; green, 0; blue, 0 }  ][line width=0.08]  [draw opacity=0] (5.36,-2.57) -- (0,0) -- (5.36,2.57) -- (3.56,0) -- cycle    ;
\draw  [fill={rgb, 255:red, 255; green, 226; blue, 187 }  ,fill opacity=1 ] (77.56,129.92) .. controls (77.56,127.71) and (79.35,125.92) .. (81.56,125.92) .. controls (83.77,125.92) and (85.56,127.71) .. (85.56,129.92) .. controls (85.56,132.13) and (83.77,133.92) .. (81.56,133.92) .. controls (79.35,133.92) and (77.56,132.13) .. (77.56,129.92) -- cycle ;
\draw  [fill={rgb, 255:red, 255; green, 226; blue, 187 }  ,fill opacity=1 ] (104.42,130.2) .. controls (104.42,127.99) and (106.21,126.2) .. (108.42,126.2) .. controls (110.63,126.2) and (112.42,127.99) .. (112.42,130.2) .. controls (112.42,132.41) and (110.63,134.2) .. (108.42,134.2) .. controls (106.21,134.2) and (104.42,132.41) .. (104.42,130.2) -- cycle ;
\draw  [fill={rgb, 255:red, 255; green, 226; blue, 187 }  ,fill opacity=1 ] (122.99,130.49) .. controls (122.99,128.28) and (124.78,126.49) .. (126.99,126.49) .. controls (129.2,126.49) and (130.99,128.28) .. (130.99,130.49) .. controls (130.99,132.7) and (129.2,134.49) .. (126.99,134.49) .. controls (124.78,134.49) and (122.99,132.7) .. (122.99,130.49) -- cycle ;
\draw  [fill={rgb, 255:red, 255; green, 226; blue, 187 }  ,fill opacity=1 ] (148.99,130.77) .. controls (148.99,128.56) and (150.78,126.77) .. (152.99,126.77) .. controls (155.2,126.77) and (156.99,128.56) .. (156.99,130.77) .. controls (156.99,132.98) and (155.2,134.77) .. (152.99,134.77) .. controls (150.78,134.77) and (148.99,132.98) .. (148.99,130.77) -- cycle ;
\draw    (91.42,113.49) -- (85,123.8) ;
\draw [shift={(83.42,126.35)}, rotate = 301.89] [fill={rgb, 255:red, 0; green, 0; blue, 0 }  ][line width=0.08]  [draw opacity=0] (5.36,-2.57) -- (0,0) -- (5.36,2.57) -- (3.56,0) -- cycle    ;
\draw    (98.85,114.06) -- (105.13,124.22) ;
\draw [shift={(106.7,126.77)}, rotate = 238.28] [fill={rgb, 255:red, 0; green, 0; blue, 0 }  ][line width=0.08]  [draw opacity=0] (5.36,-2.57) -- (0,0) -- (5.36,2.57) -- (3.56,0) -- cycle    ;
\draw    (136.85,115.49) -- (131.11,123.87) ;
\draw [shift={(129.42,126.35)}, rotate = 304.38] [fill={rgb, 255:red, 0; green, 0; blue, 0 }  ][line width=0.08]  [draw opacity=0] (5.36,-2.57) -- (0,0) -- (5.36,2.57) -- (3.56,0) -- cycle    ;
\draw    (144.56,114.92) -- (149.87,123.77) ;
\draw [shift={(151.42,126.35)}, rotate = 239.04] [fill={rgb, 255:red, 0; green, 0; blue, 0 }  ][line width=0.08]  [draw opacity=0] (5.36,-2.57) -- (0,0) -- (5.36,2.57) -- (3.56,0) -- cycle    ;
\draw  [fill={rgb, 255:red, 255; green, 226; blue, 187 }  ,fill opacity=1 ] (96.13,148.77) .. controls (96.13,146.56) and (97.92,144.77) .. (100.13,144.77) .. controls (102.34,144.77) and (104.13,146.56) .. (104.13,148.77) .. controls (104.13,150.98) and (102.34,152.77) .. (100.13,152.77) .. controls (97.92,152.77) and (96.13,150.98) .. (96.13,148.77) -- cycle ;
\draw  [fill={rgb, 255:red, 255; green, 226; blue, 187 }  ,fill opacity=1 ] (112.56,148.35) .. controls (112.56,145.98) and (114.48,144.06) .. (116.85,144.06) .. controls (119.21,144.06) and (121.13,145.98) .. (121.13,148.35) .. controls (121.13,150.71) and (119.21,152.63) .. (116.85,152.63) .. controls (114.48,152.63) and (112.56,150.71) .. (112.56,148.35) -- cycle ;
\draw  [fill={rgb, 255:red, 255; green, 226; blue, 187 }  ,fill opacity=1 ] (141.13,148.35) .. controls (141.13,146.14) and (142.92,144.35) .. (145.13,144.35) .. controls (147.34,144.35) and (149.13,146.14) .. (149.13,148.35) .. controls (149.13,150.55) and (147.34,152.35) .. (145.13,152.35) .. controls (142.92,152.35) and (141.13,150.55) .. (141.13,148.35) -- cycle ;
\draw  [fill={rgb, 255:red, 255; green, 226; blue, 187 }  ,fill opacity=1 ] (157.99,148.06) .. controls (157.99,145.85) and (159.78,144.06) .. (161.99,144.06) .. controls (164.2,144.06) and (165.99,145.85) .. (165.99,148.06) .. controls (165.99,150.27) and (164.2,152.06) .. (161.99,152.06) .. controls (159.78,152.06) and (157.99,150.27) .. (157.99,148.06) -- cycle ;
\draw    (110.56,133.77) -- (115.28,141.5) ;
\draw [shift={(116.85,144.06)}, rotate = 238.57] [fill={rgb, 255:red, 0; green, 0; blue, 0 }  ][line width=0.08]  [draw opacity=0] (5.36,-2.57) -- (0,0) -- (5.36,2.57) -- (3.56,0) -- cycle    ;
\draw    (105.99,133.49) -- (101.51,142.11) ;
\draw [shift={(100.13,144.77)}, rotate = 297.43] [fill={rgb, 255:red, 0; green, 0; blue, 0 }  ][line width=0.08]  [draw opacity=0] (5.36,-2.57) -- (0,0) -- (5.36,2.57) -- (3.56,0) -- cycle    ;
\draw    (150.85,134.92) -- (146.69,141.78) ;
\draw [shift={(145.13,144.35)}, rotate = 301.22] [fill={rgb, 255:red, 0; green, 0; blue, 0 }  ][line width=0.08]  [draw opacity=0] (5.36,-2.57) -- (0,0) -- (5.36,2.57) -- (3.56,0) -- cycle    ;
\draw    (156.27,134.35) -- (160.47,141.47) ;
\draw [shift={(161.99,144.06)}, rotate = 239.53] [fill={rgb, 255:red, 0; green, 0; blue, 0 }  ][line width=0.08]  [draw opacity=0] (5.36,-2.57) -- (0,0) -- (5.36,2.57) -- (3.56,0) -- cycle    ;
\draw   (72,74.74) .. controls (72,71.02) and (75.02,68) .. (78.74,68) -- (161.26,68) .. controls (164.98,68) and (168,71.02) .. (168,74.74) -- (168,157.26) .. controls (168,160.98) and (164.98,164) .. (161.26,164) -- (78.74,164) .. controls (75.02,164) and (72,160.98) .. (72,157.26) -- cycle ;
\draw  [fill={rgb, 255:red, 255; green, 226; blue, 187 }  ,fill opacity=1 ] (317.74,85.64) .. controls (317.74,81.22) and (321.32,77.64) .. (325.74,77.64) .. controls (330.16,77.64) and (333.74,81.22) .. (333.74,85.64) .. controls (333.74,90.06) and (330.16,93.64) .. (325.74,93.64) .. controls (321.32,93.64) and (317.74,90.06) .. (317.74,85.64) -- cycle ;
\draw  [fill={rgb, 255:red, 255; green, 226; blue, 187 }  ,fill opacity=1 ] (298.02,112.78) .. controls (298.02,109.47) and (300.71,106.78) .. (304.02,106.78) .. controls (307.34,106.78) and (310.02,109.47) .. (310.02,112.78) .. controls (310.02,116.1) and (307.34,118.78) .. (304.02,118.78) .. controls (300.71,118.78) and (298.02,116.1) .. (298.02,112.78) -- cycle ;
\draw  [fill={rgb, 255:red, 255; green, 226; blue, 187 }  ,fill opacity=1 ] (342.6,113.35) .. controls (342.6,110.04) and (345.28,107.35) .. (348.6,107.35) .. controls (351.91,107.35) and (354.6,110.04) .. (354.6,113.35) .. controls (354.6,116.67) and (351.91,119.35) .. (348.6,119.35) .. controls (345.28,119.35) and (342.6,116.67) .. (342.6,113.35) -- cycle ;
\draw    (321.43,91.78) -- (309.38,105.79) ;
\draw [shift={(307.43,108.07)}, rotate = 310.68] [fill={rgb, 255:red, 0; green, 0; blue, 0 }  ][line width=0.08]  [draw opacity=0] (5.36,-2.57) -- (0,0) -- (5.36,2.57) -- (3.56,0) -- cycle    ;
\draw    (331,92.21) -- (343.04,106.22) ;
\draw [shift={(345,108.5)}, rotate = 229.32] [fill={rgb, 255:red, 0; green, 0; blue, 0 }  ][line width=0.08]  [draw opacity=0] (5.36,-2.57) -- (0,0) -- (5.36,2.57) -- (3.56,0) -- cycle    ;
\draw  [fill={rgb, 255:red, 255; green, 226; blue, 187 }  ,fill opacity=1 ] (286.1,133.35) .. controls (286.1,131.14) and (287.89,129.35) .. (290.1,129.35) .. controls (292.3,129.35) and (294.1,131.14) .. (294.1,133.35) .. controls (294.1,135.56) and (292.3,137.35) .. (290.1,137.35) .. controls (287.89,137.35) and (286.1,135.56) .. (286.1,133.35) -- cycle ;
\draw  [fill={rgb, 255:red, 255; green, 226; blue, 187 }  ,fill opacity=1 ] (312.95,133.64) .. controls (312.95,131.43) and (314.74,129.64) .. (316.95,129.64) .. controls (319.16,129.64) and (320.95,131.43) .. (320.95,133.64) .. controls (320.95,135.85) and (319.16,137.64) .. (316.95,137.64) .. controls (314.74,137.64) and (312.95,135.85) .. (312.95,133.64) -- cycle ;
\draw  [fill={rgb, 255:red, 255; green, 226; blue, 187 }  ,fill opacity=1 ] (331.86,133.93) .. controls (331.86,131.72) and (333.65,129.93) .. (335.86,129.93) .. controls (338.07,129.93) and (339.86,131.72) .. (339.86,133.93) .. controls (339.86,136.13) and (338.07,137.93) .. (335.86,137.93) .. controls (333.65,137.93) and (331.86,136.13) .. (331.86,133.93) -- cycle ;
\draw  [fill={rgb, 255:red, 255; green, 226; blue, 187 }  ,fill opacity=1 ] (358.14,134.21) .. controls (358.14,132) and (359.93,130.21) .. (362.14,130.21) .. controls (364.35,130.21) and (366.14,132) .. (366.14,134.21) .. controls (366.14,136.42) and (364.35,138.21) .. (362.14,138.21) .. controls (359.93,138.21) and (358.14,136.42) .. (358.14,134.21) -- cycle ;
\draw    (300.29,116.93) -- (293.87,127.24) ;
\draw [shift={(292.29,129.78)}, rotate = 301.89] [fill={rgb, 255:red, 0; green, 0; blue, 0 }  ][line width=0.08]  [draw opacity=0] (5.36,-2.57) -- (0,0) -- (5.36,2.57) -- (3.56,0) -- cycle    ;
\draw    (307.71,117.5) -- (313.99,127.66) ;
\draw [shift={(315.57,130.21)}, rotate = 238.28] [fill={rgb, 255:red, 0; green, 0; blue, 0 }  ][line width=0.08]  [draw opacity=0] (5.36,-2.57) -- (0,0) -- (5.36,2.57) -- (3.56,0) -- cycle    ;
\draw    (345.71,118.93) -- (339.98,127.31) ;
\draw [shift={(338.29,129.78)}, rotate = 304.38] [fill={rgb, 255:red, 0; green, 0; blue, 0 }  ][line width=0.08]  [draw opacity=0] (5.36,-2.57) -- (0,0) -- (5.36,2.57) -- (3.56,0) -- cycle    ;
\draw    (353.43,118.35) -- (358.74,127.21) ;
\draw [shift={(360.29,129.78)}, rotate = 239.04] [fill={rgb, 255:red, 0; green, 0; blue, 0 }  ][line width=0.08]  [draw opacity=0] (5.36,-2.57) -- (0,0) -- (5.36,2.57) -- (3.56,0) -- cycle    ;
\draw  [fill={rgb, 255:red, 255; green, 226; blue, 187 }  ,fill opacity=1 ] (304.67,152.21) .. controls (304.67,150) and (306.46,148.21) .. (308.67,148.21) .. controls (310.88,148.21) and (312.67,150) .. (312.67,152.21) .. controls (312.67,154.42) and (310.88,156.21) .. (308.67,156.21) .. controls (306.46,156.21) and (304.67,154.42) .. (304.67,152.21) -- cycle ;
\draw  [fill={rgb, 255:red, 255; green, 226; blue, 187 }  ,fill opacity=1 ] (321.43,151.78) .. controls (321.43,149.42) and (323.35,147.5) .. (325.71,147.5) .. controls (328.08,147.5) and (330,149.42) .. (330,151.78) .. controls (330,154.15) and (328.08,156.07) .. (325.71,156.07) .. controls (323.35,156.07) and (321.43,154.15) .. (321.43,151.78) -- cycle ;
\draw  [fill={rgb, 255:red, 255; green, 226; blue, 187 }  ,fill opacity=1 ] (350,151.78) .. controls (350,149.57) and (351.79,147.78) .. (354,147.78) .. controls (356.21,147.78) and (358,149.57) .. (358,151.78) .. controls (358,153.99) and (356.21,155.78) .. (354,155.78) .. controls (351.79,155.78) and (350,153.99) .. (350,151.78) -- cycle ;
\draw  [fill={rgb, 255:red, 255; green, 226; blue, 187 }  ,fill opacity=1 ] (366.86,151.5) .. controls (366.86,149.29) and (368.65,147.5) .. (370.86,147.5) .. controls (373.07,147.5) and (374.86,149.29) .. (374.86,151.5) .. controls (374.86,153.71) and (373.07,155.5) .. (370.86,155.5) .. controls (368.65,155.5) and (366.86,153.71) .. (366.86,151.5) -- cycle ;
\draw    (319.43,137.21) -- (324.15,144.94) ;
\draw [shift={(325.71,147.5)}, rotate = 238.57] [fill={rgb, 255:red, 0; green, 0; blue, 0 }  ][line width=0.08]  [draw opacity=0] (5.36,-2.57) -- (0,0) -- (5.36,2.57) -- (3.56,0) -- cycle    ;
\draw    (314.86,136.93) -- (310.38,145.55) ;
\draw [shift={(309,148.21)}, rotate = 297.43] [fill={rgb, 255:red, 0; green, 0; blue, 0 }  ][line width=0.08]  [draw opacity=0] (5.36,-2.57) -- (0,0) -- (5.36,2.57) -- (3.56,0) -- cycle    ;
\draw    (359.71,138.35) -- (355.55,145.22) ;
\draw [shift={(354,147.78)}, rotate = 301.22] [fill={rgb, 255:red, 0; green, 0; blue, 0 }  ][line width=0.08]  [draw opacity=0] (5.36,-2.57) -- (0,0) -- (5.36,2.57) -- (3.56,0) -- cycle    ;
\draw    (365.14,137.78) -- (369.34,144.91) ;
\draw [shift={(370.86,147.5)}, rotate = 239.53] [fill={rgb, 255:red, 0; green, 0; blue, 0 }  ][line width=0.08]  [draw opacity=0] (5.36,-2.57) -- (0,0) -- (5.36,2.57) -- (3.56,0) -- cycle    ;
\draw   (272,75.33) .. controls (272,71.64) and (275,68.64) .. (278.69,68.64) -- (374.45,68.64) .. controls (378.15,68.64) and (381.14,71.64) .. (381.14,75.33) -- (381.14,157.31) .. controls (381.14,161) and (378.15,164) .. (374.45,164) -- (278.69,164) .. controls (275,164) and (272,161) .. (272,157.31) -- cycle ;
\draw  [fill={rgb, 255:red, 152; green, 149; blue, 149 }  ,fill opacity=0.84 ] (288.57,119.83) .. controls (308.57,79.83) and (324.86,49.12) .. (358,106.83) .. controls (391.14,164.55) and (388.29,160.4) .. (328.57,160.69) .. controls (268.86,160.97) and (268.57,159.83) .. (288.57,119.83) -- cycle ;
\draw  [fill={rgb, 255:red, 152; green, 149; blue, 149 }  ,fill opacity=0.89 ] (194.67,141.57) .. controls (203.95,122.29) and (205.24,121.29) .. (215.95,139.71) .. controls (226.67,158.14) and (226.52,161.14) .. (206.24,161.14) .. controls (185.95,161.14) and (185.38,160.86) .. (194.67,141.57) -- cycle ;

\draw (120,174) node   [align=left] {\begin{minipage}[lt]{65.28pt}\setlength\topsep{0pt}
\begin{center}
Gray-box Fuzzing
\end{center}

\end{minipage}};
\draw (220,174) node   [align=left] {\begin{minipage}[lt]{65.28pt}\setlength\topsep{0pt}
\begin{center}
\textbf{PowerFuzz}
\end{center}

\end{minipage}};
\draw (326,174) node   [align=left] {\begin{minipage}[lt]{73.44pt}\setlength\topsep{0pt}
\begin{center}
Random Testing
\end{center}

\end{minipage}};

\end{tikzpicture}

%% file: sections/2_related.tex
\section{Background and Related Work}
\label{sec:background_related}

In this section, we first provide an overview of coverage-guided fuzzing and dynamic time warping. Next, we survey related efforts and their limitations to highlight the need for the proposed work.

\subsection{Coverage-guided Fuzzing}\label{sec:background_fuzzing}

Coverage-based fuzzing is a dynamic testing approach that guides input generation using lightweight instrumentation to monitor which execution paths are triggered. Inputs that reach previously unseen basic blocks or branch edges in the control-flow graph (CFG) are retained in a corpus and prioritized for further mutation, while redundant inputs are discarded. This feedback-driven loop allows the fuzzer to systematically expand its exploration of the program's state space. AFL \cite{zalewski_afl_whitepaper} maintains a compact bitmap of edge hit counts and uses it to detect coverage-increasing inputs efficiently.

\begin{figure}[htp]
    \centering
    \input{figs/libafl}
    \caption{Block diagram of the LibAFL fuzzing framework, illustrating the coverage-guided fuzzing loop.}
    \label{fig:libafl}
\end{figure}
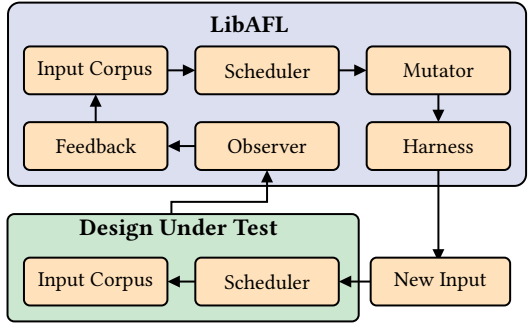

LibAFL \cite{fioraldi2022libafl} is a modular and extensible fuzzing framework that decomposes the coverage-guided fuzzing loop into a set of well-defined, reusable components. 
As shown in Figure~\ref{fig:libafl}, a LibAFL-based fuzzer operates as follows. The fuzzer maintains an \textit{input corpus}, a set of test cases selected for their ability to exercise distinct program behaviors. A \textit{scheduler} selects the next corpus entry to fuzz based on a prioritization policy, such as favoring inputs that cover rarely-reached paths. A \textit{mutator} then applies a sequence of transformations to the selected input, producing a new test case through operations such as bit flips, byte substitutions, or block insertions. The mutated input is delivered to the target through a \textit{harness}. During execution, \textit{observer} collects runtime signals from the target. In standard software fuzzing, these signals are coverage bitmaps produced by compile-time or binary-level instrumentation. The observations are passed to a \textit{feedback} module that determines whether the input has improved coverage; if so, it is added to the corpus for future mutation. This cycle repeats continuously, with the corpus growing as new paths are discovered and the fuzzer progressively penetrating deeper regions of the program. 

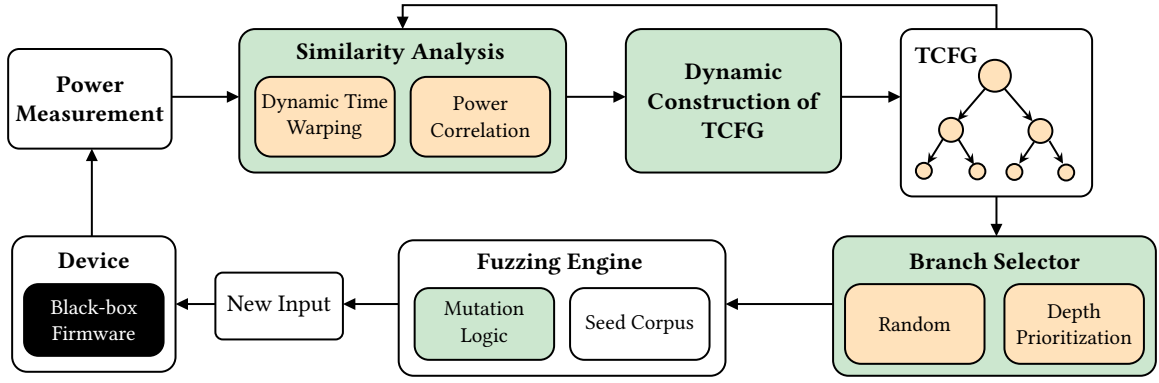
\begin{figure*}[!ht]
    \centering
    \input{figs/powerfuzz}
    \caption{Overview of PowerFuzz framework. After power measurement (trace acquisition), it performs similarity analysis using dynamic time warping and power correlation to detect execution divergences, which leads to the construction of the trace-guided control flow graph (TCFG) of the black-box firmware. A branch selector consults the TCFG to identify unexplored paths and guides the fuzzing engine's mutation logic to generate new inputs. Our contributions are highlighted in \setlength{\fboxsep}{0pt}\fbox{\textcolor[HTML]{AEF5B2}{$\blacksquare$}} color.}
    \label{fig:framework}
\end{figure*}

\subsection{Dynamic Time Warping}\label{sec:dtw}

Power traces captured from a microcontroller are time-series signals that encode the sequence of instructions executed by the firmware. Comparing two such traces to determine whether they reflect the same or different execution paths is therefore a problem of time-series similarity measurement. Naive point-by-point comparison using Euclidean distance is ill-suited for this task, as minor timing variations introduced by measurement jitter cause traces from identical executions to appear misaligned in the time domain. Dynamic Time Warping (DTW) \cite{sakoe2003dynamic} addresses this by computing an elastic alignment between two sequences that minimizes total distance while allowing non-linear stretching along the time axis, making it robust to such temporal perturbations.

Given two time-series sequences $Q = \langle q_1, q_2, \ldots, q_m \rangle$ and $C = \langle c_1, c_2, \ldots, c_n \rangle$, DTW constructs an $m \times n$ cost matrix $D$ where each entry $D(i, j)$ represents the cumulative alignment cost of matching the first $i$ elements of $Q$ with the first $j$ elements of $C$. The matrix is populated recursively as:

$$D(i,j) = d(q_i, c_j) + \min \begin{cases} D(i-1,\; j) \\ D(i,\; j-1) \\ D(i-1,\; j-1) \end{cases}$$ where $d(q_i,c_j)$ is the local distance between the $i$-th sample of trace $Q$ and the $j$-th sample of trace $C$.

The optimal warping path $\mathcal{W} = \langle w_1, w_2, \ldots, w_K \rangle$, where $w_k = (i_k, j_k)$, is then recovered by backtracking from $D(m, n)$ to $D(1, 1)$. This path represents the optimal elastic alignment between the two traces. Then for each step $w_k=(i_k,j_k)$ along the warping path, the $\delta_k=j_k-i_k$ is calculated, which quantifies how far the alignment has stretched or compressed at that point. The sequence of offsets $\Delta = \langle \delta_1, \delta_2, \ldots, \delta_K \rangle$ forms an offset profile that reflects the local temporal relationship between the two traces throughout their alignment. For identical executions, this profile remains nearly constant with minor fluctuations due to measurement jitter, whereas for true branch divergence, this introduces a sustained and monotonic shift as the traces progressively misalign.

\subsection{Related Work}
Coverage-guided fuzzing has been extensively studied for software validation, with tools such as AFL++~\cite{fioraldi2020afl++} and HonggFuzz~\cite{honggfuzz} demonstrating that feedback-driven mutation can uncover deep vulnerabilities in complex software implementations~\cite{bohme2016coverage, lemieux2018fairfuzz, gan2018collafl, chen2018angora}. However, these techniques are not directly applicable to firmware fuzzing, as they assume a standard process execution model and cannot account for the complex interactions between firmware and underlying hardware peripherals. Extending fuzzing to embedded firmware therefore requires fundamentally different strategies for target execution, observation, and feedback collection.

A significant body of work addresses firmware fuzzing through emulation and rehosting. FirmFuzz \cite{zheng2019firm} performs gray-box fuzzing via POSIX-compatible emulation, while FirmCOV \cite{kim2021firm} improves coverage through process-level virtualization combined with structured input dictionaries. Fuzzware \cite{scharnowski2022fuzzware} advances this line of work by modeling peripheral behavior to enable accurate full-system emulation of bare-metal firmware. Forming \cite{seidel2023forming} and Ember \cite{farrelly2023ember} further refine rehosting fidelity by recovering hardware abstractions from firmware binaries. While these approaches achieve high coverage, they depend on access to the firmware binary for emulation setup and instrumentation, limiting their applicability to open or extractable firmware images. The challenges of secure firmware distribution and access control that further complicate this assumption are discussed in \cite{11458680}.

Several efforts have focused on improving the runtime efficiency of firmware fuzzing. SNPSFuzzer \cite{li2022snpsfuzzer} reuses saved program states to avoid repetitive initialization overhead, and SnapFuzz \cite{andronidis2022snapfuzz} similarly leverages snapshot restoration to accelerate the fuzzing loop. Nyx \cite{schumilo2022nyx} achieves high-throughput fuzzing through hypervisor-level snapshotting. A complementary line of work combines fuzzing with symbolic execution to overcome coverage plateaus. Driller \cite{stephens2016driller} and QSYM \cite{yun2018qsym} invoke symbolic execution selectively when the fuzzer stalls, while SysFuSS \cite{tharindu2026sysfuss} extends this hybrid approach to system-level firmware fuzzing through selective symbolic execution. EM-Fuzz \cite{gao2020fuzz} targets memory-sensitive vulnerabilities in firmware through electromagnetic traces during emulation

Sperl and Böttinger~\cite{sperl2019side} have explored power side-channel measurements to improve their white-box fuzzing performance. Specifically, they collect execution traces using the target board to train the machine learning (ML)  model. During fuzzing, they use power traces for ML-based branch detection and branch-distance classification to reconstruct an approximate CFG. Since their approach requires prior knowledge of the target's instruction-level behavior, it is not applicable in a black-box setting when firmware functionality is unavailable.  In contrast, PowerFuzz enables black-box firmware fuzzing using power traces without any prior knowledge of the firmware.

All of the above approaches operate under the assumption that the firmware binary is accessible for instrumentation, rehosting, or symbolic analysis, a gray-box requirement that is frequently unmet in practice. In many real-world deployments, firmware is encrypted, obfuscated, or otherwise inaccessible. Black-box fuzzing directly targets this gap by treating the target firmware as entirely opaque. FirmXRay \cite{wen2020firmxray} and similar efforts attempt to recover partial firmware structure from network traffic or behavioral observation, but stop short of providing a closed feedback loop suitable for coverage-guided fuzzing. Most relevant to our work, Fuzz'EMup \cite{Moradihaghighietal2026} proposes leveraging electromagnetic side-channel emanations to guide black-box firmware fuzzing, demonstrating that physical side-channel signals can substitute for binary instrumentation as a fuzzing feedback source. However, Fuzz'EMup relies on electromagnetic measurements that require specialized near-field probing equipment and expensive physical setup, limiting its practical deployability. In contrast, PowerFuzz operates on power consumption measurements using low-cost current sensing hardware. \textit{To the best of our knowledge, PowerFuzz is the first framework to use power traces to enable black-box firmware fuzzing.}

%% file: figs/libafl.tex
\tikzset{every picture/.style={line width=0.75pt}} 

\begin{tikzpicture}[x=0.75pt,y=0.75pt,yscale=-1,xscale=1]

\draw  [color={rgb, 255:red, 0; green, 0; blue, 0 }  ,draw opacity=1 ][fill={rgb, 255:red, 219; green, 223; blue, 239 }  ,fill opacity=1 ] (40,21.85) .. controls (40,18.62) and (42.62,16) .. (45.85,16) -- (294.15,16) .. controls (297.38,16) and (300,18.62) .. (300,21.85) -- (300,102.15) .. controls (300,105.38) and (297.38,108) .. (294.15,108) -- (45.85,108) .. controls (42.62,108) and (40,105.38) .. (40,102.15) -- cycle ;
\draw  [color={rgb, 255:red, 0; green, 0; blue, 0 }  ,draw opacity=1 ][fill={rgb, 255:red, 255; green, 226; blue, 187 }  ,fill opacity=1 ] (48,41) .. controls (48,39.34) and (49.34,38) .. (51,38) -- (117,38) .. controls (118.66,38) and (120,39.34) .. (120,41) -- (120,59) .. controls (120,60.66) and (118.66,62) .. (117,62) -- (51,62) .. controls (49.34,62) and (48,60.66) .. (48,59) -- cycle ;
\draw    (120,50) -- (131,50) ;
\draw [shift={(134,50)}, rotate = 180] [fill={rgb, 255:red, 0; green, 0; blue, 0 }  ][line width=0.08]  [draw opacity=0] (6.25,-3) -- (0,0) -- (6.25,3) -- cycle    ;
\draw    (206,50) -- (217,50) ;
\draw [shift={(220,50)}, rotate = 180] [fill={rgb, 255:red, 0; green, 0; blue, 0 }  ][line width=0.08]  [draw opacity=0] (6.25,-3) -- (0,0) -- (6.25,3) -- cycle    ;
\draw    (123,88) -- (134,88) ;
\draw [shift={(120,88)}, rotate = 0] [fill={rgb, 255:red, 0; green, 0; blue, 0 }  ][line width=0.08]  [draw opacity=0] (6.25,-3) -- (0,0) -- (6.25,3) -- cycle    ;
\draw    (84,65) -- (84,76) ;
\draw [shift={(84,62)}, rotate = 90] [fill={rgb, 255:red, 0; green, 0; blue, 0 }  ][line width=0.08]  [draw opacity=0] (6.25,-3) -- (0,0) -- (6.25,3) -- cycle    ;
\draw    (256,73) -- (256,62) ;
\draw [shift={(256,76)}, rotate = 270] [fill={rgb, 255:red, 0; green, 0; blue, 0 }  ][line width=0.08]  [draw opacity=0] (6.25,-3) -- (0,0) -- (6.25,3) -- cycle    ;
\draw  [color={rgb, 255:red, 0; green, 0; blue, 0 }  ,draw opacity=1 ][fill={rgb, 255:red, 204; green, 231; blue, 207 }  ,fill opacity=1 ] (39.75,125.55) .. controls (39.75,123.66) and (41.28,122.13) .. (43.18,122.13) -- (212.57,122.13) .. controls (214.47,122.13) and (216,123.66) .. (216,125.55) -- (216,172.57) .. controls (216,174.47) and (214.47,176) .. (212.57,176) -- (43.18,176) .. controls (41.28,176) and (39.75,174.47) .. (39.75,172.57) -- cycle ;
\draw    (123,156) -- (134,156) ;
\draw [shift={(120,156)}, rotate = 0] [fill={rgb, 255:red, 0; green, 0; blue, 0 }  ][line width=0.08]  [draw opacity=0] (6.25,-3) -- (0,0) -- (6.25,3) -- cycle    ;
\draw    (170,103) -- (170,116) -- (122,116) -- (122,122) ;
\draw [shift={(170,100)}, rotate = 90] [fill={rgb, 255:red, 0; green, 0; blue, 0 }  ][line width=0.08]  [draw opacity=0] (6.25,-3) -- (0,0) -- (6.25,3) -- cycle    ;
\draw    (256,143) -- (256,100) ;
\draw [shift={(256,146)}, rotate = 270] [fill={rgb, 255:red, 0; green, 0; blue, 0 }  ][line width=0.08]  [draw opacity=0] (6.25,-3) -- (0,0) -- (6.25,3) -- cycle    ;
\draw    (209,156) -- (222,156) ;
\draw [shift={(206,156)}, rotate = 0] [fill={rgb, 255:red, 0; green, 0; blue, 0 }  ][line width=0.08]  [draw opacity=0] (6.25,-3) -- (0,0) -- (6.25,3) -- cycle    ;
\draw  [color={rgb, 255:red, 0; green, 0; blue, 0 }  ,draw opacity=1 ][fill={rgb, 255:red, 255; green, 226; blue, 187 }  ,fill opacity=1 ] (134,41) .. controls (134,39.34) and (135.34,38) .. (137,38) -- (203,38) .. controls (204.66,38) and (206,39.34) .. (206,41) -- (206,59) .. controls (206,60.66) and (204.66,62) .. (203,62) -- (137,62) .. controls (135.34,62) and (134,60.66) .. (134,59) -- cycle ;
\draw  [color={rgb, 255:red, 0; green, 0; blue, 0 }  ,draw opacity=1 ][fill={rgb, 255:red, 255; green, 226; blue, 187 }  ,fill opacity=1 ] (220,41) .. controls (220,39.34) and (221.34,38) .. (223,38) -- (289,38) .. controls (290.66,38) and (292,39.34) .. (292,41) -- (292,59) .. controls (292,60.66) and (290.66,62) .. (289,62) -- (223,62) .. controls (221.34,62) and (220,60.66) .. (220,59) -- cycle ;
\draw  [color={rgb, 255:red, 0; green, 0; blue, 0 }  ,draw opacity=1 ][fill={rgb, 255:red, 255; green, 226; blue, 187 }  ,fill opacity=1 ] (220,79) .. controls (220,77.34) and (221.34,76) .. (223,76) -- (289,76) .. controls (290.66,76) and (292,77.34) .. (292,79) -- (292,97) .. controls (292,98.66) and (290.66,100) .. (289,100) -- (223,100) .. controls (221.34,100) and (220,98.66) .. (220,97) -- cycle ;
\draw  [color={rgb, 255:red, 0; green, 0; blue, 0 }  ,draw opacity=1 ][fill={rgb, 255:red, 255; green, 226; blue, 187 }  ,fill opacity=1 ] (134,79) .. controls (134,77.34) and (135.34,76) .. (137,76) -- (203,76) .. controls (204.66,76) and (206,77.34) .. (206,79) -- (206,97) .. controls (206,98.66) and (204.66,100) .. (203,100) -- (137,100) .. controls (135.34,100) and (134,98.66) .. (134,97) -- cycle ;
\draw  [color={rgb, 255:red, 0; green, 0; blue, 0 }  ,draw opacity=1 ][fill={rgb, 255:red, 255; green, 226; blue, 187 }  ,fill opacity=1 ] (48,79) .. controls (48,77.34) and (49.34,76) .. (51,76) -- (117,76) .. controls (118.66,76) and (120,77.34) .. (120,79) -- (120,97) .. controls (120,98.66) and (118.66,100) .. (117,100) -- (51,100) .. controls (49.34,100) and (48,98.66) .. (48,97) -- cycle ;
\draw  [color={rgb, 255:red, 0; green, 0; blue, 0 }  ,draw opacity=1 ][fill={rgb, 255:red, 255; green, 226; blue, 187 }  ,fill opacity=1 ] (222,147) .. controls (222,145.34) and (223.34,144) .. (225,144) -- (289,144) .. controls (290.66,144) and (292,145.34) .. (292,147) -- (292,165) .. controls (292,166.66) and (290.66,168) .. (289,168) -- (225,168) .. controls (223.34,168) and (222,166.66) .. (222,165) -- cycle ;
\draw  [color={rgb, 255:red, 0; green, 0; blue, 0 }  ,draw opacity=1 ][fill={rgb, 255:red, 255; green, 226; blue, 187 }  ,fill opacity=1 ] (134,147) .. controls (134,145.34) and (135.34,144) .. (137,144) -- (203,144) .. controls (204.66,144) and (206,145.34) .. (206,147) -- (206,165) .. controls (206,166.66) and (204.66,168) .. (203,168) -- (137,168) .. controls (135.34,168) and (134,166.66) .. (134,165) -- cycle ;
\draw  [color={rgb, 255:red, 0; green, 0; blue, 0 }  ,draw opacity=1 ][fill={rgb, 255:red, 255; green, 226; blue, 187 }  ,fill opacity=1 ] (48,147) .. controls (48,145.34) and (49.34,144) .. (51,144) -- (117,144) .. controls (118.66,144) and (120,145.34) .. (120,147) -- (120,165) .. controls (120,166.66) and (118.66,168) .. (117,168) -- (51,168) .. controls (49.34,168) and (48,166.66) .. (48,165) -- cycle ;

\draw (83.63,50) node  [font=\small] [align=left] {\begin{minipage}[lt]{49.47pt}\setlength\topsep{0pt}
\begin{center}
Input Corpus
\end{center}

\end{minipage}};
\draw (161,26.5) node   [align=left] {\begin{minipage}[lt]{170pt}\setlength\topsep{0pt}
\begin{center}
\textbf{LibAFL}
\end{center}

\end{minipage}};
\draw (170.13,50) node  [font=\small] [align=left] {\begin{minipage}[lt]{48.79pt}\setlength\topsep{0pt}
\begin{center}
Scheduler
\end{center}

\end{minipage}};
\draw (256,50) node  [font=\small] [align=left] {\begin{minipage}[lt]{48.96pt}\setlength\topsep{0pt}
\begin{center}
Mutator
\end{center}

\end{minipage}};
\draw (84,88) node  [font=\small] [align=left] {\begin{minipage}[lt]{48.96pt}\setlength\topsep{0pt}
\begin{center}
Feedback
\end{center}

\end{minipage}};
\draw (170,88) node  [font=\small] [align=left] {\begin{minipage}[lt]{48.96pt}\setlength\topsep{0pt}
\begin{center}
Observer
\end{center}

\end{minipage}};
\draw (256,88) node  [font=\small] [align=left] {\begin{minipage}[lt]{48.96pt}\setlength\topsep{0pt}
\begin{center}
Harness
\end{center}

\end{minipage}};
\draw (84,156) node  [font=\small] [align=left] {\begin{minipage}[lt]{48.96pt}\setlength\topsep{0pt}
\begin{center}
Input Corpus
\end{center}

\end{minipage}};
\draw (125.5,129.94) node   [align=left] {\begin{minipage}[lt]{117.47pt}\setlength\topsep{0pt}
\begin{center}
\textbf{Design Under Test}
\end{center}

\end{minipage}};
\draw (170,156) node  [font=\small] [align=left] {\begin{minipage}[lt]{48.96pt}\setlength\topsep{0pt}
\begin{center}
Scheduler
\end{center}

\end{minipage}};
\draw (257,156) node  [font=\small] [align=left] {\begin{minipage}[lt]{47.6pt}\setlength\topsep{0pt}
\begin{center}
New Input
\end{center}

\end{minipage}};

\end{tikzpicture}

%% file: figs/powerfuzz.tex
\tikzset{every picture/.style={line width=0.75pt}} 

\begin{tikzpicture}[x=0.75pt,y=0.75pt,yscale=-1,xscale=1]

\draw   (98,197.53) .. controls (98,193.37) and (101.37,190) .. (105.53,190) -- (172.47,190) .. controls (176.63,190) and (180,193.37) .. (180,197.53) -- (180,250.47) .. controls (180,254.63) and (176.63,258) .. (172.47,258) -- (105.53,258) .. controls (101.37,258) and (98,254.63) .. (98,250.47) -- cycle ;
\draw  [fill={rgb, 255:red, 0; green, 0; blue, 0 }  ,fill opacity=1 ] (104,221.2) .. controls (104,217.22) and (107.22,214) .. (111.2,214) -- (166.8,214) .. controls (170.78,214) and (174,217.22) .. (174,221.2) -- (174,242.8) .. controls (174,246.78) and (170.78,250) .. (166.8,250) -- (111.2,250) .. controls (107.22,250) and (104,246.78) .. (104,242.8) -- cycle ;
\draw   (96,101.54) .. controls (96,98.48) and (98.48,96) .. (101.54,96) -- (172.46,96) .. controls (175.52,96) and (178,98.48) .. (178,101.54) -- (178,140.46) .. controls (178,143.52) and (175.52,146) .. (172.46,146) -- (101.54,146) .. controls (98.48,146) and (96,143.52) .. (96,140.46) -- cycle ;
\draw  [fill={rgb, 255:red, 204; green, 231; blue, 207 }  ,fill opacity=1 ] (212,93.97) .. controls (212,89.57) and (215.57,86) .. (219.97,86) -- (368.03,86) .. controls (372.43,86) and (376,89.57) .. (376,93.97) -- (376,150.03) .. controls (376,154.43) and (372.43,158) .. (368.03,158) -- (219.97,158) .. controls (215.57,158) and (212,154.43) .. (212,150.03) -- cycle ;
\draw  [fill={rgb, 255:red, 255; green, 226; blue, 187 }  ,fill opacity=1 ] (298,118) .. controls (298,113.58) and (301.58,110) .. (306,110) -- (360,110) .. controls (364.42,110) and (368,113.58) .. (368,118) -- (368,142) .. controls (368,146.42) and (364.42,150) .. (360,150) -- (306,150) .. controls (301.58,150) and (298,146.42) .. (298,142) -- cycle ;

\draw  [fill={rgb, 255:red, 255; green, 226; blue, 187 }  ,fill opacity=1 ] (220,118) .. controls (220,113.58) and (223.58,110) .. (228,110) -- (282,110) .. controls (286.42,110) and (290,113.58) .. (290,118) -- (290,142) .. controls (290,146.42) and (286.42,150) .. (282,150) -- (228,150) .. controls (223.58,150) and (220,146.42) .. (220,142) -- cycle ;

\draw  [fill={rgb, 255:red, 255; green, 226; blue, 187 }  ,fill opacity=1 ] (583.21,109.43) .. controls (583.21,105.01) and (586.8,101.43) .. (591.21,101.43) .. controls (595.63,101.43) and (599.21,105.01) .. (599.21,109.43) .. controls (599.21,113.85) and (595.63,117.43) .. (591.21,117.43) .. controls (586.8,117.43) and (583.21,113.85) .. (583.21,109.43) -- cycle ;
\draw  [fill={rgb, 255:red, 255; green, 226; blue, 187 }  ,fill opacity=1 ] (563.21,136.57) .. controls (563.21,133.26) and (565.9,130.57) .. (569.21,130.57) .. controls (572.53,130.57) and (575.21,133.26) .. (575.21,136.57) .. controls (575.21,139.89) and (572.53,142.57) .. (569.21,142.57) .. controls (565.9,142.57) and (563.21,139.89) .. (563.21,136.57) -- cycle ;
\draw  [fill={rgb, 255:red, 255; green, 226; blue, 187 }  ,fill opacity=1 ] (608.07,137.14) .. controls (608.07,133.83) and (610.76,131.14) .. (614.07,131.14) .. controls (617.39,131.14) and (620.07,133.83) .. (620.07,137.14) .. controls (620.07,140.46) and (617.39,143.14) .. (614.07,143.14) .. controls (610.76,143.14) and (608.07,140.46) .. (608.07,137.14) -- cycle ;
\draw    (586.57,115.57) -- (574.53,129.58) ;
\draw [shift={(572.57,131.86)}, rotate = 310.68] [fill={rgb, 255:red, 0; green, 0; blue, 0 }  ][line width=0.08]  [draw opacity=0] (5.36,-2.57) -- (0,0) -- (5.36,2.57) -- (3.56,0) -- cycle    ;
\draw    (596.14,116) -- (608.19,130.01) ;
\draw [shift={(610.14,132.29)}, rotate = 229.32] [fill={rgb, 255:red, 0; green, 0; blue, 0 }  ][line width=0.08]  [draw opacity=0] (5.36,-2.57) -- (0,0) -- (5.36,2.57) -- (3.56,0) -- cycle    ;
\draw  [fill={rgb, 255:red, 255; green, 226; blue, 187 }  ,fill opacity=1 ] (551.57,157.14) .. controls (551.57,154.93) and (553.36,153.14) .. (555.57,153.14) .. controls (557.78,153.14) and (559.57,154.93) .. (559.57,157.14) .. controls (559.57,159.35) and (557.78,161.14) .. (555.57,161.14) .. controls (553.36,161.14) and (551.57,159.35) .. (551.57,157.14) -- cycle ;
\draw  [fill={rgb, 255:red, 255; green, 226; blue, 187 }  ,fill opacity=1 ] (578.43,157.43) .. controls (578.43,155.22) and (580.22,153.43) .. (582.43,153.43) .. controls (584.64,153.43) and (586.43,155.22) .. (586.43,157.43) .. controls (586.43,159.64) and (584.64,161.43) .. (582.43,161.43) .. controls (580.22,161.43) and (578.43,159.64) .. (578.43,157.43) -- cycle ;
\draw  [fill={rgb, 255:red, 255; green, 226; blue, 187 }  ,fill opacity=1 ] (597,157.71) .. controls (597,155.51) and (598.79,153.71) .. (601,153.71) .. controls (603.21,153.71) and (605,155.51) .. (605,157.71) .. controls (605,159.92) and (603.21,161.71) .. (601,161.71) .. controls (598.79,161.71) and (597,159.92) .. (597,157.71) -- cycle ;
\draw  [fill={rgb, 255:red, 255; green, 226; blue, 187 }  ,fill opacity=1 ] (623,158) .. controls (623,155.79) and (624.79,154) .. (627,154) .. controls (629.21,154) and (631,155.79) .. (631,158) .. controls (631,160.21) and (629.21,162) .. (627,162) .. controls (624.79,162) and (623,160.21) .. (623,158) -- cycle ;
\draw    (565.43,140.71) -- (559.01,151.02) ;
\draw [shift={(557.43,153.57)}, rotate = 301.89] [fill={rgb, 255:red, 0; green, 0; blue, 0 }  ][line width=0.08]  [draw opacity=0] (5.36,-2.57) -- (0,0) -- (5.36,2.57) -- (3.56,0) -- cycle    ;
\draw    (572.86,141.29) -- (579.14,151.45) ;
\draw [shift={(580.71,154)}, rotate = 238.28] [fill={rgb, 255:red, 0; green, 0; blue, 0 }  ][line width=0.08]  [draw opacity=0] (5.36,-2.57) -- (0,0) -- (5.36,2.57) -- (3.56,0) -- cycle    ;
\draw    (610.86,142.71) -- (605.12,151.1) ;
\draw [shift={(603.43,153.57)}, rotate = 304.38] [fill={rgb, 255:red, 0; green, 0; blue, 0 }  ][line width=0.08]  [draw opacity=0] (5.36,-2.57) -- (0,0) -- (5.36,2.57) -- (3.56,0) -- cycle    ;
\draw    (618.57,142.14) -- (623.89,151) ;
\draw [shift={(625.43,153.57)}, rotate = 239.04] [fill={rgb, 255:red, 0; green, 0; blue, 0 }  ][line width=0.08]  [draw opacity=0] (5.36,-2.57) -- (0,0) -- (5.36,2.57) -- (3.56,0) -- cycle    ;
\draw   (544,91.9) .. controls (544,88.64) and (546.64,86) .. (549.9,86) -- (634.1,86) .. controls (637.36,86) and (640,88.64) .. (640,91.9) -- (640,164.1) .. controls (640,167.36) and (637.36,170) .. (634.1,170) -- (549.9,170) .. controls (546.64,170) and (544,167.36) .. (544,164.1) -- cycle ;
\draw  [fill={rgb, 255:red, 204; green, 231; blue, 207 }  ,fill opacity=1 ] (406,93.97) .. controls (406,89.57) and (409.57,86) .. (413.97,86) -- (506.03,86) .. controls (510.43,86) and (514,89.57) .. (514,93.97) -- (514,150.03) .. controls (514,154.43) and (510.43,158) .. (506.03,158) -- (413.97,158) .. controls (409.57,158) and (406,154.43) .. (406,150.03) -- cycle ;
\draw  [fill={rgb, 255:red, 204; green, 231; blue, 207 }  ,fill opacity=1 ] (510,197.75) .. controls (510,193.47) and (513.47,190) .. (517.75,190) -- (664.25,190) .. controls (668.53,190) and (672,193.47) .. (672,197.75) -- (672,252.25) .. controls (672,256.53) and (668.53,260) .. (664.25,260) -- (517.75,260) .. controls (513.47,260) and (510,256.53) .. (510,252.25) -- cycle ;
\draw  [fill={rgb, 255:red, 255; green, 226; blue, 187 }  ,fill opacity=1 ] (516,222) .. controls (516,217.58) and (519.58,214) .. (524,214) -- (578,214) .. controls (582.42,214) and (586,217.58) .. (586,222) -- (586,246) .. controls (586,250.42) and (582.42,254) .. (578,254) -- (524,254) .. controls (519.58,254) and (516,250.42) .. (516,246) -- cycle ;
\draw  [fill={rgb, 255:red, 255; green, 226; blue, 187 }  ,fill opacity=1 ] (596,222) .. controls (596,217.58) and (599.58,214) .. (604,214) -- (658,214) .. controls (662.42,214) and (666,217.58) .. (666,222) -- (666,246) .. controls (666,250.42) and (662.42,254) .. (658,254) -- (604,254) .. controls (599.58,254) and (596,250.42) .. (596,246) -- cycle ;
\draw   (292,199.53) .. controls (292,195.37) and (295.37,192) .. (299.53,192) -- (448.47,192) .. controls (452.63,192) and (456,195.37) .. (456,199.53) -- (456,252.47) .. controls (456,256.63) and (452.63,260) .. (448.47,260) -- (299.53,260) .. controls (295.37,260) and (292,256.63) .. (292,252.47) -- cycle ;
\draw  [fill={rgb, 255:red, 204; green, 231; blue, 207 }  ,fill opacity=1 ] (300,223.2) .. controls (300,219.22) and (303.22,216) .. (307.2,216) -- (362.8,216) .. controls (366.78,216) and (370,219.22) .. (370,223.2) -- (370,244.8) .. controls (370,248.78) and (366.78,252) .. (362.8,252) -- (307.2,252) .. controls (303.22,252) and (300,248.78) .. (300,244.8) -- cycle ;
\draw   (378,223.2) .. controls (378,219.22) and (381.22,216) .. (385.2,216) -- (440.8,216) .. controls (444.78,216) and (448,219.22) .. (448,223.2) -- (448,244.8) .. controls (448,248.78) and (444.78,252) .. (440.8,252) -- (385.2,252) .. controls (381.22,252) and (378,248.78) .. (378,244.8) -- cycle ;
\draw   (200,211.95) .. controls (200,209.89) and (201.66,208.23) .. (203.71,208.23) -- (260.29,208.23) .. controls (262.34,208.23) and (264,209.89) .. (264,211.95) -- (264,238.05) .. controls (264,240.11) and (262.34,241.77) .. (260.29,241.77) -- (203.71,241.77) .. controls (201.66,241.77) and (200,240.11) .. (200,238.05) -- cycle ;
\draw [line width=0.75]    (138,190) -- (138,149) ;
\draw [shift={(138,146)}, rotate = 90] [fill={rgb, 255:red, 0; green, 0; blue, 0 }  ][line width=0.08]  [draw opacity=0] (6.25,-3) -- (0,0) -- (6.25,3) -- cycle    ;
\draw [line width=0.75]    (178,120) -- (209,120) ;
\draw [shift={(212,120)}, rotate = 180] [fill={rgb, 255:red, 0; green, 0; blue, 0 }  ][line width=0.08]  [draw opacity=0] (6.25,-3) -- (0,0) -- (6.25,3) -- cycle    ;
\draw [line width=0.75]    (376,120) -- (403,120) ;
\draw [shift={(406,120)}, rotate = 180] [fill={rgb, 255:red, 0; green, 0; blue, 0 }  ][line width=0.08]  [draw opacity=0] (6.25,-3) -- (0,0) -- (6.25,3) -- cycle    ;
\draw [line width=0.75]    (514,120) -- (541,120) ;
\draw [shift={(544,120)}, rotate = 180] [fill={rgb, 255:red, 0; green, 0; blue, 0 }  ][line width=0.08]  [draw opacity=0] (6.25,-3) -- (0,0) -- (6.25,3) -- cycle    ;
\draw [line width=0.75]    (200,224) -- (183,224) ;
\draw [shift={(180,224)}, rotate = 360] [fill={rgb, 255:red, 0; green, 0; blue, 0 }  ][line width=0.08]  [draw opacity=0] (6.25,-3) -- (0,0) -- (6.25,3) -- cycle    ;
\draw [line width=0.75]    (292,224) -- (267,224) ;
\draw [shift={(264,224)}, rotate = 360] [fill={rgb, 255:red, 0; green, 0; blue, 0 }  ][line width=0.08]  [draw opacity=0] (6.25,-3) -- (0,0) -- (6.25,3) -- cycle    ;
\draw [line width=0.75]    (592,170) -- (592,187) ;
\draw [shift={(592,190)}, rotate = 270] [fill={rgb, 255:red, 0; green, 0; blue, 0 }  ][line width=0.08]  [draw opacity=0] (6.25,-3) -- (0,0) -- (6.25,3) -- cycle    ;
\draw [line width=0.75]    (510,224) -- (459,224) ;
\draw [shift={(456,224)}, rotate = 360] [fill={rgb, 255:red, 0; green, 0; blue, 0 }  ][line width=0.08]  [draw opacity=0] (6.25,-3) -- (0,0) -- (6.25,3) -- cycle    ;
\draw [line width=0.75]    (592,86) -- (592,74) -- (294,74) -- (294,83) ;
\draw [shift={(294,86)}, rotate = 270] [fill={rgb, 255:red, 0; green, 0; blue, 0 }  ][line width=0.08]  [draw opacity=0] (6.25,-3) -- (0,0) -- (6.25,3) -- cycle    ;

\draw (139,232) node  [font=\small,color={rgb, 255:red, 255; green, 255; blue, 255 }  ,opacity=1 ] [align=left] {\begin{minipage}[lt]{47.6pt}\setlength\topsep{0pt}
\begin{center}
Black-box Firmware
\end{center}

\end{minipage}};
\draw (139,202) node   [align=left] {\begin{minipage}[lt]{55.76pt}\setlength\topsep{0pt}
\begin{center}
\textbf{Device}
\end{center}

\end{minipage}};
\draw (137,121) node   [align=left] {\begin{minipage}[lt]{55.76pt}\setlength\topsep{0pt}
\begin{center}
\textbf{Power}\\\textbf{Measurement}
\end{center}

\end{minipage}};
\draw (293,98) node   [align=left] {\begin{minipage}[lt]{110.16pt}\setlength\topsep{0pt}
\begin{center}
\textbf{Similarity Analysis}
\end{center}

\end{minipage}};
\draw (460,122) node   [align=left] {\begin{minipage}[lt]{73.44pt}\setlength\topsep{0pt}
\begin{center}
\textbf{Dynamic Construction of TCFG}
\end{center}

\end{minipage}};
\draw (551,234) node  [font=\small] [align=left] {\begin{minipage}[lt]{47.6pt}\setlength\topsep{0pt}
\begin{center}
Random
\end{center}

\end{minipage}};
\draw (591,202) node   [align=left] {\begin{minipage}[lt]{110.16pt}\setlength\topsep{0pt}
\begin{center}
\textbf{Branch Selector}
\end{center}

\end{minipage}};
\draw (631,234) node  [font=\small] [align=left] {\begin{minipage}[lt]{47.6pt}\setlength\topsep{0pt}
\begin{center}
Depth Prioritization
\end{center}

\end{minipage}};
\draw (373,204) node   [align=left] {\begin{minipage}[lt]{110.16pt}\setlength\topsep{0pt}
\begin{center}
\textbf{Fuzzing Engine}
\end{center}

\end{minipage}};
\draw (413,234) node  [font=\small] [align=left] {\begin{minipage}[lt]{47.6pt}\setlength\topsep{0pt}
\begin{center}
Seed Corpus
\end{center}

\end{minipage}};
\draw (232,225) node   [align=left] {\begin{minipage}[lt]{43.78pt}\setlength\topsep{0pt}
\begin{center}
New Input
\end{center}

\end{minipage}};
\draw (567,99) node   [align=left] {\begin{minipage}[lt]{31.28pt}\setlength\topsep{0pt}
\begin{center}
\textbf{TCFG}
\end{center}

\end{minipage}};
\draw (333,130) node  [font=\small] [align=left] {\begin{minipage}[lt]{47.6pt}\setlength\topsep{0pt}
\begin{center}
Power Correlation
\end{center}

\end{minipage}};
\draw (255,130) node  [font=\small] [align=left] {\begin{minipage}[lt]{47.6pt}\setlength\topsep{0pt}
\begin{center}
Dynamic Time Warping
\end{center}

\end{minipage}};
\draw (335,234) node  [font=\small] [align=left] {\begin{minipage}[lt]{47.6pt}\setlength\topsep{0pt}
\begin{center}
Mutation Logic
\end{center}

\end{minipage}};

\end{tikzpicture}

%% file: sections/3_methodology.tex
\section{Power-based Black-Box Firmware Fuzzing}
\label{sec:methodology}

Figure~\ref{fig:framework} provides an overview of power-based black-box firmware fuzzing (\textbf{PowerFuzz}) framework that operates as a closed feedback loop between a black-box embedded device and a coverage-guided fuzzing engine.  When the fuzzing engine dispatches an input to the target device, trace acquisition is triggered synchronously with input delivery, capturing the MCU's power consumption as a time-series signal $P = \langle p_1, p_2, \ldots, p_n \rangle$, where each sample $p_i$ represents the instantaneous power draw of the MCU at time step $i$. Since different instructions and branch decisions produce measurably distinct power consumption patterns, this trace encodes the firmware's execution behavior in response to the given input without requiring any internal visibility into the firmware binary. The captured trace is then passed to the similarity analysis module, which applies power correlation and dynamic time warping to compare it against previously observed traces and detect execution divergences. Detected divergences are used by the TCFG updater to incrementally refine the firmware's control flow behavior. The branch selector then consults the current TCFG to identify promising unexplored branches and communicates this structural insight to the fuzzing engine, which biases its mutation logic toward inputs likely to drive execution into those branches. PowerFuzz progressively builds structural knowledge of the firmware and exploits it to guide input generation without any visibility of the firmware binary. Algorithm \ref{alg:powerfuzz} highlights the three major components in the PowerFuzz framework: similarity analysis, dynamic construction of TCFG, and branch selection for test generation. The remainder of this section describes these components in detail.

\begin{algorithm}[htp]
    \caption{Overall Framework of PowerFuzz}
    \label{alg:powerfuzz}
    \begin{algorithmic}[1]
        \State Initialize Seed Corpus $\mathcal{C} \leftarrow$ random inputs
        \While{budget not exhausted}
            \State Send $new\_input$ to target device
            \State $P_{curr} \leftarrow$ average of $N$ captured power traces
            \State $dp \leftarrow$ \Call{SimilarityAnalysis}{$\mathcal{G},\ P_{curr}$}
            \State $F \leftarrow$ \Call{UpdateTCFG}{$\mathcal{G},\ P_{curr},\ new\_input$}
            \If{$F \neq 0$}
                \State Add $(new\_input)$ to corpus $\mathcal{C}$
            \Else
                \State Discard $new\_input$
            \EndIf
            
            \If{$|\mathcal{G}| <$ depth\_threshold}
            \State $v^* \leftarrow$ \Call{RandomSelect}{$\mathcal{G}$}
            \Else
                \State $v^* \leftarrow$ \Call{DepthPrioritization}{$\mathcal{G}$}
            \EndIf
    
            \State $s \leftarrow$ \Call{SelectSeed}{$\mathcal{C},\ v^*.data$}
            \State $new\_input \leftarrow$ \Call{Mutate}{$s$}
        \EndWhile
    \end{algorithmic}
\end{algorithm}

\subsection{Similarity Analysis for Identification of Executed Branches}\label{sec:ebi}


A basic block is a straight-line sequence of instructions with a single entry point and a single exit point, meaning no branches occur within the block except at its boundaries. Consequently, when two different inputs execute the same basic block, the corresponding power traces differ only due to data-dependent variations. As observed in our experiments, such data-dependent fluctuations are typically negligible compared to operation-dependent variations that arise when inputs drive the firmware into distinct execution paths with fundamentally different instruction sequences. This property forms the physical foundation of PowerFuzz: branch decisions in the firmware manifest as structurally distinct changes in the power trace, making it possible to identify newly covered branches purely from power measurements.

Since the proposed work is implemented on embedded systems, actual power consumption measurements are collected using a state-of-the-art physical power side-channel acquisition setup. Specifically, we connect trace-capturing probes across a shunt resistor, which transforms operation-dependent current fluctuations into measurable voltage fluctuations. The captured power traces are then used for subsequent processing within PowerFuzz.

\begin{figure}[!ht]
    \centering
    \input{figs/dtw_divergence_detection}
    \vspace{-0.2in}
    \caption{DTW-based divergence detection between two power traces. Traces 1 and 2 correspond to two inputs that initially follow the same execution path before diverging. The resulting DTW offset profile remains approximately constant during the shared execution region and begins growing after timestamp 550, identifying the approximate temporal window within which the branch divergence occurs.}
    \label{fig:dtw_divergence}
\end{figure}
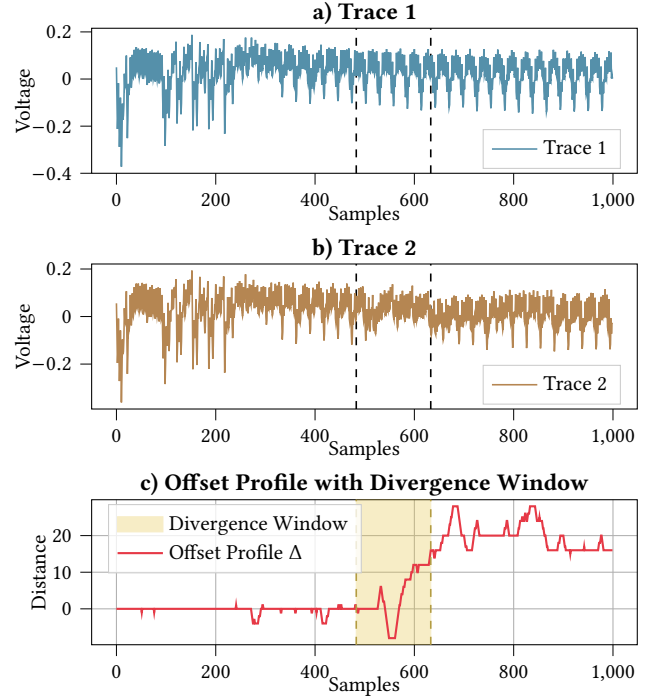

\subsubsection{Branch Deviation Detection using Dynamic Time Warping:}
The core challenge of PowerFuzz is determining, from a pair of power traces, whether two inputs drove the firmware through the same execution path or diverged at some branch point. We address this through a two-stage similarity analysis combining dynamic time warping (DTW) and Pearson correlation, applied sequentially to progressively narrow down the location of any divergence.
In the first stage, DTW (Section~\ref{sec:dtw}) is applied to compare the newly captured trace against the previously captured trace. To distinguish true execution divergences from transient noise, we apply a sliding-window analysis over the offset profile $\Delta$. For a window of size $w$ centered at step $k$, we compute the mean offset $\mu_k$ as:

$$\mu_k = \frac{1}{w} \sum_{l=k}^{k+w-1} \delta_l$$

A divergence is detected at position $k^*$ if the difference between consecutive window means exceeds a threshold $\tau$. This approach ensures that PowerFuzz responds only to sustained divergences indicative of a genuine branch transition, and not to the localized perturbations characteristic of timing jitter. 


Figure~\ref{fig:dtw_divergence} illustrates this behavior: Trace 1 and 2 correspond to two inputs that initially follow the same execution path before diverging, and the resulting offset profile begins growing steadily around after timestamp 550, identifying the approximate divergence window within which the divergence occurs. The DTW stage efficiently filters out trace pairs whose differences are attributable to jitter noise, and localizes the divergence to a coarse temporal region for further analysis.

\subsubsection{Precise Divergence Localization using Pearson Correlation:}
Once the approximate deviation window is identified by DTW, we apply a sliding-window followed by a growing-window Pearson correlation analysis to pinpoint the precise divergence point. Formally, the Pearson correlation coefficient between two trace segments $X = [X_1, \ldots, X_n]$ and $Y = [Y_1, \ldots, Y_n]$ is defined as:

\begin{equation}
Corr(X,Y) =
\frac{\displaystyle\sum_{i=1}^{n} (X_i - \bar{X})(Y_i - \bar{Y})}
{\sqrt{\displaystyle\sum_{i=1}^{n} (X_i - \bar{X})^{2}} \;
 \sqrt{\displaystyle\sum_{i=1}^{n} (Y_i - \bar{Y})^{2}}},
\label{eq:pearson}
\end{equation}

where $\bar{X}$ and $\bar{Y}$ are the sample means of $X$ and $Y$, respectively, and $Corr \in [-1, 1]$ quantifies the linear similarity between the two segments. A value close to $1$ indicates highly similar power behavior, while lower values indicate divergence.

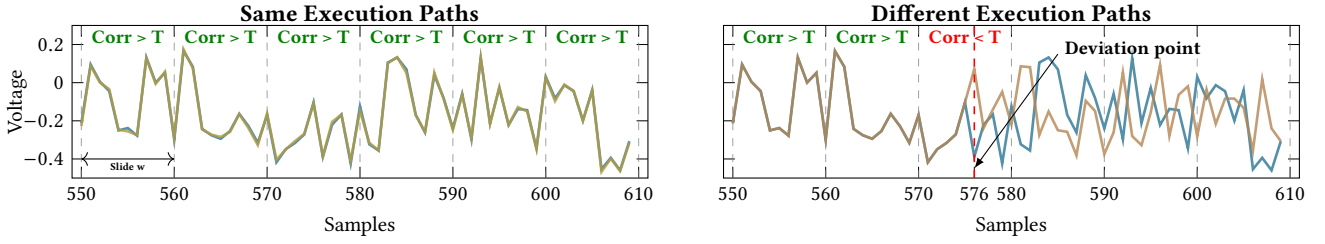
\begin{figure*}[!htp]
    \centering
    \input{figs/corr_detection}
    \vspace{-0.1in}
    \caption{Sliding-window Pearson correlation analysis for precise branch divergence localization. (Left) Two power traces corresponding to inputs that follow the same execution path, producing consistently high correlation values ($\text{Corr} > T$) across all sliding windows. (Right) Two power traces that share an initial execution path before diverging, where the correlation drops below the threshold ($\text{Corr} < T$) at timestamp 576, identifying the precise deviation point of the newly covered branch.}
    \label{fig:correlation}
\end{figure*}

Figure~\ref{fig:correlation} illustrates this analysis on two trace pairs. When the windowed segments correspond to the same execution path, the traces exhibit nearly identical power fluctuations, yielding a high correlation value ($Corr > T$, where T denotes the divergence threshold). When the segments span a branch divergence point, the structural difference in the power patterns produces a noticeably lower correlation ($\text{Corr} < T$), as shown at timestamp 576 in Figure~\ref{fig:correlation}.


The full procedure is formalized in Algorithm~\ref{alg:branch_identifier}. First, DTW algorithm is called with the two traces and it returns the start index ($DW_{start})$ and the end index ($DW_{end})$ of the deviation window. A sliding window of fixed length $Slide\_w$ is first applied over the traces to efficiently identify the approximate location of the divergence (lines 8--23). Once the coarse deviation region is identified, a growing-window analysis is applied starting from a window of one sample at the beginning of the deviation region, incrementally expanding the window until the correlation drops below $T$ (lines 24--33). This two-stage design balances computational efficiency, the sliding window rapidly eliminates non-divergent regions, with localization precision, as the growing window narrows the divergence to a precise sample index.

The identified branch point is then used to provide feedback to the fuzzer: inputs that do not cover any new branch are discarded, while inputs associated with a newly discovered branch are added to the seed corpus. However, as the number of previously identified branches grows, determining whether a new trace corresponds to a novel branch requires running Algorithm~\ref{alg:branch_identifier} against all previously captured branch traces, which becomes computationally expensive. To address this, we propose a TCFG-based approach in the following subsection that reduces the number of required correlation computations by exploiting the structural relationships between already-identified branches

\begin{flushleft}
\begin{algorithm}[htp]
\caption{Divergent Point Identifier}
\label{alg:branch_identifier}
\begin{algorithmic}[1]
\Procedure{GetDivergentPoint}{$Pre\_trace, Curr\_trace$} \label{proc:get_divergent}
\Statex \textbf{/* Inputs */}
\State Previously stored power trace - $Pre\_trace$
\State Current execution power trace - $Curr\_trace$
\State Sliding-window size - $Slide\_w$
\State Sliding-window correlation threshold - $T$
\State Growing-window correlation threshold - $T_g$
\vspace{0.3em}

\Statex \textbf{/* Initialization */}
\State $DW_{start}, \ DW_{end} \gets DTW(Pre\_trace, \ Curr\_trace)$
\State $Start\_idx \gets DW_{start}$
\State $Deviation \gets -1$
\vspace{0.3em}

\Statex \textbf{/* Sliding-Window Full-Trace Correlation */}
\While {$Start\_idx < DW_{end} - Slide\_w$}
    \If{$DW_{end} < Start\_idx + 2*Slide\_w$}
        \State $End\_idx \gets Min\_len$
    \Else
        \State $End\_idx \gets Start\_idx + Slide\_w$
    \EndIf
    \State $W_{pre} \gets Pre\_trace[Start\_idx : End\_idx]$
    \State $W_{curr} \gets Curr\_trace[Start\_idx : End\_idx]$
    \State $Corr \gets Corr(W_{pre}, W_{curr})$

    \If{$Corr < T$}
        \State $Deviation \gets Start\_idx $
        \State \textbf{break}
    \EndIf
    \State $Start\_idx \gets Start\_idx + Slide\_w$
\EndWhile
\vspace{0.3em}

\Statex \textbf{/* Growing-Window Correlation */}
\If{$Deviation \neq -1$}
    \For{$End\_idx\_g = Deviation$ \textbf{to} $Deviation + Slide\_w$}
        \State $GW_{pre} \gets Pre\_trace[Deviation : End\_idx\_g]$
        \State $GW_{curr} \gets Curr\_trace[Deviation : End\_idx\_g]$
        \State $Corr_g \gets Corr(GW_{pre}, GW_{curr})$

        \If{$Corr_g < T_g$}
            \State $Deviation\_point \gets End\_idx\_g$
            \State \Return $Deviation\_point$
        \EndIf
    \EndFor
\EndIf

\State \Return length($Pre\_trace$) \Comment{Traces are similar}
\EndProcedure
\end{algorithmic}
\end{algorithm}
\end{flushleft}

\subsection{Dynamic Construction of TCFG}\label{sec:branch_selection}

\begin{listing}[htp]  
\caption{Example firmware with a branch condition}
\label{lst:ex_branch}
\begin{lstlisting}[style=vscodeLightC]
if (device_status & FLAG_INITIALIZED) {         // B1
    int new_temp = (acc * 3) + (in * 7) ;
    if (fw_version >= MIN_SUPPORTED_VERSION) {  // B2
        temp = new_temp;
        temp = ((temp) | (temp >> 2)) ;
        temp = (temp + (in * 17) + (in * 31)) ;
        acc ^= temp;
    } else {
        return ERROR_FW_UNSUPPORTED;
    }
} else {
    int new_temp = (acc * 5) + (in * 11) ;
    if (hw_ready & HW_READY_MASK) {             // B3
        temp = new_temp;
        temp = ((temp >> 2) ^ (temp)) ;
        acc = (acc + temp) ;
    } else {
        return ERROR_HW_NOT_READY;
    }
}
\end{lstlisting}
\vspace{0.15in} 
\end{listing}

\begin{figure*}[htp]
    \centering
    \input{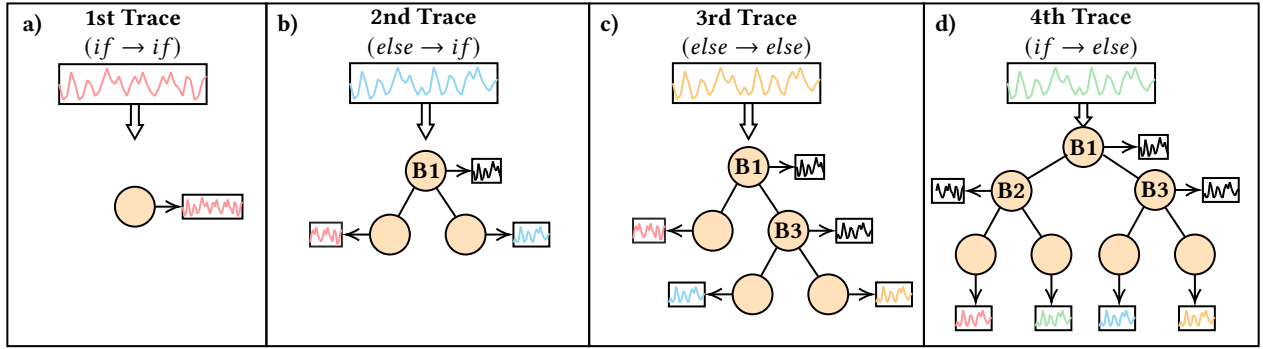}
    \caption{An overview of control-flow graph update based on power traces collected during the execution of an example firmware shown in Listing~\ref{lst:ex_branch}. The corresponding execution path for each trace is also mentioned below the title. The first trace initializes the TCFG as a single root node storing the complete power trace. When the 2nd trace is encountered, the \textsc{GetDivergentPoint} detects a divergence partway through the root node's trace segment at branch point $B1$. Then the root is split with a new node $B1$ storing the shared prefix up to the divergence point, a new left node storing the remaining trace suffix of the original root, and a new right node storing the remaining trace suffix of the new trace. (c) and (d) show the TCFG after the third and fourth traces are processed, following the same procedure.}
    \label{fig:cfg_gen2}
\end{figure*}

The TCFG is a graph $\mathcal{G} = (\mathcal{V}, \mathcal{E})$, where each node $v \in \mathcal{V}$ represents a basic block, a maximal straight-line sequence of instructions with no internal branches, and each directed edge $(v_i, v_j) \in \mathcal{E}$ represents a possible execution transition between two basic blocks. 

PowerFuzz dynamically constructs the TCFG of the firmware from power traces, using the branch identification method described in Section~\ref{sec:ebi}. As the fuzzing campaign progresses and more execution paths are exercised, the generated TCFG grows to closely approximate the actual firmware's control flow graph (CFG). Note that TCFG will be same as CFG if the underlying fuzzer can cover all the branches in the firmware. The dynamic nature of TCFG serves two complementary purposes: (i) it guides the fuzzer toward unexplored branches by providing structural knowledge of the firmware's execution behavior, and (ii) it organizes previously captured power traces in a structured and searchable form, reducing the number of pairwise comparisons required during fuzzing.

\subsubsection{Node Structure}
Each node $v \in \mathcal{V}$ in the PowerFuzz TCFG stores four fields: (1) a trace segment $v.data$, representing the portion of the power trace corresponding to the basic block executed at that node; (2) the input $v.input$ that corresponds to reach this node; (3) a left child pointer $v.left$; and (4) a right child pointer $v.right$. The binary child structure directly reflects the nature of branch instructions in firmware binaries. At any branch point, execution follows exactly one of two possible paths: the branch-taken path or the branch-not-taken path corresponding to the left and right children, respectively.

\subsubsection{TCFG Construction}
Algorithm~\ref{alg:cfg_update} describes the TCFG construction and update procedure. To explain the TCFG construction clearly, we use an illustrative example using the firmware Listing~\ref{lst:ex_branch} and Figure~\ref{fig:cfg_gen2}. Assume that four inputs are dispatched in sequence, following the paths \textit{if}$\rightarrow$\textit{if}, \textit{else}$\rightarrow$\textit{if}, \textit{else}$\rightarrow$\textit{else}, and \textit{if}$\rightarrow$\textit{else}, through branch points $B1$, $B2$, and $B3$.

When the first trace arrives, the TCFG is initialized with that trace stored as the root node. As shown in Figure~\ref{fig:cfg_gen2}(a), the root node holds the complete power trace of the \textit{if}$\rightarrow$\textit{if} execution path with no children, representing the entire firmware execution observed so far. For each subsequent power trace $P_{new}$, the algorithm recursively traverses the TCFG from the root to determine where the new trace diverges from the existing structure. At each node $v$, \textsc{GetDivergentPoint} (Algorithm~\ref{alg:branch_identifier}) is called between $v.data$ and the corresponding segment of $T_{new}$. Two cases arise depending on the result:

\textit{Case 1: New trace matches the current node ($dp = |v.data|$ line 7).} The new trace is fully consistent with the current node's trace segment, meaning the new input follows the same execution path through this basic block. The remaining trace segment $T_{rem}$ is then compared against both children of $v$ by computing $dp_{left}$ and $dp_{right}$ using \textsc{GetDivergentPoint}. If $dp_{left} \neq 0$, the remaining trace shares execution history with the left child and the algorithm recurses into $v.left$ with $T_{rem}$. If $dp_{right} \neq 0$. If $dp_{right} \neq 0$, it recurses into $v.right$ instead. In the example, this case is encountered when processing the third and fourth traces, which both match the shared $B1$ prefix before diverging deeper into the tree at $B2$ and $B3$ respectively.

\textit{Case 2: New trace diverges within the current node ($dp < |v.data|$ line 17).} The new trace diverges partway through the current node's basic block, revealing a previously unidentified branch within an already-captured basic block. The node $v$ is split at the divergence point $dp$: a new parent node $v_{new}$ is created storing the shared prefix $v.data[:dp]$; the current node's trace is trimmed to the remaining suffix $v.data[dp:]$; and a new sibling node $v_{sib}$ is created storing the diverging trace segment $T_{new}[dp:]$. The trimmed current node is attached as the left child of $v_{new}$, preserving the previously identified execution path, and $v_{sib}$ is attached as the right child, representing the newly discovered branch direction. If $v$ was previously the root, $v_{new}$ becomes the new root; otherwise, $v_{new}$ replaces $v$ in its parent's left or right pointer. 

This case is first encountered when the second trace (\textit{else}$\rightarrow$\textit{if}) arrives. \textsc{GetDivergentPoint} detects a divergence within the root at branch point $B1$, where the outer \texttt{if} and \texttt{else} paths produce measurably different power patterns. As shown in Figure~\ref{fig:cfg_gen2}(b), the root is split into a new parent node $B1$ storing the shared prefix up to the divergence point, with the original root's remaining trace attached as the left child and the second trace's remaining segment attached as the right child. When the third trace (\textit{else}$\rightarrow$\textit{else}) arrives, it matches $B1$ and recurses into the right child, since both the second and third traces enter the outer \texttt{else} branch. A divergence is then detected within that child at $B3$, where the inner \texttt{if} and \texttt{else} paths differ, causing a further split and producing the structure shown in Figure~\ref{fig:cfg_gen2}(c). The fourth trace (\textit{if}$\rightarrow$\textit{else}) matches $B1$, recurses into the left child, sharing the outer \texttt{if} path with the first trace, and triggers a split at $B2$, yielding the final TCFG shown in Figure~\ref{fig:cfg_gen2}(d).

After processing $P_{new}$, if \textsc{NewCoverage} is returned, the corresponding input is retained in the fuzzing engine's seed corpus for further mutation. If no TCFG update occurs, the input is discarded. This coverage signal replaces the conventional coverage bitmap of standard coverage-guided fuzzing, closing the feedback loop entirely from power trace observations. After all four traces, the generated TCFG structurally mirrors the actual control flow of Listing~\ref{lst:ex_branch}, with $B1$, $B2$, and $B3$ correctly identified as branch points purely from power trace observations.

\begin{algorithm}[t]
\caption{TCFG Construction and Update}
\label{alg:cfg_update}
\begin{algorithmic}[1]

\Require TCFG $\mathcal{G}$, new trace $T_{new}$
\Ensure Updated $\mathcal{G}$

\If{$\mathcal{G}$ is empty}
    \State Initialize $\mathcal{G}$ with $T_{new}$ as root
    \State \Return \textsc{NewCoverage}
\EndIf
\State \Call{UpdateRecursive}{$\mathcal{G}.root,\ T_{new}$}

\Statex
\Function{UpdateRecursive}{$v,\ T$}
    \State $dp \leftarrow$ \Call{GetDivergentPoint}{$v.data,\ T$}
    \If{$dp = |v.data|$}
        \Comment{Trace matches current node length}
        \State $T_{rem} \leftarrow T[|v.data|:]$
        \State $dp_{left} \leftarrow$ \Call{GetDivergentPoint}{$v.left.data,\ T_{rem}$}
        \State $dp_{right} \leftarrow$ \Call{GetDivergentPoint}{$v.right.data,\ T_{rem}$}
        \If{$dp_{left} \neq 0$}
            \State \Call{UpdateRecursive}{$v.left,\ T_{rem}$}
        \ElsIf{$dp_{right} \neq 0$}
            \State \Call{UpdateRecursive}{$v.right,\ T_{rem}$}
        \EndIf
    \ElsIf{$dp < |v.data|$}
        \Comment{Divergence within current node; split}
        \State $v_{new} \leftarrow \textsc{NewNode}(v.data[:dp])$
        \State $v_{sib} \leftarrow \textsc{NewNode}(T[dp:])$
        \State $v.data \leftarrow v.data[dp:]$
        \State $v_{new}.left \leftarrow v$
        \Comment{Existing path becomes left child}
        \State $v_{new}.right \leftarrow v_{sib}$
        \Comment{New path becomes right child}
        \If{$v.parent \neq$ \textbf{null}}
            \State Replace $v$ with $v_{new}$ in $v.parent$
        \Else
            \State $\mathcal{G}.root \leftarrow v_{new}$
        \EndIf
    \EndIf
\EndFunction

\end{algorithmic}
\end{algorithm}


\subsection{Branch Selection for Test Generation}
TCFG constructed by PowerFuzz not only serves as a control flow transition of the firmware, but also provides actionable guidance to the fuzzing engine regarding which regions of the firmware remain unexplored. The branch selector consults the current TCFG at each iteration and selects a target branch for the fuzzing engine to explore, replacing the role of the coverage bitmap in conventional coverage-guided fuzzing. PowerFuzz implements two branch selection strategies: random selection and depth-prioritized selection.

\subsubsection{Random Branch Selection:} In the early stages of the fuzzing campaign, 
the TCFG is sparse and contains few identified branches. At this stage, there is insufficient 
structural information to meaningfully differentiate between branches. PowerFuzz therefore begins with a random branch 
selection strategy, in which an unexplored branch is selected uniformly at random from the 
current TCFG frontier. The corresponding trace segment stored at the selected node is 
retrieved and provided to the fuzzing engine as a reference, which uses it to bias mutation 
toward inputs likely to reach that branch. This strategy ensures broad initial exploration 
of the firmware's execution space and populates the TCFG with enough structural information 
to support informed selection in later stages.

\subsubsection{Depth-Prioritized Branch Selection:} As the TCFG grows and deeper nodes 
are identified, PowerFuzz transitions to a depth-prioritized selection strategy. The 
intuition behind this transition is that deeper nodes in the TCFG correspond to nested 
conditional logic, multi-stage computations, or state-dependent behaviors that are 
structurally harder to reach and thus more likely to harbor complex or vulnerable firmware 
behavior. Prioritizing these nodes directs the fuzzer's mutation effort toward the most 
structurally significant unexplored regions of the firmware.

To formalize this, let $\mathcal{G} = (\mathcal{V}, \mathcal{E})$ be the current TCFG. We define the depth $d(v)$ of a node $v \in \mathcal{V}$ as the length 
of the shortest path from the root to $v$. When a newly generated input produces a trace 
that results in a TCFG update, a new node $v_{new}$ is inserted at depth $d(v_{new})$. We 
define a scalar feedback value $F$ as:

\begin{align}
F=
\begin{cases}
0, & \text{if no new node is added to the TCFG,} \\
d(v_{new}), & \text{if a new node } v_{new} \text{ is inserted.}
\end{cases}
\end{align}

\noindent A value of $F = 0$ indicates no coverage improvement, while larger positive values indicate that the input has explored deeper regions of the firmware's TCFG. The scalar feedback value $F$ is used directly as the mutation effort assigned to its corresponding input by the fuzzing engine, replacing the binary coverage indicator in the feedback stage of the LibAFL loop described in Section~\ref{sec:background_fuzzing}. This enables the scheduler to assign proportionally higher mutation effort  to inputs that introduce nodes at greater depths, while inputs that produce no TCFG update are assigned zero effort and discarded from the corpus.

%% file: figs/dtw_divergence_detection.tex
\begin{tikzpicture}

\definecolor{crimson2305770}{RGB}{230,57,70}
\definecolor{darkgray176}{RGB}{176,176,176}
\definecolor{darkslateblue7891133}{RGB}{85,144,170}
\definecolor{lightgray204}{RGB}{204,204,204}
\definecolor{peru17915666}{RGB}{179,156,66}
\definecolor{sandybrown23019765}{RGB}{230,197,65}
\definecolor{sienna1408284}{RGB}{181,134,82}

\begin{groupplot}[group style={group size=1 by 3, vertical sep=1.2cm}, height=3.5cm, width=8.8cm]
\nextgroupplot[
xlabel={\small Samples},
xlabel shift={-2ex},
ylabel={\small Voltage},
ylabel shift={-2ex},
legend cell align={left},
tick label style={font=\small},
legend style={
  fill opacity=0.8,
  draw opacity=1,
  text opacity=1,
  at={(0.97,0.03)},
  anchor=south east,
  draw=lightgray204
},
tick align=outside,
tick pos=left,
title={\textbf{a) Trace 1}},
title style={yshift=-2ex},
x grid style={darkgray176},
xmin=-49.95, xmax=1048.95,
xtick style={color=black},
y grid style={darkgray176},
ymin=-0.400048828125, ymax=0.215478515625,
ytick style={color=black}
]
\draw[semithick, black, dashed] (axis cs:483, \pgfkeysvalueof{/pgfplots/ymin}) -- (axis cs:483, \pgfkeysvalueof{/pgfplots/ymax});
\draw[semithick, black, dashed] (axis cs:633, \pgfkeysvalueof{/pgfplots/ymin}) -- (axis cs:633, \pgfkeysvalueof{/pgfplots/ymax});
\addplot [semithick, darkslateblue7891133]
table {%
0 0.0498046875
1 -0.0791015625
2 -0.2119140625
3 -0.15234375
4 -0.017578125
5 -0.173828125
6 -0.2880859375
7 -0.2177734375
8 -0.0791015625
9 -0.2158203125
10 -0.3720703125
11 -0.2705078125
12 -0.1044921875
13 -0.1728515625
14 -0.15625
15 -0.109375
16 0.0224609375
17 -0.0380859375
18 -0.03125
19 -0.0078125
20 0.099609375
21 -0.099609375
22 -0.2509765625
23 -0.1728515625
24 -0.0205078125
25 -0.068359375
26 -0.044921875
27 -0.0185546875
28 0.091796875
29 -0.01171875
30 -0.0361328125
31 -0.0087890625
32 0.1025390625
33 0.0107421875
34 0.0048828125
35 0.0205078125
36 0.1240234375
37 0.0185546875
38 0.0126953125
39 0.0283203125
40 0.1357421875
41 0.048828125
42 0.033203125
43 0.041015625
44 0.12890625
45 0.0302734375
46 0.0048828125
47 0.0205078125
48 0.1240234375
49 0.041015625
50 0.0224609375
51 0.033203125
52 0.1298828125
53 0.03125
54 0
55 0.021484375
56 0.126953125
57 0.0458984375
58 0.0205078125
59 0.0302734375
60 0.1298828125
61 0.021484375
62 -0.025390625
63 0
64 0.1103515625
65 0.0205078125
66 -0.015625
67 0.009765625
68 0.1171875
69 0.0146484375
70 -0.009765625
71 0.013671875
72 0.1240234375
73 0.03125
74 -0.0009765625
75 0.0126953125
76 0.1123046875
77 0.0126953125
78 -0.0107421875
79 0.0107421875
80 0.1123046875
81 0.0146484375
82 -0.005859375
83 0.0146484375
84 0.1201171875
85 0.0087890625
86 -0.0185546875
87 0.005859375
88 0.1083984375
89 0.0283203125
90 0.0009765625
91 0.0087890625
92 0.1005859375
93 -0.0693359375
94 -0.166015625
95 -0.1201171875
96 0.0126953125
97 -0.12109375
98 -0.2841796875
99 -0.19921875
100 -0.044921875
101 -0.107421875
102 -0.1455078125
103 -0.0908203125
104 0.0390625
105 -0.1162109375
106 -0.166015625
107 -0.1142578125
108 0.0126953125
109 -0.0458984375
110 -0.0009765625
111 0.0126953125
112 0.1083984375
113 0.0302734375
114 0.0263671875
115 0.02734375
116 0.1171875
117 0.0615234375
118 0.0947265625
119 0.08203125
120 0.158203125
121 0.0029296875
122 -0.1884765625
123 -0.125
124 0.00390625
125 -0.0498046875
126 -0.087890625
127 -0.041015625
128 0.0751953125
129 -0.080078125
130 -0.1494140625
131 -0.103515625
132 0.0146484375
133 -0.0107421875
134 0.048828125
135 0.0478515625
136 0.13671875
137 0.017578125
138 -0.0029296875
139 0.0029296875
140 0.1064453125
141 0.0380859375
142 0.072265625
143 0.0654296875
144 0.1494140625
145 0.048828125
146 0.037109375
147 0.046875
148 0.1435546875
149 0.0830078125
150 0.1123046875
151 0.103515625
152 0.1875
153 -0.01171875
154 -0.2177734375
155 -0.1533203125
156 -0.0107421875
157 -0.06640625
158 -0.0859375
159 -0.0439453125
160 0.072265625
161 -0.1220703125
162 -0.2001953125
163 -0.1279296875
164 0.021484375
165 -0.017578125
166 0.046875
167 0.0615234375
168 0.1572265625
169 0.021484375
170 -0.0205078125
171 -0.005859375
172 0.095703125
173 0.0029296875
174 0.0244140625
175 0.0283203125
176 0.1220703125
177 -0.0185546875
178 -0.060546875
179 -0.037109375
180 0.0732421875
181 0.033203125
182 0.076171875
183 0.064453125
184 0.1484375
185 -0.005859375
186 -0.212890625
187 -0.140625
188 -0.0009765625
189 -0.0400390625
190 -0.0869140625
191 -0.0361328125
192 0.0791015625
193 -0.08984375
194 -0.17578125
195 -0.1220703125
196 0.005859375
197 -0.02734375
198 0.0341796875
199 0.0390625
200 0.12890625
201 0.0107421875
202 -0.0126953125
203 -0.0078125
204 0.087890625
205 0.001953125
206 0.0380859375
207 0.0380859375
208 0.1298828125
209 0.0185546875
210 0.009765625
211 0.0205078125
212 0.1142578125
213 0.0595703125
214 0.0986328125
215 0.087890625
216 0.169921875
217 -0.0283203125
218 -0.232421875
219 -0.1689453125
220 -0.0224609375
221 -0.064453125
222 -0.0615234375
223 -0.025390625
224 0.0869140625
225 -0.0361328125
226 -0.056640625
227 -0.033203125
228 0.0771484375
229 0.0029296875
230 -0.0234375
231 0.0029296875
232 0.107421875
233 -0.041015625
234 -0.1064453125
235 -0.0673828125
236 0.0546875
237 -0.0009765625
238 0.0341796875
239 0.0400390625
240 0.1337890625
241 0.0537109375
242 0.0703125
243 0.060546875
244 0.1416015625
245 0.05078125
246 0.0615234375
247 0.0634765625
248 0.1455078125
249 0.0478515625
250 0.041015625
251 0.0439453125
252 0.134765625
253 0.0263671875
254 -0.001953125
255 0.01171875
256 0.1181640625
257 0.0556640625
258 0.0908203125
259 0.0869140625
260 0.17578125
261 0.0439453125
262 0.021484375
263 0.0166015625
264 0.1103515625
265 0.0146484375
266 0.0439453125
267 0.0439453125
268 0.134765625
269 0.091796875
270 0.130859375
271 0.1064453125
272 0.17578125
273 0.056640625
274 0.0537109375
275 0.052734375
276 0.1396484375
277 0.0615234375
278 0.091796875
279 0.0810546875
280 0.166015625
281 0.0634765625
282 0.0908203125
283 0.0732421875
284 0.1494140625
285 0.0087890625
286 -0.0078125
287 -0.0009765625
288 0.1015625
289 0.0302734375
290 0.0576171875
291 0.0517578125
292 0.1337890625
293 0.0361328125
294 0.0458984375
295 0.04296875
296 0.1357421875
297 0.0615234375
298 0.080078125
299 0.0634765625
300 0.1494140625
301 0.0283203125
302 0.0087890625
303 0.005859375
304 0.0966796875
305 0.025390625
306 0.0556640625
307 0.044921875
308 0.1279296875
309 0.037109375
310 0.05078125
311 0.0439453125
312 0.126953125
313 0.017578125
314 0.01171875
315 0.0234375
316 0.123046875
317 0.0546875
318 0.0673828125
319 0.064453125
320 0.1533203125
321 0.0400390625
322 0.046875
323 0.052734375
324 0.1513671875
325 0.048828125
326 0.05078125
327 0.0419921875
328 0.123046875
329 -0.0068359375
330 -0.01171875
331 0.0078125
332 0.1162109375
333 -0.033203125
334 -0.0771484375
335 -0.048828125
336 0.060546875
337 0.0048828125
338 0.05078125
339 0.041015625
340 0.1259765625
341 0.01953125
342 0.0166015625
343 0.017578125
344 0.1162109375
345 0.0400390625
346 0.0625
347 0.056640625
348 0.146484375
349 0.037109375
350 0.037109375
351 0.0458984375
352 0.1435546875
353 0.048828125
354 0.0546875
355 0.0419921875
356 0.12890625
357 -0.0244140625
358 -0.033203125
359 -0.009765625
360 0.1015625
361 -0.0517578125
362 -0.0966796875
363 -0.0634765625
364 0.0546875
365 0.0009765625
366 0.0458984375
367 0.0400390625
368 0.1220703125
369 0.0283203125
370 0.013671875
371 0.0234375
372 0.1142578125
373 0.04296875
374 0.056640625
375 0.052734375
376 0.1396484375
377 0.0458984375
378 0.0458984375
379 0.0439453125
380 0.1416015625
381 0.056640625
382 0.044921875
383 0.037109375
384 0.123046875
385 -0.0068359375
386 -0.021484375
387 -0.0029296875
388 0.1064453125
389 -0.0595703125
390 -0.1044921875
391 -0.0732421875
392 0.0458984375
393 -0.0419921875
394 -0.0087890625
395 0.001953125
396 0.103515625
397 -0.0283203125
398 -0.04296875
399 -0.0244140625
400 0.08203125
401 -0.01171875
402 0.005859375
403 0.0126953125
404 0.111328125
405 0.0283203125
406 0.044921875
407 0.0400390625
408 0.12109375
409 0.046875
410 0.078125
411 0.0615234375
412 0.1455078125
413 0.0068359375
414 -0.0244140625
415 -0.0068359375
416 0.1005859375
417 -0.05078125
418 -0.0849609375
419 -0.068359375
420 0.0400390625
421 -0.0185546875
422 0.0390625
423 0.0322265625
424 0.1259765625
425 0.0263671875
426 0.0439453125
427 0.041015625
428 0.1328125
429 0.037109375
430 0.0380859375
431 0.0361328125
432 0.1279296875
433 0.0361328125
434 0.05078125
435 0.041015625
436 0.1259765625
437 0.0478515625
438 0.0654296875
439 0.0537109375
440 0.134765625
441 0.005859375
442 -0.033203125
443 -0.0107421875
444 0.0966796875
445 -0.056640625
446 -0.09375
447 -0.076171875
448 0.033203125
449 -0.0146484375
450 0.0390625
451 0.0400390625
452 0.1220703125
453 0.021484375
454 0.0322265625
455 0.0302734375
456 0.123046875
457 0.0341796875
458 0.041015625
459 0.03125
460 0.1220703125
461 0.0380859375
462 0.0498046875
463 0.041015625
464 0.125
465 0.046875
466 0.0673828125
467 0.046875
468 0.1298828125
469 -0.005859375
470 -0.0458984375
471 -0.0224609375
472 0.0869140625
473 -0.0576171875
474 -0.1025390625
475 -0.0830078125
476 0.0263671875
477 -0.0234375
478 0.0322265625
479 0.03125
480 0.1220703125
481 0.013671875
482 0.029296875
483 0.0263671875
484 0.1181640625
485 0.013671875
486 0.0302734375
487 0.0244140625
488 0.111328125
489 0.0146484375
490 0.025390625
491 0.01953125
492 0.111328125
493 0.0302734375
494 0.046875
495 0.0400390625
496 0.115234375
497 -0.0087890625
498 -0.05078125
499 -0.025390625
500 0.083984375
501 -0.0537109375
502 -0.1044921875
503 -0.08984375
504 0.0185546875
505 -0.01953125
506 0.0283203125
507 0.0302734375
508 0.1201171875
509 0.03125
510 0.03125
511 0.0283203125
512 0.1181640625
513 0.0400390625
514 0.0478515625
515 0.0419921875
516 0.126953125
517 0.033203125
518 0.037109375
519 0.02734375
520 0.1123046875
521 0.03125
522 0.0419921875
523 0.0341796875
524 0.1181640625
525 -0.0263671875
526 -0.060546875
527 -0.03515625
528 0.0751953125
529 -0.0693359375
530 -0.111328125
531 -0.0908203125
532 0.009765625
533 -0.0380859375
534 0.0166015625
535 0.0205078125
536 0.111328125
537 0.0205078125
538 0.033203125
539 0.02734375
540 0.11328125
541 0.0341796875
542 0.0517578125
543 0.0419921875
544 0.126953125
545 0.03515625
546 0.0458984375
547 0.02734375
548 0.1142578125
549 0.01953125
550 0.0380859375
551 0.0283203125
552 0.1123046875
553 -0.021484375
554 -0.060546875
555 -0.037109375
556 0.0732421875
557 -0.0693359375
558 -0.115234375
559 -0.0947265625
560 0.015625
561 -0.0341796875
562 0.0185546875
563 0.0224609375
564 0.111328125
565 0.0283203125
566 0.0341796875
567 0.0302734375
568 0.1162109375
569 0.0400390625
570 0.0556640625
571 0.048828125
572 0.12890625
573 0.037109375
574 0.0341796875
575 0.02734375
576 0.1044921875
577 0.0185546875
578 0.01953125
579 0.017578125
580 0.1025390625
581 -0.0283203125
582 -0.0771484375
583 -0.0556640625
584 0.056640625
585 -0.0791015625
586 -0.1259765625
587 -0.1015625
588 0.0068359375
589 -0.033203125
590 0.0126953125
591 0.015625
592 0.1044921875
593 0.0166015625
594 0.0302734375
595 0.0263671875
596 0.11328125
597 0.037109375
598 0.0556640625
599 0.0439453125
600 0.126953125
601 0.029296875
602 0.029296875
603 0.0205078125
604 0.10546875
605 0.009765625
606 0.021484375
607 0.015625
608 0.1025390625
609 -0.0283203125
610 -0.0595703125
611 -0.0400390625
612 0.0654296875
613 -0.0986328125
614 -0.1328125
615 -0.1123046875
616 0
617 -0.0556640625
618 0.00390625
619 0.0048828125
620 0.0966796875
621 0.005859375
622 0.029296875
623 0.0205078125
624 0.111328125
625 0.0390625
626 0.0654296875
627 0.0439453125
628 0.126953125
629 0.01953125
630 0.01953125
631 0.0107421875
632 0.0986328125
633 0.013671875
634 0.01953125
635 0.015625
636 0.1015625
637 -0.021484375
638 -0.0595703125
639 -0.0380859375
640 0.072265625
641 -0.080078125
642 -0.12109375
643 -0.1025390625
644 0.005859375
645 -0.0458984375
646 0.009765625
647 0.009765625
648 0.103515625
649 0
650 0.0244140625
651 0.017578125
652 0.107421875
653 0.0361328125
654 0.0712890625
655 0.052734375
656 0.1259765625
657 0.0263671875
658 0.033203125
659 0.0185546875
660 0.1044921875
661 0.013671875
662 0.02734375
663 0.0107421875
664 0.09765625
665 -0.025390625
666 -0.0546875
667 -0.03515625
668 0.072265625
669 -0.08203125
670 -0.12109375
671 -0.109375
672 0.0009765625
673 -0.0517578125
674 0.005859375
675 0.0068359375
676 0.099609375
677 -0.0068359375
678 0.0107421875
679 0.0068359375
680 0.0986328125
681 0.0263671875
682 0.05859375
683 0.04296875
684 0.1240234375
685 0.0224609375
686 0.0244140625
687 0.017578125
688 0.1015625
689 0.015625
690 0.0224609375
691 0.0146484375
692 0.09375
693 -0.0283203125
694 -0.068359375
695 -0.044921875
696 0.0634765625
697 -0.076171875
698 -0.1201171875
699 -0.1025390625
700 -0.001953125
701 -0.0439453125
702 0.009765625
703 0.0107421875
704 0.1005859375
705 0.015625
706 0.0322265625
707 0.0185546875
708 0.10546875
709 0.0390625
710 0.064453125
711 0.0517578125
712 0.1279296875
713 0.0380859375
714 0.0263671875
715 0.017578125
716 0.0986328125
717 0.0146484375
718 0.0166015625
719 0.0087890625
720 0.0947265625
721 -0.0322265625
722 -0.076171875
723 -0.05078125
724 0.0556640625
725 -0.0810546875
726 -0.13671875
727 -0.1123046875
728 -0.0048828125
729 -0.052734375
730 0.0009765625
731 0.0048828125
732 0.09375
733 0.009765625
734 0.0205078125
735 0.0166015625
736 0.095703125
737 0.0322265625
738 0.064453125
739 0.0478515625
740 0.126953125
741 0.0146484375
742 0.025390625
743 0.009765625
744 0.091796875
745 -0.0068359375
746 0.0087890625
747 0.001953125
748 0.087890625
749 -0.0322265625
750 -0.064453125
751 -0.046875
752 0.060546875
753 -0.0810546875
754 -0.1240234375
755 -0.1044921875
756 0.001953125
757 -0.0439453125
758 -0.0009765625
759 0.0029296875
760 0.0927734375
761 0.0068359375
762 0.0263671875
763 0.0205078125
764 0.1044921875
765 0.0263671875
766 0.0625
767 0.044921875
768 0.125
769 0.0283203125
770 0.044921875
771 0.02734375
772 0.1044921875
773 0.001953125
774 0.0126953125
775 0.0029296875
776 0.087890625
777 -0.0341796875
778 -0.064453125
779 -0.0439453125
780 0.0576171875
781 -0.0908203125
782 -0.1337890625
783 -0.1142578125
784 -0.0048828125
785 -0.052734375
786 0.0009765625
787 -0.0068359375
788 0.087890625
789 -0.0068359375
790 0.017578125
791 0.0107421875
792 0.0986328125
793 0.02734375
794 0.0634765625
795 0.044921875
796 0.123046875
797 0.0341796875
798 0.0576171875
799 0.0361328125
800 0.1142578125
801 0.0166015625
802 0.0205078125
803 0.005859375
804 0.0927734375
805 -0.0546875
806 -0.0859375
807 -0.0654296875
808 0.0478515625
809 -0.109375
810 -0.1416015625
811 -0.12109375
812 -0.0107421875
813 -0.060546875
814 -0.001953125
815 -0.0009765625
816 0.0830078125
817 -0.0087890625
818 0.0068359375
819 0.0029296875
820 0.091796875
821 0.0263671875
822 0.060546875
823 0.0400390625
824 0.1162109375
825 0.037109375
826 0.0556640625
827 0.0390625
828 0.1123046875
829 0.017578125
830 0.017578125
831 0.0009765625
832 0.0869140625
833 -0.037109375
834 -0.0771484375
835 -0.0537109375
836 0.0546875
837 -0.0947265625
838 -0.146484375
839 -0.125
840 -0.013671875
841 -0.060546875
842 -0.0078125
843 -0.0029296875
844 0.0859375
845 -0.005859375
846 0.015625
847 0.0087890625
848 0.0966796875
849 0.0341796875
850 0.0693359375
851 0.0517578125
852 0.1240234375
853 0.04296875
854 0.0634765625
855 0.041015625
856 0.1162109375
857 0.0126953125
858 0.009765625
859 0.0009765625
860 0.0771484375
861 -0.0517578125
862 -0.087890625
863 -0.0634765625
864 0.0478515625
865 -0.099609375
866 -0.1396484375
867 -0.1279296875
868 -0.0146484375
869 -0.076171875
870 -0.021484375
871 -0.0166015625
872 0.076171875
873 -0.0205078125
874 -0.0009765625
875 -0.0078125
876 0.0830078125
877 0.0234375
878 0.0634765625
879 0.046875
880 0.1240234375
881 0.046875
882 0.060546875
883 0.0400390625
884 0.115234375
885 0.01953125
886 0.017578125
887 0.0068359375
888 0.08984375
889 -0.037109375
890 -0.0732421875
891 -0.0517578125
892 0.0546875
893 -0.083984375
894 -0.130859375
895 -0.11328125
896 -0.0126953125
897 -0.0537109375
898 -0.0048828125
899 -0.001953125
900 0.0869140625
901 -0.0029296875
902 0.0126953125
903 0.0068359375
904 0.0869140625
905 0.0234375
906 0.060546875
907 0.044921875
908 0.1201171875
909 0.04296875
910 0.0556640625
911 0.0302734375
912 0.1083984375
913 0.013671875
914 0.017578125
915 0.005859375
916 0.0869140625
917 -0.0380859375
918 -0.087890625
919 -0.0654296875
920 0.04296875
921 -0.095703125
922 -0.1376953125
923 -0.119140625
924 -0.0126953125
925 -0.0615234375
926 -0.0078125
927 -0.0068359375
928 0.0859375
929 -0.0029296875
930 0.0185546875
931 0.009765625
932 0.095703125
933 0.013671875
934 0.056640625
935 0.0380859375
936 0.1162109375
937 0.0322265625
938 0.0556640625
939 0.03515625
940 0.1044921875
941 0.021484375
942 0.0263671875
943 0.0126953125
944 0.09375
945 -0.033203125
946 -0.06640625
947 -0.0556640625
948 0.052734375
949 -0.095703125
950 -0.1357421875
951 -0.1201171875
952 -0.01171875
953 -0.05859375
954 -0.0048828125
955 -0.0078125
956 0.0830078125
957 -0.0068359375
958 0.0126953125
959 0.0048828125
960 0.091796875
961 0.0205078125
962 0.0556640625
963 0.037109375
964 0.1142578125
965 0.02734375
966 0.0517578125
967 0.0302734375
968 0.1064453125
969 0.005859375
970 0.0205078125
971 0.0078125
972 0.08984375
973 -0.0302734375
974 -0.064453125
975 -0.0478515625
976 0.052734375
977 -0.09375
978 -0.1357421875
979 -0.1181640625
980 -0.01171875
981 -0.060546875
982 -0.0068359375
983 -0.0087890625
984 0.078125
985 -0.009765625
986 0.0126953125
987 0.0029296875
988 0.091796875
989 0.0234375
990 0.0654296875
991 0.037109375
992 0.1171875
993 0.0302734375
994 0.0625
995 0.0361328125
996 0.1123046875
997 0.0029296875
998 0.0185546875
999 0.001953125
};
\addlegendentry{\small Trace 1}

\nextgroupplot[
xlabel={\small Samples},
xlabel shift={-2ex},
ylabel={\small Voltage},
ylabel shift={-2ex},
legend cell align={left},
tick label style={font=\small},
legend cell align={left},
legend style={
  fill opacity=0.8,
  draw opacity=1,
  text opacity=1,
  at={(0.97,0.03)},
  anchor=south east,
  draw=lightgray204
},
tick align=outside,
tick pos=left,
title={\textbf{b) Trace 2}},
title style={yshift=-2ex},
x grid style={darkgray176},
xmin=-49.95, xmax=1048.95,
xtick style={color=black},
y grid style={darkgray176},
ymin=-0.3890625, ymax=0.22109375,
ytick style={color=black}
]
\draw[semithick, black, dashed] (axis cs:483, \pgfkeysvalueof{/pgfplots/ymin}) -- (axis cs:483, \pgfkeysvalueof{/pgfplots/ymax});
\draw[semithick, black, dashed] (axis cs:633, \pgfkeysvalueof{/pgfplots/ymin}) -- (axis cs:633, \pgfkeysvalueof{/pgfplots/ymax});
\addplot [semithick, sienna1408284]
table {%
0 0.0556640625
1 -0.0634765625
2 -0.1953125
3 -0.1455078125
4 -0.0087890625
5 -0.1552734375
6 -0.271484375
7 -0.203125
8 -0.05859375
9 -0.1923828125
10 -0.361328125
11 -0.2568359375
12 -0.0966796875
13 -0.1552734375
14 -0.1494140625
15 -0.091796875
16 0.03515625
17 -0.017578125
18 -0.025390625
19 0.001953125
20 0.1064453125
21 -0.076171875
22 -0.2412109375
23 -0.1611328125
24 -0.01171875
25 -0.0625
26 -0.0498046875
27 -0.01953125
28 0.0908203125
29 -0.0107421875
30 -0.037109375
31 -0.005859375
32 0.09765625
33 0.009765625
34 0
35 0.01953125
36 0.1240234375
37 0.033203125
38 0.0224609375
39 0.0341796875
40 0.138671875
41 0.0537109375
42 0.037109375
43 0.0458984375
44 0.138671875
45 0.041015625
46 0.0166015625
47 0.0263671875
48 0.12890625
49 0.048828125
50 0.033203125
51 0.0439453125
52 0.1416015625
53 0.044921875
54 0.0087890625
55 0.02734375
56 0.1357421875
57 0.0439453125
58 0.025390625
59 0.037109375
60 0.1376953125
61 0.013671875
62 -0.0263671875
63 -0.001953125
64 0.1123046875
65 0.021484375
66 -0.0009765625
67 0.0205078125
68 0.1240234375
69 0.0263671875
70 -0.001953125
71 0.0244140625
72 0.1318359375
73 0.0517578125
74 0.0126953125
75 0.0263671875
76 0.115234375
77 0.021484375
78 -0.0087890625
79 0.015625
80 0.1181640625
81 0.03125
82 0
83 0.0146484375
84 0.1220703125
85 0.0146484375
86 -0.0205078125
87 0.0087890625
88 0.119140625
89 0.02734375
90 -0.01171875
91 -0.0029296875
92 0.0908203125
93 -0.0751953125
94 -0.17578125
95 -0.1181640625
96 0.013671875
97 -0.1083984375
98 -0.2841796875
99 -0.1943359375
100 -0.0419921875
101 -0.0810546875
102 -0.123046875
103 -0.0634765625
104 0.0576171875
105 -0.087890625
106 -0.1455078125
107 -0.0966796875
108 0.0244140625
109 -0.021484375
110 0.0205078125
111 0.03125
112 0.11328125
113 0.0458984375
114 0.033203125
115 0.0361328125
116 0.1220703125
117 0.078125
118 0.103515625
119 0.0888671875
120 0.1572265625
121 0.005859375
122 -0.1923828125
123 -0.126953125
124 0.0078125
125 -0.046875
126 -0.107421875
127 -0.05859375
128 0.060546875
129 -0.080078125
130 -0.16015625
131 -0.1064453125
132 0.0185546875
133 -0.00390625
134 0.044921875
135 0.0517578125
136 0.1376953125
137 0.025390625
138 0.0009765625
139 0.013671875
140 0.11328125
141 0.0458984375
142 0.0810546875
143 0.07421875
144 0.1533203125
145 0.0673828125
146 0.052734375
147 0.0634765625
148 0.1552734375
149 0.09375
150 0.119140625
151 0.111328125
152 0.193359375
153 0.013671875
154 -0.2216796875
155 -0.150390625
156 -0.0166015625
157 -0.0546875
158 -0.0908203125
159 -0.041015625
160 0.0732421875
161 -0.103515625
162 -0.2041015625
163 -0.134765625
164 0.0146484375
165 -0.0087890625
166 0.0517578125
167 0.068359375
168 0.1669921875
169 0.0400390625
170 -0.0205078125
171 -0.0078125
172 0.09375
173 0.001953125
174 0.0146484375
175 0.025390625
176 0.12109375
177 -0.017578125
178 -0.0703125
179 -0.0419921875
180 0.06640625
181 0.0263671875
182 0.0703125
183 0.06640625
184 0.1494140625
185 -0.0205078125
186 -0.220703125
187 -0.1474609375
188 -0.0048828125
189 -0.0439453125
190 -0.0849609375
191 -0.03515625
192 0.078125
193 -0.0966796875
194 -0.177734375
195 -0.1259765625
196 0.0078125
197 -0.01171875
198 0.0498046875
199 0.05078125
200 0.1357421875
201 0.0205078125
202 -0.005859375
203 -0.001953125
204 0.09375
205 0.009765625
206 0.037109375
207 0.03125
208 0.125
209 0.0244140625
210 0.01171875
211 0.0234375
212 0.123046875
213 0.072265625
214 0.0966796875
215 0.0888671875
216 0.1689453125
217 -0.0341796875
218 -0.236328125
219 -0.1669921875
220 -0.021484375
221 -0.068359375
222 -0.0703125
223 -0.0302734375
224 0.08203125
225 -0.037109375
226 -0.048828125
227 -0.02734375
228 0.0810546875
229 0.0185546875
230 -0.0048828125
231 0.01953125
232 0.12109375
233 -0.0322265625
234 -0.0908203125
235 -0.0556640625
236 0.0537109375
237 -0.005859375
238 0.0263671875
239 0.0341796875
240 0.1279296875
241 0.056640625
242 0.07421875
243 0.05859375
244 0.140625
245 0.041015625
246 0.0556640625
247 0.056640625
248 0.1494140625
249 0.0498046875
250 0.0595703125
251 0.0498046875
252 0.140625
253 0.0146484375
254 0.001953125
255 0.0166015625
256 0.1240234375
257 0.05859375
258 0.0927734375
259 0.0888671875
260 0.177734375
261 0.0400390625
262 0.025390625
263 0.0244140625
264 0.1181640625
265 0.013671875
266 0.046875
267 0.0458984375
268 0.13671875
269 0.044921875
270 0.0625
271 0.05078125
272 0.130859375
273 0.0595703125
274 0.0947265625
275 0.078125
276 0.154296875
277 0.0380859375
278 0.0380859375
279 0.0380859375
280 0.1240234375
281 0.0400390625
282 0.0654296875
283 0.0498046875
284 0.130859375
285 0.013671875
286 0.013671875
287 0.0078125
288 0.1044921875
289 0.03125
290 0.0419921875
291 0.0380859375
292 0.1240234375
293 0.0244140625
294 0.0185546875
295 0.0283203125
296 0.126953125
297 0.052734375
298 0.05859375
299 0.044921875
300 0.123046875
301 0
302 -0.0341796875
303 -0.017578125
304 0.0810546875
305 0.0224609375
306 0.0517578125
307 0.041015625
308 0.1220703125
309 0.02734375
310 0.0546875
311 0.052734375
312 0.142578125
313 0.021484375
314 0.00390625
315 0.0126953125
316 0.10546875
317 0.029296875
318 0.0498046875
319 0.0498046875
320 0.1376953125
321 0.0439453125
322 0.0556640625
323 0.0478515625
324 0.1298828125
325 0.046875
326 0.0732421875
327 0.0615234375
328 0.1455078125
329 0.0263671875
330 0.0146484375
331 0.0224609375
332 0.126953125
333 -0.025390625
334 -0.0966796875
335 -0.0654296875
336 0.046875
337 0.0029296875
338 0.04296875
339 0.046875
340 0.134765625
341 0.0439453125
342 0.0478515625
343 0.0458984375
344 0.134765625
345 0.0341796875
346 0.0458984375
347 0.0439453125
348 0.1298828125
349 0.0361328125
350 0.0419921875
351 0.0361328125
352 0.1220703125
353 0.0478515625
354 0.0751953125
355 0.0625
356 0.142578125
357 0.0263671875
358 0.0107421875
359 0.0263671875
360 0.1201171875
361 -0.0341796875
362 -0.115234375
363 -0.0830078125
364 0.03125
365 -0.0283203125
366 0.00390625
367 0.001953125
368 0.0908203125
369 -0.01953125
370 -0.029296875
371 -0.009765625
372 0.0927734375
373 0.021484375
374 0.0400390625
375 0.0400390625
376 0.1298828125
377 0.025390625
378 0.0322265625
379 0.0400390625
380 0.1376953125
381 0.0458984375
382 0.0361328125
383 0.0341796875
384 0.1240234375
385 0.0107421875
386 0.001953125
387 0.017578125
388 0.1201171875
389 -0.0283203125
390 -0.0732421875
391 -0.052734375
392 0.052734375
393 0.0126953125
394 0.056640625
395 0.046875
396 0.12109375
397 0.021484375
398 0.0068359375
399 0.0166015625
400 0.11328125
401 0.0419921875
402 0.056640625
403 0.052734375
404 0.1357421875
405 0.0390625
406 0.0517578125
407 0.0537109375
408 0.146484375
409 0.0517578125
410 0.0439453125
411 0.0302734375
412 0.1201171875
413 -0.0029296875
414 -0.02734375
415 -0.005859375
416 0.1005859375
417 -0.0390625
418 -0.1005859375
419 -0.0732421875
420 0.037109375
421 -0.0029296875
422 0.041015625
423 0.03515625
424 0.11328125
425 0.013671875
426 -0.0107421875
427 0.001953125
428 0.0986328125
429 0.02734375
430 0.03515625
431 0.0390625
432 0.125
433 0.0283203125
434 0.0380859375
435 0.0439453125
436 0.1396484375
437 0.052734375
438 0.046875
439 0.041015625
440 0.1171875
441 -0.0068359375
442 -0.02734375
443 -0.0068359375
444 0.099609375
445 -0.041015625
446 -0.09375
447 -0.0751953125
448 0.0361328125
449 -0.0048828125
450 0.0439453125
451 0.0341796875
452 0.1171875
453 0.01171875
454 -0.00390625
455 -0.0009765625
456 0.09765625
457 0.01953125
458 0.0283203125
459 0.029296875
460 0.119140625
461 0.0322265625
462 0.041015625
463 0.044921875
464 0.138671875
465 0.0517578125
466 0.046875
467 0.0400390625
468 0.125
469 -0.0087890625
470 -0.025390625
471 -0.0068359375
472 0.095703125
473 -0.06640625
474 -0.1181640625
475 -0.091796875
476 0.017578125
477 -0.0341796875
478 0.0166015625
479 0.0126953125
480 0.09765625
481 -0.001953125
482 -0.015625
483 -0.0048828125
484 0.0869140625
485 0.015625
486 0.03125
487 0.029296875
488 0.119140625
489 0.017578125
490 0.0224609375
491 0.0224609375
492 0.1220703125
493 0.0478515625
494 0.0556640625
495 0.0458984375
496 0.126953125
497 0.0029296875
498 -0.01953125
499 -0.005859375
500 0.0986328125
501 -0.044921875
502 -0.1015625
503 -0.0771484375
504 0.03125
505 -0.0576171875
506 -0.029296875
507 -0.0185546875
508 0.0810546875
509 -0.0458984375
510 -0.060546875
511 -0.060546875
512 0.0302734375
513 -0.0146484375
514 0.0380859375
515 0.0166015625
516 0.0908203125
517 -0.041015625
518 -0.0576171875
519 -0.056640625
520 0.0322265625
521 -0.0673828125
522 -0.0732421875
523 -0.0595703125
524 0.0419921875
525 -0.0302734375
526 0.0029296875
527 -0.0166015625
528 0.05859375
529 -0.021484375
530 0.01171875
531 0.0068359375
532 0.09765625
533 -0.0107421875
534 -0.0126953125
535 -0.009765625
536 0.091796875
537 0.017578125
538 0.0498046875
539 0.03125
540 0.1103515625
541 0.0068359375
542 0.0107421875
543 0.0078125
544 0.0966796875
545 0.017578125
546 0.0341796875
547 0.025390625
548 0.1083984375
549 0.01953125
550 0.0439453125
551 0.0390625
552 0.1240234375
553 0.0625
554 0.0947265625
555 0.0673828125
556 0.1357421875
557 0.0166015625
558 -0.0029296875
559 -0.0126953125
560 0.0703125
561 -0.0009765625
562 0.025390625
563 0.0126953125
564 0.0908203125
565 -0.0283203125
566 -0.046875
567 -0.037109375
568 0.064453125
569 -0.0146484375
570 0.01171875
571 -0.001953125
572 0.0927734375
573 -0.01171875
574 0.0107421875
575 0.0126953125
576 0.1083984375
577 0.0126953125
578 0.0126953125
579 0.0009765625
580 0.087890625
581 -0.0224609375
582 -0.0048828125
583 -0.001953125
584 0.0947265625
585 0.009765625
586 -0.001953125
587 0.0009765625
588 0.095703125
589 -0.005859375
590 -0.005859375
591 0.005859375
592 0.107421875
593 0.03515625
594 0.0693359375
595 0.0517578125
596 0.1298828125
597 0.0009765625
598 -0.0146484375
599 -0.0244140625
600 0.056640625
601 -0.0263671875
602 0.0126953125
603 0.0029296875
604 0.0869140625
605 -0.0166015625
606 -0.005859375
607 -0.005859375
608 0.0830078125
609 -0.0009765625
610 0.02734375
611 0.013671875
612 0.095703125
613 0.009765625
614 0.017578125
615 -0.00390625
616 0.078125
617 -0.0087890625
618 0.0205078125
619 0.0107421875
620 0.0947265625
621 0.0087890625
622 0.0244140625
623 0.0166015625
624 0.0986328125
625 0.029296875
626 0.0576171875
627 0.033203125
628 0.107421875
629 -0.013671875
630 -0.037109375
631 -0.041015625
632 0.0498046875
633 -0.0673828125
634 -0.0791015625
635 -0.0673828125
636 0.0322265625
637 -0.0458984375
638 -0.021484375
639 -0.0234375
640 0.06640625
641 -0.0595703125
642 -0.0693359375
643 -0.076171875
644 0.0078125
645 -0.064453125
646 -0.0166015625
647 -0.0185546875
648 0.07421875
649 -0.033203125
650 -0.029296875
651 -0.0283203125
652 0.0751953125
653 -0.0107421875
654 -0.021484375
655 -0.0322265625
656 0.0556640625
657 -0.0771484375
658 -0.099609375
659 -0.0908203125
660 0.0166015625
661 -0.0458984375
662 -0.0185546875
663 -0.0185546875
664 0.072265625
665 -0.0361328125
666 -0.0263671875
667 -0.0361328125
668 0.048828125
669 -0.06640625
670 -0.06640625
671 -0.0556640625
672 0.044921875
673 -0.0361328125
674 -0.0048828125
675 -0.0068359375
676 0.0791015625
677 -0.005859375
678 -0.00390625
679 -0.0009765625
680 0.087890625
681 -0.0087890625
682 -0.0126953125
683 -0.009765625
684 0.0849609375
685 -0.0380859375
686 -0.060546875
687 -0.0390625
688 0.064453125
689 -0.0732421875
690 -0.1162109375
691 -0.0986328125
692 0.005859375
693 -0.0439453125
694 0.001953125
695 -0.0166015625
696 0.068359375
697 -0.0390625
698 -0.0419921875
699 -0.03515625
700 0.0615234375
701 -0.0205078125
702 -0.001953125
703 -0.0048828125
704 0.08203125
705 -0.017578125
706 -0.0087890625
707 -0.0068359375
708 0.091796875
709 -0.0146484375
710 -0.013671875
711 -0.0126953125
712 0.08203125
713 -0.05078125
714 -0.0556640625
715 -0.0361328125
716 0.072265625
717 -0.0849609375
718 -0.1201171875
719 -0.1044921875
720 0.00390625
721 -0.0458984375
722 0.0029296875
723 -0.0087890625
724 0.0654296875
725 -0.029296875
726 -0.0322265625
727 -0.02734375
728 0.06640625
729 -0.0205078125
730 0.0087890625
731 -0.001953125
732 0.0830078125
733 -0.017578125
734 -0.0078125
735 -0.0029296875
736 0.091796875
737 0.001953125
738 -0.0087890625
739 -0.0126953125
740 0.0810546875
741 -0.0361328125
742 -0.0712890625
743 -0.04296875
744 0.068359375
745 -0.0673828125
746 -0.140625
747 -0.1142578125
748 -0.00390625
749 -0.07421875
750 -0.064453125
751 -0.046875
752 0.0546875
753 -0.03125
754 -0.009765625
755 -0.0068359375
756 0.0849609375
757 -0.0048828125
758 0.0087890625
759 0.0107421875
760 0.1064453125
761 0.0068359375
762 0.013671875
763 -0.0048828125
764 0.0751953125
765 -0.0517578125
766 -0.0615234375
767 -0.0546875
768 0.037109375
769 -0.0263671875
770 0.013671875
771 0.0029296875
772 0.087890625
773 -0.017578125
774 0
775 -0.0078125
776 0.083984375
777 -0.0322265625
778 -0.056640625
779 -0.0419921875
780 0.05859375
781 -0.0107421875
782 0.0048828125
783 0.0029296875
784 0.09375
785 0.001953125
786 0.0068359375
787 0.001953125
788 0.087890625
789 0.01953125
790 0.0400390625
791 0.029296875
792 0.1123046875
793 -0.01171875
794 -0.029296875
795 -0.0126953125
796 0.0908203125
797 -0.0791015625
798 -0.1318359375
799 -0.09765625
800 0.01953125
801 -0.0283203125
802 0.017578125
803 0.0166015625
804 0.0947265625
805 0
806 0.0009765625
807 0.0009765625
808 0.08984375
809 0.0146484375
810 0.025390625
811 0.0185546875
812 0.099609375
813 0.0048828125
814 -0.001953125
815 -0.0029296875
816 0.08203125
817 0.0205078125
818 0.044921875
819 0.025390625
820 0.109375
821 -0.005859375
822 -0.0234375
823 -0.0107421875
824 0.09375
825 -0.072265625
826 -0.1328125
827 -0.099609375
828 0.017578125
829 -0.0322265625
830 0.0166015625
831 0.013671875
832 0.1005859375
833 -0.0009765625
834 0.0029296875
835 0.0009765625
836 0.0908203125
837 0.0146484375
838 0.0380859375
839 0.0263671875
840 0.1103515625
841 -0.00390625
842 0
843 -0.0068359375
844 0.08203125
845 0.0185546875
846 0.0498046875
847 0.037109375
848 0.1064453125
849 -0.009765625
850 -0.033203125
851 -0.0185546875
852 0.0859375
853 -0.0751953125
854 -0.1337890625
855 -0.107421875
856 0.01171875
857 -0.046875
858 -0.0009765625
859 -0.0009765625
860 0.08984375
861 -0.013671875
862 -0.005859375
863 -0.009765625
864 0.0810546875
865 0.013671875
866 0.0361328125
867 0.0283203125
868 0.111328125
869 0.009765625
870 0.0029296875
871 -0.001953125
872 0.083984375
873 0.017578125
874 0.0458984375
875 0.033203125
876 0.1123046875
877 -0.015625
878 -0.0341796875
879 -0.0205078125
880 0.0830078125
881 -0.076171875
882 -0.146484375
883 -0.109375
884 0.0048828125
885 -0.0703125
886 -0.0439453125
887 -0.0400390625
888 0.05078125
889 -0.068359375
890 -0.0634765625
891 -0.0458984375
892 0.048828125
893 -0.041015625
894 -0.0244140625
895 -0.0166015625
896 0.078125
897 -0.015625
898 -0.0068359375
899 -0.005859375
900 0.0986328125
901 -0.0048828125
902 0.0009765625
903 0
904 0.09375
905 -0.0400390625
906 -0.0634765625
907 -0.0419921875
908 0.064453125
909 -0.0888671875
910 -0.140625
911 -0.119140625
912 -0.0107421875
913 -0.06640625
914 -0.0283203125
915 -0.0263671875
916 0.064453125
917 -0.02734375
918 -0.025390625
919 -0.015625
920 0.0791015625
921 -0.025390625
922 -0.01171875
923 -0.0087890625
924 0.0859375
925 -0.0107421875
926 -0.0048828125
927 0.005859375
928 0.09765625
929 0.0146484375
930 0.00390625
931 0.005859375
932 0.0966796875
933 -0.0283203125
934 -0.05859375
935 -0.0380859375
936 0.06640625
937 -0.08203125
938 -0.1357421875
939 -0.115234375
940 -0.0078125
941 -0.0654296875
942 -0.0234375
943 -0.02734375
944 0.0625
945 -0.0361328125
946 -0.0322265625
947 -0.0224609375
948 0.076171875
949 -0.0185546875
950 -0.017578125
951 -0.0107421875
952 0.0830078125
953 -0.0185546875
954 -0.005859375
955 0.0009765625
956 0.099609375
957 0.0009765625
958 0.009765625
959 0.0087890625
960 0.0986328125
961 -0.0263671875
962 -0.044921875
963 -0.0302734375
964 0.0791015625
965 -0.0869140625
966 -0.1376953125
967 -0.1181640625
968 -0.0068359375
969 -0.0712890625
970 -0.0234375
971 -0.025390625
972 0.0615234375
973 -0.0361328125
974 -0.021484375
975 -0.0166015625
976 0.0830078125
977 -0.0146484375
978 -0.001953125
979 -0.0068359375
980 0.08984375
981 -0.0068359375
982 0.013671875
983 0.015625
984 0.1142578125
985 0
986 -0.001953125
987 -0.0029296875
988 0.091796875
989 -0.0439453125
990 -0.060546875
991 -0.0390625
992 0.072265625
993 -0.0849609375
994 -0.1376953125
995 -0.1171875
996 -0.0078125
997 -0.072265625
998 -0.029296875
999 -0.0263671875
};
\addlegendentry{\small Trace 2}

\nextgroupplot[
xlabel={\small Samples},
xlabel shift={-2ex},
ylabel={\small Distance},
ylabel shift={-1ex},
legend cell align={left},
tick label style={font=\small},
legend cell align={left},
legend style={
  fill opacity=0.8,
  draw opacity=1,
  text opacity=1,
  at={(0.03,0.97)},
  anchor=north west,
  draw=lightgray204
},
tick align=outside,
tick pos=left,
title={\textbf{c) Offset Profile with Divergence Window}},
title style={yshift=-2ex},
x grid style={darkgray176},
xmajorgrids,
xmin=-49.95, xmax=1048.95,
xtick style={color=black},
y grid style={darkgray176},
ymajorgrids,
ymin=-9.8, ymax=29.8,
ytick style={color=black}
]
\path [draw=sandybrown23019765, fill=sandybrown23019765, opacity=0.3]
(axis cs:483,-9.8)
--(axis cs:483,29.8)
--(axis cs:633,29.8)
--(axis cs:633,-9.8)
--cycle;
\addlegendimage{area legend, draw=sandybrown23019765, fill=sandybrown23019765, opacity=0.3}
\addlegendentry{\small Divergence Window}

\addplot [thick, crimson2305770]
table {%
0 0
1 0
2 0
3 0
4 0
5 0
6 0
7 0
8 0
9 0
10 0
11 0
12 0
13 0
14 0
15 0
16 0
17 0
18 0
19 0
20 0
21 0
22 0
23 0
24 0
25 0
26 0
27 0
28 0
29 0
30 0
31 0
32 0
33 0
34 0
35 0
36 0
37 0
38 0
39 0
40 0
41 0
42 0
43 0
44 0
45 0
46 0
47 0
48 0
49 0
50 0
51 -1
52 0
53 0
54 0
55 0
56 0
57 0
58 0
59 0
60 0
61 0
62 0
63 0
64 0
65 0
66 0
67 0
68 0
69 0
70 0
71 0
72 0
73 0
74 0
75 0
76 -1
77 0
78 0
79 0
80 0
81 0
82 0
83 0
84 0
85 0
86 0
87 0
88 0
89 0
90 0
91 0
92 0
93 0
94 0
95 0
96 0
97 0
98 0
99 0
100 0
101 0
102 0
103 0
104 0
105 0
106 0
107 0
108 0
109 0
110 0
111 0
112 0
113 0
114 0
115 0
116 0
117 0
118 0
119 0
120 0
121 0
122 0
123 0
124 0
125 0
126 0
127 0
128 0
129 0
130 0
131 0
132 0
133 0
134 0
135 0
136 0
137 0
138 0
139 0
140 0
141 0
142 0
143 0
144 0
145 0
146 0
147 0
148 0
149 0
150 0
151 0
152 0
153 0
154 0
155 0
156 0
157 0
158 0
159 0
160 0
161 0
162 0
163 0
164 0
165 0
166 0
167 0
168 0
169 0
170 0
171 0
172 0
173 0
174 0
175 0
176 0
177 0
178 0
179 0
180 0
181 0
182 0
183 0
184 0
185 0
186 0
187 0
188 0
189 0
190 0
191 0
192 0
193 0
194 0
195 0
196 0
197 0
198 0
199 0
200 0
201 0
202 0
203 0
204 0
205 0
206 0
207 0
208 0
209 0
210 0
211 0
212 0
213 0
214 0
215 0
216 0
217 0
218 0
219 0
220 0
221 0
222 0
223 0
224 0
225 0
226 0
227 0
228 0
229 0
230 0
231 0
232 0
233 0
234 0
235 0
236 0
237 0
238 0
239 0
240 0
241 1
242 0
243 0
244 0
245 0
246 0
247 0
248 0
249 0
250 0
251 0
252 0
253 0
254 0
255 0
256 0
257 0
258 0
259 0
260 0
261 0
262 0
263 0
264 0
265 0
266 0
267 0
268 0
269 0
270 0
271 0
272 -1
273 -2
274 -3
275 -4
276 -4
277 -3
278 -3
279 -4
280 -4
281 -4
282 -4
283 -4
284 -4
285 -4
286 -3
287 -2
288 -2
289 -2
290 -1
291 0
292 0
293 0
294 1
295 1
296 0
297 0
298 0
299 0
300 0
301 0
302 0
303 0
304 0
305 0
306 0
307 0
308 0
309 0
310 0
311 0
312 0
313 0
314 0
315 0
316 0
317 0
318 0
319 0
320 0
321 0
322 0
323 0
324 0
325 0
326 0
327 0
328 0
329 0
330 0
331 -1
332 -1
333 0
334 0
335 0
336 0
337 0
338 0
339 0
340 0
341 0
342 0
343 0
344 0
345 0
346 0
347 0
348 0
349 0
350 0
351 0
352 0
353 0
354 0
355 0
356 0
357 0
358 0
359 0
360 -1
361 -1
362 0
363 0
364 0
365 0
366 0
367 0
368 0
369 0
370 0
371 0
372 0
373 0
374 0
375 0
376 0
377 0
378 0
379 0
380 0
381 0
382 0
383 0
384 0
385 0
386 0
387 0
388 0
389 0
390 0
391 0
392 0
393 -1
394 -1
395 0
396 0
397 0
398 0
399 0
400 0
401 0
402 0
403 -1
404 0
405 0
406 0
407 0
408 0
409 0
410 0
411 0
412 -1
413 -2
414 -3
415 -4
416 -4
417 -4
418 -4
419 -4
420 -4
421 -4
422 -4
423 -3
424 -2
425 -2
426 -2
427 -1
428 0
429 0
430 0
431 0
432 0
433 0
434 0
435 0
436 0
437 0
438 0
439 0
440 0
441 0
442 0
443 0
444 0
445 0
446 0
447 1
448 1
449 0
450 0
451 0
452 1
453 1
454 0
455 0
456 0
457 0
458 0
459 0
460 0
461 0
462 0
463 0
464 0
465 0
466 0
467 0
468 0
469 0
470 0
471 0
472 0
473 0
474 0
475 0
476 0
477 0
478 0
479 0
480 0
481 1
482 1
483 0
484 0
485 0
486 -1
487 -1
488 0
489 0
490 0
491 0
492 0
493 0
494 0
495 0
496 0
497 0
498 0
499 0
500 0
501 0
502 0
503 0
504 0
505 0
506 0
507 0
508 0
509 0
510 0
511 0
512 0
513 0
514 0
515 0
516 0
517 0
518 0
519 0
520 0
521 0
522 0
523 0
524 0
525 0
526 0
527 1
528 2
529 3
530 4
531 5
532 6
533 6
534 5
535 4
536 3
537 2
538 2
539 2
540 1
541 0
542 -1
543 -2
544 -3
545 -4
546 -5
547 -6
548 -7
549 -8
550 -8
551 -8
552 -8
553 -8
554 -8
555 -8
556 -8
557 -8
558 -8
559 -8
560 -8
561 -8
562 -7
563 -6
564 -5
565 -4
566 -3
567 -2
568 -1
569 0
570 1
571 1
572 2
573 3
574 4
575 4
576 4
577 5
578 6
579 6
580 6
581 7
582 8
583 8
584 8
585 8
586 8
587 8
588 8
589 8
590 8
591 8
592 8
593 9
594 10
595 10
596 10
597 11
598 12
599 12
600 12
601 12
602 12
603 12
604 12
605 11
606 10
607 10
608 11
609 12
610 12
611 12
612 12
613 12
614 12
615 12
616 12
617 12
618 12
619 12
620 12
621 12
622 12
623 12
624 12
625 12
626 12
627 12
628 12
629 12
630 13
631 14
632 15
633 16
634 16
635 16
636 16
637 16
638 16
639 15
640 14
641 14
642 15
643 16
644 16
645 16
646 16
647 16
648 16
649 16
650 16
651 16
652 16
653 17
654 18
655 18
656 18
657 19
658 20
659 20
660 20
661 20
662 20
663 20
664 20
665 20
666 20
667 20
668 20
669 21
670 22
671 23
672 23
673 24
674 25
675 25
676 26
677 27
678 28
679 28
680 28
681 28
682 28
683 28
684 28
685 28
686 28
687 28
688 27
689 26
690 25
691 24
692 23
693 22
694 21
695 20
696 20
697 20
698 20
699 20
700 20
701 19
702 18
703 17
704 17
705 17
706 16
707 16
708 16
709 16
710 16
711 16
712 16
713 16
714 16
715 16
716 16
717 17
718 18
719 19
720 20
721 21
722 22
723 22
724 22
725 23
726 24
727 24
728 23
729 22
730 21
731 20
732 20
733 20
734 20
735 20
736 20
737 20
738 20
739 20
740 20
741 20
742 20
743 20
744 20
745 20
746 20
747 20
748 20
749 20
750 20
751 20
752 20
753 20
754 20
755 20
756 20
757 20
758 20
759 20
760 20
761 20
762 20
763 20
764 20
765 20
766 20
767 20
768 20
769 20
770 20
771 20
772 20
773 20
774 20
775 20
776 20
777 20
778 20
779 20
780 20
781 21
782 22
783 22
784 22
785 23
786 24
787 24
788 23
789 22
790 21
791 20
792 20
793 20
794 20
795 20
796 20
797 20
798 20
799 20
800 20
801 20
802 20
803 20
804 20
805 20
806 20
807 20
808 20
809 21
810 22
811 22
812 23
813 24
814 24
815 24
816 24
817 25
818 25
819 24
820 24
821 24
822 25
823 26
824 26
825 25
826 24
827 24
828 25
829 26
830 26
831 26
832 27
833 28
834 28
835 28
836 28
837 28
838 28
839 28
840 28
841 28
842 28
843 28
844 27
845 26
846 25
847 24
848 24
849 24
850 24
851 24
852 25
853 24
854 24
855 24
856 24
857 24
858 23
859 22
860 22
861 21
862 20
863 20
864 20
865 19
866 18
867 17
868 17
869 16
870 16
871 16
872 16
873 16
874 16
875 16
876 16
877 16
878 16
879 16
880 16
881 16
882 16
883 16
884 16
885 16
886 16
887 16
888 16
889 17
890 18
891 18
892 18
893 19
894 20
895 20
896 20
897 20
898 20
899 20
900 20
901 19
902 18
903 17
904 17
905 16
906 16
907 16
908 16
909 16
910 16
911 16
912 16
913 16
914 15
915 16
916 16
917 16
918 16
919 16
920 16
921 16
922 16
923 16
924 16
925 16
926 16
927 16
928 16
929 16
930 16
931 16
932 16
933 16
934 16
935 16
936 16
937 16
938 16
939 16
940 16
941 16
942 17
943 16
944 16
945 16
946 16
947 16
948 16
949 16
950 16
951 16
952 16
953 16
954 16
955 16
956 16
957 16
958 16
959 16
960 16
961 16
962 16
963 16
964 16
965 16
966 16
967 16
968 16
969 16
970 16
971 17
972 18
973 18
974 18
975 19
976 20
977 20
978 19
979 18
980 17
981 16
982 16
983 16
984 16
985 16
986 16
987 16
988 16
989 16
990 16
991 16
992 16
993 16
994 16
995 16
996 16
997 16
998 16
999 16
};
\addlegendentry{\small Offset Profile $\Delta$}
\addplot [semithick, peru17915666, dashed, forget plot]
table {%
483 -9.8
483 29.8
};
\addplot [semithick, peru17915666, dashed, forget plot]
table {%
633 -9.8
633 29.8
};
\end{groupplot}

\end{tikzpicture}

%% file: figs/corr_detection.tex
\begin{tikzpicture}

\definecolor{cadetblue85144170}{RGB}{85,144,170}
\definecolor{darkgray176}{RGB}{176,176,176}
\definecolor{gray}{RGB}{128,128,128}
\definecolor{green}{RGB}{0,128,0}
\definecolor{peru17915666}{RGB}{179,156,66}
\definecolor{peru18113482}{RGB}{181,134,82}

\begin{groupplot}[group style={group size=2 by 1}, height=3.6cm, width=9.2cm]
\nextgroupplot[
xlabel={\small Samples},
ylabel={\small Voltage},
ylabel shift={-2ex},
legend cell align={left},
tick label style={font=\small},
tick pos=both,
title={\textbf{Same Execution Paths}},
title style={yshift=-2ex},
x grid style={darkgray176},
xmin=-1, xmax=61,
xtick style={color=black},
xtick={0,10,20,30,40,50,60},
xticklabels={550,560,570,580,590,600,610},
y grid style={darkgray176},
ymin=-0.5, ymax=0.3,
ytick style={color=black}
]
\addplot [gray, opacity=0.7, dashed]
table {%
0 -0.5
0 0.3
};
\addplot [gray, opacity=0.7, dashed]
table {%
10 -0.5
10 0.3
};
\addplot [gray, opacity=0.7, dashed]
table {%
20 -0.5
20 0.3
};
\addplot [gray, opacity=0.7, dashed]
table {%
30 -0.5
30 0.3
};
\addplot [gray, opacity=0.7, dashed]
table {%
40 -0.5
40 0.3
};
\addplot [gray, opacity=0.7, dashed]
table {%
50 -0.5
50 0.3
};
\addplot [gray, opacity=0.7, dashed]
table {%
60 -0.5
60 0.3
};
\addplot [line width=1pt, cadetblue85144170]
table {%
0 -0.212729940576319
1 0.0952240862844642
2 0.00493880513656758
3 -0.0442065105619782
4 -0.250255070688829
5 -0.237855641351109
6 -0.277754285319178
7 0.131633045886314
8 0.000514866175754899
9 0.0514210519858422
10 -0.298778010169531
11 0.165804566462956
12 0.0837676384553259
13 -0.242280307478716
14 -0.275588701380859
15 -0.294185744267296
16 -0.253716292862989
17 -0.163175894386564
18 -0.228279535008427
19 -0.315571218995251
20 -0.169753802169603
21 -0.417410646915338
22 -0.349087883121343
23 -0.316188178716501
24 -0.271581468775066
25 -0.103304446769807
26 -0.388508574492835
27 -0.220159229548793
28 -0.166919379356211
29 -0.423235011581377
30 -0.124169123869173
31 -0.323046878438104
32 -0.355819283022311
33 0.105596904978004
34 0.13222735165114
35 0.0698973339301095
36 -0.168326329028399
37 -0.261293133415645
38 0.0389084804592271
39 -0.0800694185927389
40 -0.240045057915272
41 -0.0583384892763876
42 -0.297345848633563
43 0.128099910826802
44 -0.212118289910815
45 -0.0275270092988334
46 -0.22185547054527
47 -0.137488446865759
48 -0.144077538450658
49 -0.344220685162429
50 0.0303902027933423
51 -0.082421057078797
52 -0.0130331761264708
53 -0.0448638669474563
54 -0.199143633601107
55 -0.039061903143512
56 -0.453671521889172
57 -0.393941421356895
58 -0.459669215041602
59 -0.30668734309608
};
\addplot [line width=1pt, peru17915666, opacity=0.8]
table {%
0 -0.227515160479993
1 0.0880256442005171
2 0.000332417426969706
3 -0.033635288299789
4 -0.246818887793145
5 -0.255486042904736
6 -0.27451344562523
7 0.12778222308215
8 -0.00625435382730469
9 0.0575378148742509
10 -0.288468014944571
11 0.175117367654118
12 0.0753754632230996
13 -0.245372431237228
14 -0.272276067066824
15 -0.284430292996073
16 -0.258508035241442
17 -0.165032484153202
18 -0.239342884748488
19 -0.327533285236058
20 -0.161628543945661
21 -0.40384824662963
22 -0.349807984337146
23 -0.30615284973758
24 -0.26796510852459
25 -0.109755644315858
26 -0.384894618437751
27 -0.204778863884133
28 -0.16727763974731
29 -0.407588575023237
30 -0.150366574910071
31 -0.314827853394352
32 -0.354948812339929
33 0.102606831473346
34 0.133144959416495
35 0.0500216447841005
36 -0.170523047906774
37 -0.257722007700528
38 0.0536874209066422
39 -0.0852521207754753
40 -0.248129993944204
41 -0.063356059712233
42 -0.288191827456542
43 0.131387421923399
44 -0.217415891948486
45 -0.0223943349676998
46 -0.220884695051789
47 -0.12780199696043
48 -0.151098069389432
49 -0.347497306628407
50 0.0264691212620207
51 -0.0970562065601182
52 -0.0100719733558251
53 -0.0422533142256574
54 -0.199092499034682
55 -0.0414077744772634
56 -0.467825229309676
57 -0.398147874584549
58 -0.46309636020687
59 -0.314710115788296
};
\draw (axis cs:5,0.21) node[
  scale=0.8,
  anchor=base,
  text=green,
  rotate=0.0
]{\bfseries Corr > T};
\draw (axis cs:15,0.21) node[
  scale=0.8,
  anchor=base,
  text=green,
  rotate=0.0
]{\bfseries Corr > T};
\draw (axis cs:25,0.21) node[
  scale=0.8,
  anchor=base,
  text=green,
  rotate=0.0
]{\bfseries Corr > T};
\draw (axis cs:35,0.21) node[
  scale=0.8,
  anchor=base,
  text=green,
  rotate=0.0
]{\bfseries Corr > T};
\draw (axis cs:45,0.21) node[
  scale=0.8,
  anchor=base,
  text=green,
  rotate=0.0
]{\bfseries Corr > T};
\draw (axis cs:55,0.21) node[
  scale=0.8,
  anchor=base,
  text=green,
  rotate=0.0
]{\bfseries Corr > T};
\draw[<->,draw=black] (axis cs:10,-0.4) -- (axis cs:0,-0.4);
\draw (axis cs:5,-0.4) node[
  scale=0.45,
  anchor=north,
  text=black,
  rotate=0.0
]{\bfseries Slide w};

\nextgroupplot[
xlabel={\small Samples},
legend cell align={left},
tick label style={font=\small},
scaled y ticks=manual:{}{\pgfmathparse{#1}},
tick pos=both,
title={\textbf{Different Execution Paths}},
title style={yshift=-2ex},
x grid style={darkgray176},
xmin=-1, xmax=61,
xtick style={color=black},
xtick={0,10,20,30,40,50,60,26},
xticklabels={550,560,570,580,590,600,610,576},
y grid style={darkgray176},
ymin=-0.5, ymax=0.3,
ytick style={color=black},
yticklabels={}
]
\addplot [gray, opacity=0.7, dashed]
table {%
-1.4210854715202e-14 -0.5
-1.4210854715202e-14 0.3
};
\addplot [gray, opacity=0.7, dashed]
table {%
9.99999999999998 -0.5
9.99999999999998 0.3
};
\addplot [gray, opacity=0.7, dashed]
table {%
20 -0.5
20 0.3
};
\addplot [gray, opacity=0.7, dashed]
table {%
30 -0.5
30 0.3
};
\addplot [gray, opacity=0.7, dashed]
table {%
40 -0.5
40 0.3
};
\addplot [gray, opacity=0.7, dashed]
table {%
50 -0.5
50 0.3
};
\addplot [gray, opacity=0.7, dashed]
table {%
60 -0.5
60 0.3
};
\addplot [line width=1pt, cadetblue85144170]
table {%
0 -0.212729940576319
1 0.0952240862844642
2 0.00493880513656758
3 -0.0442065105619782
4 -0.250255070688829
5 -0.237855641351109
6 -0.277754285319178
7 0.131633045886314
8 0.000514866175754899
9 0.0514210519858422
10 -0.298778010169531
11 0.165804566462956
12 0.0837676384553259
13 -0.242280307478716
14 -0.275588701380859
15 -0.294185744267296
16 -0.253716292862989
17 -0.163175894386564
18 -0.228279535008427
19 -0.315571218995251
20 -0.169753802169603
21 -0.417410646915338
22 -0.349087883121343
23 -0.316188178716501
24 -0.271581468775066
25 -0.103304446769807
26 -0.388508574492835
27 -0.220159229548793
28 -0.166919379356211
29 -0.423235011581377
30 -0.124169123869173
31 -0.323046878438104
32 -0.355819283022311
33 0.105596904978004
34 0.13222735165114
35 0.0698973339301095
36 -0.168326329028399
37 -0.261293133415645
38 0.0389084804592271
39 -0.0800694185927389
40 -0.240045057915272
41 -0.0583384892763876
42 -0.297345848633563
43 0.128099910826802
44 -0.212118289910815
45 -0.0275270092988334
46 -0.22185547054527
47 -0.137488446865759
48 -0.144077538450658
49 -0.344220685162429
50 0.0303902027933423
51 -0.082421057078797
52 -0.0130331761264708
53 -0.0448638669474563
54 -0.199143633601107
55 -0.039061903143512
56 -0.453671521889172
57 -0.393941421356895
58 -0.459669215041602
59 -0.30668734309608
};
\addplot [line width=1pt, peru18113482, opacity=0.8]
table {%
0 -0.212729940576319
1 0.0952240862844642
2 0.00493880513656758
3 -0.0442065105619782
4 -0.250255070688829
5 -0.237855641351109
6 -0.277754285319178
7 0.131633045886314
8 0.000514866175754899
9 0.0514210519858422
10 -0.298778010169531
11 0.165804566462956
12 0.0837676384553259
13 -0.242280307478716
14 -0.275588701380859
15 -0.294185744267296
16 -0.253716292862989
17 -0.163175894386564
18 -0.228279535008427
19 -0.315571218995251
20 -0.169753802169603
21 -0.417410646915338
22 -0.349087883121343
23 -0.316188178716501
24 -0.271581468775066
25 -0.103304446769807
26 0.0714548519562596
27 -0.238398533989622
28 -0.140604689128317
29 -0.0484905205524111
30 -0.218185198810353
31 0.0858910413604803
32 0.0812236474710556
33 -0.274108852087318
34 -0.151375747053807
35 -0.249560845091615
36 -0.257579752811266
37 -0.381556526322734
38 -0.0952178330100516
39 -0.148660488385569
40 -0.374260624375005
41 -0.260676767881694
42 0.0541329429833268
43 -0.280219054666514
44 -0.327552563954388
45 -0.155273619861219
46 0.0928252270553003
47 -0.27897236424425
48 -0.0639322262970607
49 -0.0191901923356412
50 -0.2811812280038
51 -0.0358918256940702
52 -0.216108433640373
53 -0.0838470847032103
54 -0.0832351446195527
55 -0.132112657962621
56 -0.354855114972796
57 0.017651247794619
58 -0.239609967514132
59 -0.306740744800073
};
\addplot [semithick, red, dashed]
table {%
26 -0.5
26 0.3
};
\draw (axis cs:5,0.21) node[
  scale=0.8,
  anchor=base,
  text=green,
  rotate=0.0
]{\bfseries Corr > T};
\draw (axis cs:15,0.21) node[
  scale=0.8,
  anchor=base,
  text=green,
  rotate=0.0
]{\bfseries Corr > T};
\draw (axis cs:25,0.21) node[
  scale=0.8,
  anchor=base,
  text=red,
  rotate=0.0
]{\bfseries Corr < T};
\draw[-latex,draw=black] (axis cs:35,0.15) -- (axis cs:26,-0.45);
\draw (axis cs:35,0.15) node[
  scale=0.8,
  anchor=base west,
  text=black,
  rotate=0.0
]{\bfseries Deviation point};
\end{groupplot}

\end{tikzpicture}

%% file: sections/4_experiments.tex
\section{Experiments}
\label{sec:experiments}

We evaluate PowerFuzz across two dimensions. First, we compare PowerFuzz against Fuzz'EMup~\cite{Moradihaghighietal2026}, the closest state-of-the-art black-box firmware fuzzer with a set of large-scale real-world firmware applications to assess its fuzzing effectiveness in a fully black-box setting. Second, we evaluate PowerFuzz across three distinct embedded hardware architectures to validate its applicability across platforms. All experiments are conducted on physical embedded devices with no access to the firmware binary, source code, or hardware specification.


\begin{table*}[!ht]
\centering
\caption{Firmware Code Coverage Results using the ChipWhisperer Pro board with STM32F3 target board comparing Gray-Box Fuzzing, Fuzz'EMup, and PowerFuzz, showing improvement over Fuzz'EMup.}
\vspace{-0.1in}
\label{tab:firmware_coverage_extended}
\begin{tabular*}{\textwidth}{@{\extracolsep{\fill}}lcrrrrrrrrr}
\hline
\textbf{Firmware} & \textbf{Total} & \multicolumn{2}{c}{\textbf{Gray-Box}} & \multicolumn{2}{c}{\textbf{Random}} & \multicolumn{2}{c}{\textbf{Fuzz’EMup}} & \multicolumn{3}{c}{\textbf{PowerFuzz (Ours)}} \\ 
\cline{3-11}
 & \textbf{Branches} & \textbf{Covered} & \textbf{\%} & \textbf{Covered} & \textbf{\%} & \textbf{Covered} & \textbf{\%} & \textbf{Covered} & \textbf{\%} & \textbf{vs Fuzz'EMup} \\ 
\hline
Stepper           & 2143 & 2017 & 94.1\% & 1321  & 61.6\% & 1244 & 58.1\% & 1730 & 80.7\% & +22.7\% \\
CNC               & 1763 & 1631 & 92.5\% & 993  & 56.3\% & 1180 & 66.9\% & 1467 & 83.2\% & +16.3\% \\
GPS               & 1356 & 1210 & 89.2\% & 885  & 65.3\% & 920  & 67.8\% & 1045 & 77.1\% & +9.2\%  \\
Soldering Station & 1871 & 1617 & 86.4\% & 1045  & 55.9\% & 1085 & 58.0\% & 1438 & 76.9\% & +18.9\% \\
\hline
\end{tabular*}
\end{table*}

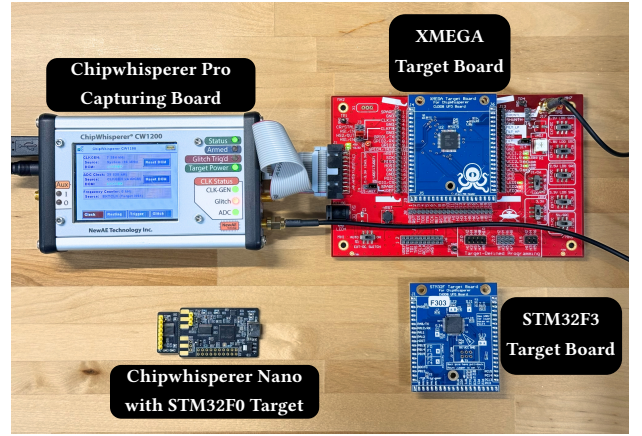
\begin{figure}[!ht]
    \centering
    \input{figs/experimental_setup}
    \caption{Experimental setup used for evaluation. The ChipWhisperer Nano is a self-contained board that integrates both the capturing infrastructure and the STM32F0 target. The ChipWhisperer Pro serves exclusively as a capturing board and it connects with two target boards: STM32F3 and XMEGA. Each target board is mounted on the CW308 UFO (red PCB) that provides an interface to the capturing board.}
    \label{fig:experimental_setup}
\end{figure}

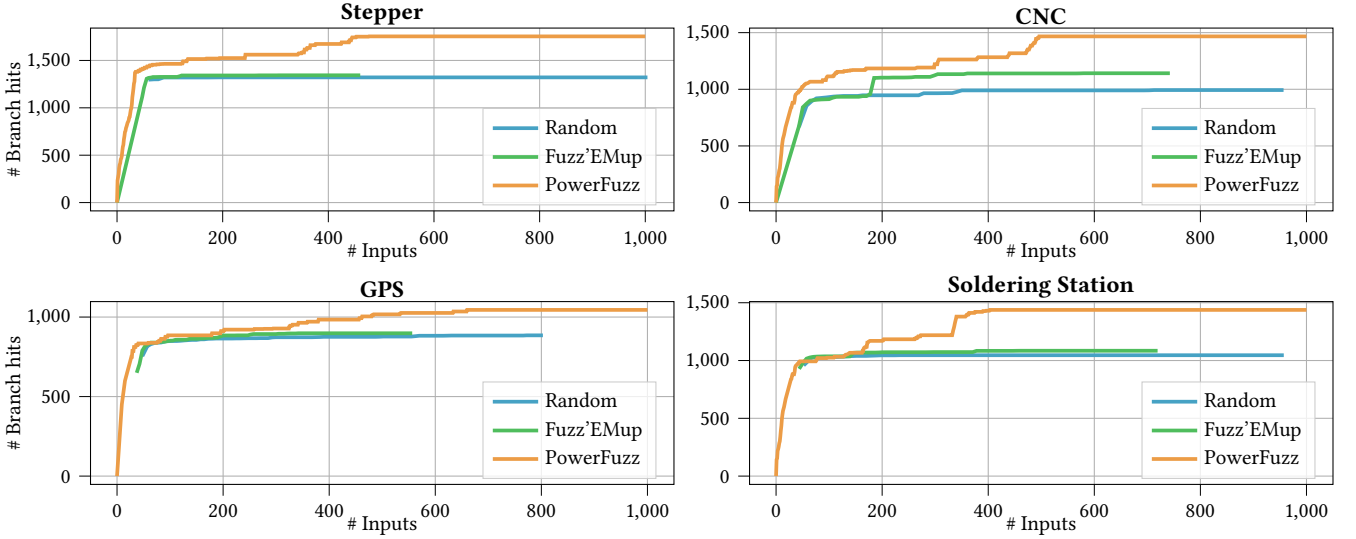
\begin{figure*}[!ht]
    \centering
    \input{figs/comparison_with_sota}
    \vspace{-0.2in}
    \caption{Branch coverage vs.\ number of inputs for PowerFuzz, Fuzz'EMup, and the random baseline across four large-scale firmware benchmarks on the ChipWhisperer Pro with STM32F3 target.}
    \label{fig:comparison_with_sota}
\end{figure*}

\subsection{Experiment Setup}

\subsubsection{Hardware Platforms}

We evaluate PowerFuzz on three ChipWhisperer -based embedded platforms that cover two distinct MCU architectures and three device configurations:
\begin{itemize}
\item \textbf{ChipWhisperer Nano with  STM32F0 }: A Cortex-M0-based microcontroller operating at 7.37 MHz, representing a lightweight 32-bit ARM platform.
\item \textbf{ChipWhisperer Pro with STM32F3 }: A Cortex-M4-based microcontroller operating at 7.37 MHz, representing a  32-bit ARM platform with hardware floating-point support.
\item \textbf{ChipWhisperer Pro with XMega }: An AVR-based microcontroller, representing a fundamentally different architecture from the ARM-based targets.
\end{itemize}

\noindent This selection spans two instruction set architectures (ARM Cortex-M and AVR), two word sizes (32-bit and 8-bit), and varying pipeline depths, providing a diverse and representative evaluation of PowerFuzz's cross-architecture applicability. Figure~\ref{fig:experimental_setup} shows our experimental setup.

\subsubsection{Power Trace Acquisition.} Power traces are captured using the ChipWhisperer platform's built-in measurement infrastructure~\cite{chipwhisperer}. Trace acquisition is triggered synchronously with input delivery to the target device, ensuring consistent temporal alignment across all captured traces. The sampling rate and capture window length are configured per platform to ensure that the full firmware response to each input is captured within a single trace. Also, for each iteration, power traces are captured $N=10$ times, and the average is taken to remove the measurement noise.

\subsubsection{Fuzzing Configuration.} The PowerFuzz fuzzing engine is built on top of LibAFL~\cite{fioraldi2022libafl} and communicates with the target device over a serial interface. The seed corpus is initialized with set of random inputs. The DTW threshold, sliding window size $Slide\_w$, and correlation threshold $T$ are calibrated empirically per platform from a set of reference trace pairs collected under identical execution conditions. The branch selector operates in random mode until the TCFG reaches a 100 node count, after which it transitions to depth-prioritized selection.


\subsubsection{Benchmark Firmware.}\label{sec:larger_benchmarks} We evaluate PowerFuzz on ten firmware benchmarks divided into two groups. The first group consists of four large-scale, application-level firmware images adopted from Fuzz'EMup~\cite{Moradihaghighietal2026}, used for direct state-of-the-art comparison:

\begin{enumerate}[nosep]
    \item \textbf{GPS Receiver:} A DMA-driven NMEA 0183 parser for mapping serial sentences to structured GPS data.
    \item \textbf{Stepper Controller:} A three-axis UART controller featuring trajectory planning and timer-based pulse generation.
    \item \textbf{CNC (Grbl-based):} A Cortex-M Grbl port for G-code interpretation and synchronized multi-axis motion planning.
    \item \textbf{Soldering Station:} A TS100-style thermal controller utilizing PID logic and PWM-modulated power delivery.
\end{enumerate}

The second group consists of six smaller benchmark firmware applications used to evaluate PowerFuzz across different hardware architectures. These benchmarks are taken from Fuzzbench~\cite{metzman2021fuzzbench}.
\begin{enumerate}
    \item \textit{cjson}: A lightweight C library for parsing JSON data. It features a high density of conditional branches and recursion, making it a target for testing fuzzer coverage efficiency.
    \item \textit{zlib}: A standard compression/decompression library. Its execution flow is highly dependent on input bit-patterns and internal lookup tables.
    \item \textit{TinyFFT}: An implementation of the Fast Fourier Transform for microcontrollers, representing signal processing workloads.
    \item \textit{microECDSA}: A micro-implementation of the Elliptic Curve Digital Signature Algorithm, focusing on cryptographic control logic.
    \item \textit{miniAES}: A minimal Advanced Encryption Standard implementation. While its execution is highly regular, it is a standard benchmark for side-channel leakage analysis.
    \item \textit{TinyMaix}: A lightweight neural network inference library for ARM Cortex-M microcontrollers, representing modern edge-AI workloads.
\end{enumerate}

\subsection{TCFG Structural Accuracy Analysis}\label{sec:tcfg_accuracy}

While Sections 4.2 and 4.3 establish that PowerFuzz achieves high branch coverage in a fully black-box setting, they do not directly address whether the inferred TCFG accurately reflects the true control flow structure of the firmware. In this section, we evaluate the structural accuracy of the TCFG by comparing it against the ground-truth CFG extracted from the firmware binary, quantifying how faithfully power-trace-driven divergence detection identifies real branch transitions. We evaluate TCFG structural accuracy using four larger benchmark firmware applications from Section~\ref{sec:larger_benchmarks}.

\subsubsection{Ground-Truth CFG Extraction} For each firmware binary compiled for the STM32F3 target, we extract the ground-truth CFG using angr~\cite{wang2017angr} binary analysis framework. Specifically, we apply angr's \texttt{CFGEmulated} analysis technique to perform the symbolic execution to recover basic block boundaries and inter-block edges from the compiled binary. This ground-truth CFG consists of a set of basic block nodes $N_{true}$ and directed branch edges $E_{true}$.

\subsubsection{TCFG-to-CFG Correspondence} The TCFG nodes produced by PowerFuzz store power trace segments rather than binary addresses, so a correspondence must be established between TCFG nodes and ground-truth CFG nodes. To do this, we use the input stored at each TCFG node ($v.input$) and re-execute it on the firmware in QEMU with basic block tracing enabled, using the same QEMU-based infrastructure described in Section 4.2. Then using intruction-level QEMU trace, we determine the  corresponding path in the real CFG and mark the covered nodes. We do this iteratively for all the inputs we collected throughout the fuzzing run. After that we can count the number of actual nodes covered in the CFG using a node travesal in the CFG. We say a TCFG node $v$ is a \textit{true positive} (TP) if the corresponding QEMU execution reveals a basic block transition at that point that exists as an edge in $E_{true}$, and a \textit{false positive} (FP) if no such transition exists in $E_{true}$.

Table~\ref{tab:tcfg_accuracy} reports the structural accuracy of the TCFG for each of the four benchmark firmware applications on the STM32F3 platform. For each benchmark, we report the number of nodes in the inferred TCFG and in the ground-truth CFG, along with node-level precision and recall, defined as:

\begin{equation}
\text{Node Precision} = \frac{TP_{nodes}}{TP_{nodes} + FP_{nodes}}, \quad \text{Node Recall} = \frac{TP_{nodes}}{|N_{true}|}
\end{equation}

\begin{table}[h]
\centering
\caption{TCFG structural accuracy compared against ground-truth CFGs extracted via angr, across four benchmark firmware applications on the STM32F3 platform.}
\label{tab:tcfg_accuracy}
\begin{tabular}{lcccc}
\toprule
\textbf{Firmware} & \textbf{TCFG} & \textbf{CFG} & \textbf{Node} & \textbf{Node} \\
& \textbf{Nodes} & \textbf{Nodes} & \textbf{Precison} & \textbf{Recall} \\
\midrule
Stepper        & 1730 & 1694 & 97.91\% & 79.04\% \\
CNC         & 1467 & 1431 & 97.55\% & 81.17\% \\
GPS      & 1045 & 1018 & 97.41\% & 75.07\% \\
Soldering Station   & 1438 & 1406 & 97.77\% & 75.15\% \\
\midrule
\textbf{Average} &  & & \textbf{97.66\%} & \textbf{77.61\%} \\
\bottomrule
\end{tabular}%
\end{table}

PowerFuzz achieves node precision of 97.66\% and node recall of 77.61\% on average across the four benchmarks. The consistently high precision, exceeding 97\% on every benchmark, indicates that the vast majority of TCFG nodes constructed by PowerFuzz correspond to genuine basic block boundaries in the firmware: the two-stage DTW-then-Pearson-correlation divergence detector rarely declares a branch split where none exists, confirming that the sliding-window and growing-window analysis effectively suppresses spurious detections caused by timing jitter or measurement noise. Recall is comparatively lower and more variable across benchmarks, ranging from 75.07\% on GPS to 81.17\% on CNC. This gap between precision and recall is consistent with the asymmetric nature of the detection mechanism: PowerFuzz is conservative by design, requiring a sustained correlation drop across both the sliding and growing window stages before committing to a new TCFG split, which biases the system toward missing subtle divergences rather than over-fragmenting the structure.

\subsection{Comparison with State-of-the-Art}\label{sec:comparison}
We compare PowerFuzz against Fuzz'EMup~\cite{Moradihaghighietal2026}, the closest state-of-the-art black-box firmware fuzzer, and a random baseline fuzzer, across the four large-scale firmware benchmarks described in Section~\ref{sec:experiments}. All three methods are evaluated on the ChipWhisperer Pro board with the STM32F3 target under identical conditions. Gray-box fuzzing results, obtained through binary instrumentation of the same firmware images, are additionally reported in Table~\ref{tab:firmware_coverage_extended} as an upper-bound reference, representing the best achievable branch coverage with full internal 
visibility.

Table~\ref{tab:firmware_coverage_extended} reports the total branch coverage achieved by each method across the four benchmarks. Gray-box fuzzing achieves coverage ranging from 86.4\% to 94.1\%, establishing a practical upper bound for each firmware. PowerFuzz achieves branch coverage of 80.7\%, 83.2\%, 77.1\%, and 76.9\% on the Stepper Motor Controller, CNC Controller, GPS Receiver, and Soldering Station firmware, respectively, consistently outperforming Fuzz'EMup across all four benchmarks. Fuzz'EMup achieves coverage between 58.0\% and 67.8\%, falling significantly short of both the gray-box upper bound and PowerFuzz. Notably, the coverage gap between PowerFuzz and gray-box fuzzing remains within 13.5\% across all benchmarks, demonstrating that power-trace-driven TCFG guidance can closely approximate the effectiveness of binary instrumentation in a fully black-box setting.

Figure~\ref{fig:comparison_with_sota} shows the branch coverage growth curves for random testing, Fuzz'EMup, and PowerFuzz across the four firmware benchmarks. PowerFuzz consistently reaches higher coverage plateaus than both Fuzz'EMup and the random baseline, and does so with fewer inputs in most benchmarks. Across all four firmware images, PowerFuzz's coverage curve rises steeply in the early phase of the process, driven by random branch selection populating the initial TCFG and continues to grow steadily into deeper firmware regions as depth-prioritized selection takes over. In contrast, Fuzz'EMup and the random baseline both plateau earlier and at lower coverage levels, reflecting the absence of TCFG-guided mutation effort allocation. The CNC Controller and Soldering Station benchmarks exhibit the largest performance gap between PowerFuzz and Fuzz'EMup, which we attribute to the complex state machine logic and multi-stage conditional structures present in these firmware images, precisely the scenarios where depth-prioritized branch selection provides the greatest advantage by directing mutation effort toward deeper, harder-to-reach execution paths.

\begin{table}[htbp]
\centering
\renewcommand{\arraystretch}{1} 
\caption{Novelty hit rate comparison between PowerFuzz, Fuzz'EMup, and random testing 
across four large-scale firmware benchmarks.}
\vspace{-0.1in}
\label{tab:novelty}
\begin{tabular}{lrrr}
\hline
\textbf{Firmware} & \textbf{Random} & \textbf{Fuzz’EMup} & \textbf{PowerFuzz} \\ \hline
Stepper           & 1.3\%               & 3.3\%                  & 6.6\% \\ 
CNC               & 4.4\%               & 5.2\%                  & 10.0\% \\ 
GPS               & 4.7\%               & 6.8\%                  & 7.1\% \\ 
Soldering Station & 2.1\%               & 2.4\%                  & 8.9\% \\ \hline
\end{tabular}
\end{table}

\begin{figure*}[ht]
    \centering
    \input{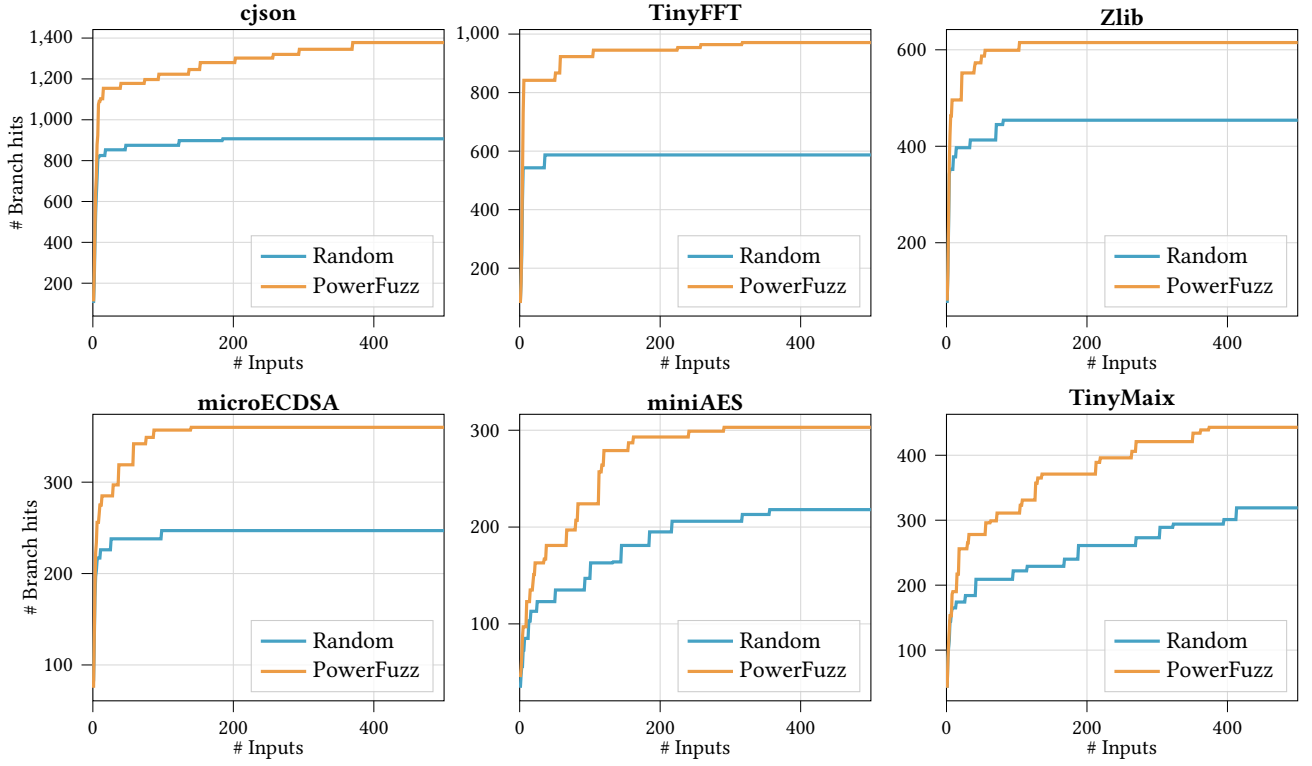}
    \caption{Branch coverage vs.\ number of inputs for PowerFuzz and random testing across six benchmark firmware applications on the with STM32F3 target. PowerFuzz reaches higher coverage plateaus significantly faster than random testing across all benchmarks.}
    \label{fig:cross_arch_curves}
\end{figure*}

\begin{table*}[t]
\centering
\caption{Branch coverage comparison between PowerFuzz (PF) and random testing across three targets (Nano/STM32F0, Pro/STM32F3, Pro/XMega) on six benchmark firmware applications.}
\label{tab:cross_arch}
\vspace{-0.1in}
\begin{tabular}{lrrrrrrrrrr}
\hline
\textbf{Firmware} & \textbf{Total} 
& \multicolumn{2}{c}{\textbf{Nano (STM32F0)}} 
& \multicolumn{2}{c}{\textbf{Pro (STM32F3)}} 
& \multicolumn{2}{c}{\textbf{Pro (XMEGA)}} 
& \multicolumn{2}{c}{\textbf{Random Testing}} \\
\cline{3-4} \cline{5-6} \cline{7-8} \cline{9-10}
 &  & \textbf{Covered} & \textbf{Percentage} 
    & \textbf{Covered} & \textbf{Percentage} 
    & \textbf{Covered} & \textbf{Percentage} 
    & \textbf{Covered} & \textbf{Percentage} \\
\hline
cjson      & 1578 & 1387 & 87.9\% & 1354 & 85.8\% & 1370 & 86.8\% & 1003 & 63.6\% \\
zlib       & 707  & 612  & 86.6\% & 623  & 88.1\% & 615  & 87.0\% & 456  & 64.5\% \\
TinyFFT    & 1150 & 974  & 84.7\% & 962  & 83.7\% & 965  & 83.9\% & 702  & 61.1\% \\
microECDSA  & 445  & 368  & 82.7\% & 368  & 82.7\% & 368  & 82.7\% & 261  & 58.7\% \\
miniAES    & 337  & 302  & 89.6\% & 315  & 93.5\% & 310  & 92.0\% & 221  & 65.6\% \\
TinyMaix   & 511  & 444  & 86.9\% & 461  & 90.2\% & 450  & 88.1\% & 323  & 63.2\% \\
\hline
\end{tabular}
\end{table*}

\noindent\textbf{Novelty Rate:} Table~\ref{tab:novelty} reports the novelty hit rate of 
each method across the four firmware benchmarks, measured as the fraction of generated 
inputs that exercise at least one previously unseen branch relative to all prior inputs. 
To ensure a fair and instrumentation-independent evaluation, novelty is assessed by 
re-executing each input on the firmware in QEMU after the fuzzing process and marking 
it as novel if it reveals at least one new branch. This calculation is conducted separately from 
the fuzzing execution itself. PowerFuzz achieves novelty rates of 6.6\%, 10\%, 7.1\%, 
and 8.9\% on the Stepper Motor Controller, CNC Controller, GPS Receiver, and Soldering 
Station firmware, respectively, outperforms Fuzz'EMup by up to 3.7$\times$ and the 
random baseline by up to 5.0$\times$ across all benchmarks. While absolute novelty 
rates are low for all methods, the consistent relative advantage of 
PowerFuzz confirms that TCFG-guided depth-prioritized mutation effectively steers 
input generation toward unexplored firmware behaviors.


\subsection{Evaluation across Hardware Architectures}

We evaluate PowerFuzz across three distinct embedded hardware platforms using the six smaller benchmark firmware applications described in Section~\ref{sec:experiments}. For each benchmark, PowerFuzz is compared against a random baseline fuzzer with no TCFG guidance under identical conditions. The goal of this evaluation is to demonstrate that the power-trace-driven TCFG construction mechanism of PowerFuzz generalizes across fundamentally different MCU architectures, including both 32-bit ARM Cortex-M and 8-bit AVR XMEGA platforms, without any architecture-specific tuning.

Table~\ref{tab:cross_arch} reports the branch coverage achieved by PowerFuzz and random testing across all six benchmarks and three platforms. PowerFuzz consistently outperforms random testing across every firmware and platform combination, achieving coverage between 82.7\% and 93.5\% compared to 58.7\%--65.6\% for random testing. The coverage advantage of PowerFuzz over random testing ranges from approximately 20 to 30 percentage points across all benchmarks, confirming that TCFG-guided mutation provides a substantial and consistent benefit regardless of the underlying hardware architecture. Notably, the coverage achieved by PowerFuzz remains stable across the three platforms for each firmware, for instance, microECDSA achieves exactly 82.7\% on all three platforms, indicating that the DTW-based similarity analysis and Pearson correlation produce consistent branch identification results across different power trace characteristics induced by different MCU architectures.

Figure~\ref{fig:cross_arch_curves} shows the branch coverage growth curves for PowerFuzz and random testing across all six benchmarks on the STM32F3 platform, which is representative of the behavior observed across all three platforms. PowerFuzz reaches its coverage plateau significantly faster than random testing in all six benchmarks, with the gap between the two methods widening progressively as the fuzzing process advances. The most pronounced efficiency gains are observed on cjson and zlib, where PowerFuzz achieves nearly 87\% and 88\% coverage respectively within the first 200 inputs, while random testing plateaus well below 65\% even after 500 inputs. This behavior reflects the high branch density of these benchmarks. cjson's deeply nested conditional logic and zlib's bit-pattern-dependent execution paths are precisely the structures that depth-prioritized branch selection is designed to exploit. For microECDSA and miniAES, both methods converge more quickly due to the more regular and structured execution patterns of cryptographic firmware, though PowerFuzz still achieves a consistently higher final coverage level.

%% file: figs/experimental_setup.tex
\tikzset{every picture/.style={line width=0.75pt}} 

\begin{tikzpicture}[x=0.75pt,y=0.75pt,yscale=-1,xscale=1]

\draw (185,133.71) node  {\includegraphics[width=232.5pt,height=162.44pt]{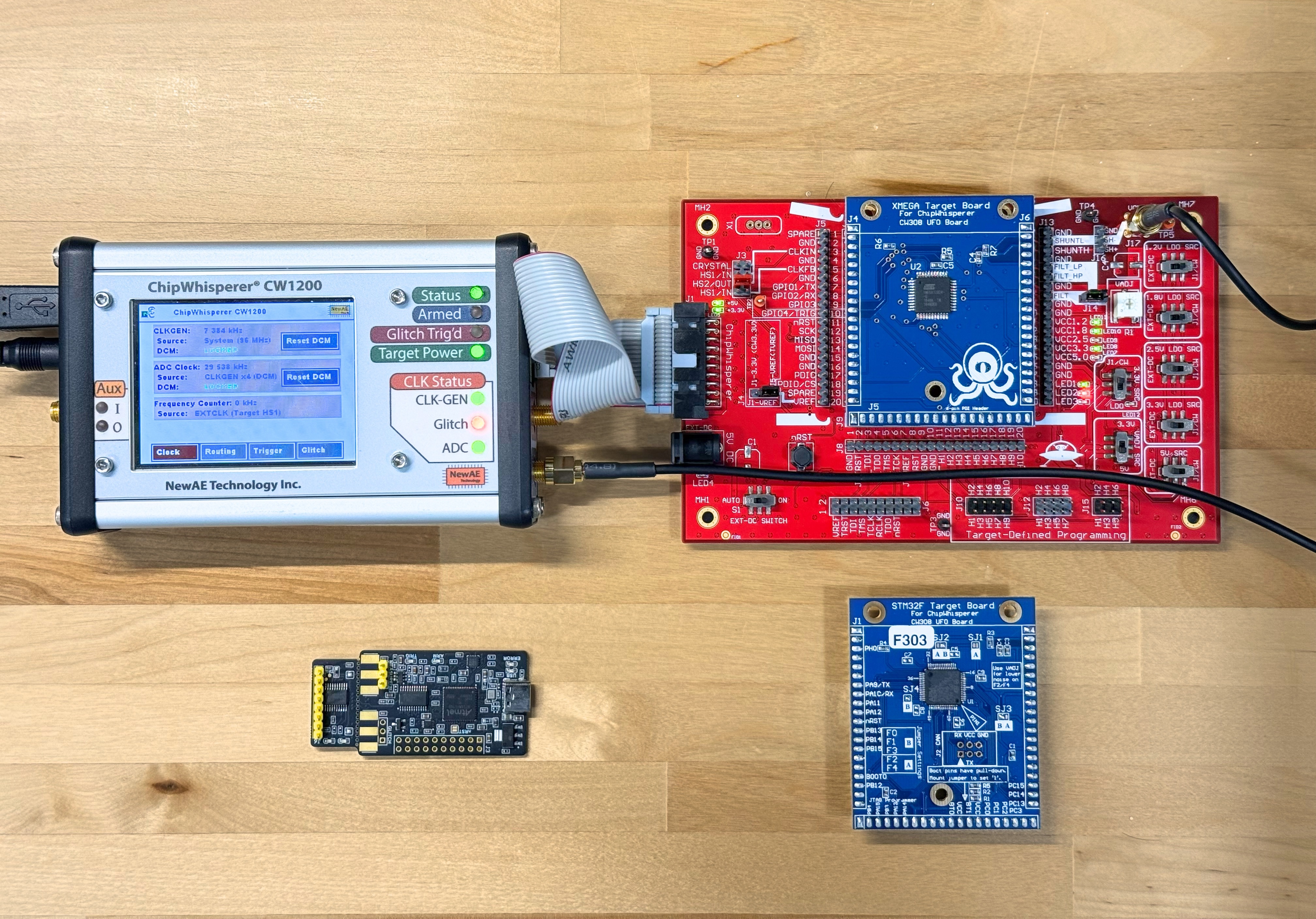}};
\draw  [fill={rgb, 255:red, 0; green, 0; blue, 0 }  ,fill opacity=1 ] (48,59.2) .. controls (48,56.33) and (50.33,54) .. (53.2,54) -- (146.8,54) .. controls (149.67,54) and (152,56.33) .. (152,59.2) -- (152,74.8) .. controls (152,77.67) and (149.67,80) .. (146.8,80) -- (53.2,80) .. controls (50.33,80) and (48,77.67) .. (48,74.8) -- cycle ;

\draw  [fill={rgb, 255:red, 0; green, 0; blue, 0 }  ,fill opacity=1 ] (80,213.2) .. controls (80,210.33) and (82.33,208) .. (85.2,208) -- (178.8,208) .. controls (181.67,208) and (184,210.33) .. (184,213.2) -- (184,228.8) .. controls (184,231.67) and (181.67,234) .. (178.8,234) -- (85.2,234) .. controls (82.33,234) and (80,231.67) .. (80,228.8) -- cycle ;

\draw  [fill={rgb, 255:red, 0; green, 0; blue, 0 }  ,fill opacity=1 ] (276,181.6) .. controls (276,177.4) and (279.4,174) .. (283.6,174) -- (328.4,174) .. controls (332.6,174) and (336,177.4) .. (336,181.6) -- (336,204.4) .. controls (336,208.6) and (332.6,212) .. (328.4,212) -- (283.6,212) .. controls (279.4,212) and (276,208.6) .. (276,204.4) -- cycle ;

\draw  [fill={rgb, 255:red, 0; green, 0; blue, 0 }  ,fill opacity=1 ] (220,39.6) .. controls (220,35.4) and (223.4,32) .. (227.6,32) -- (272.4,32) .. controls (276.6,32) and (280,35.4) .. (280,39.6) -- (280,62.4) .. controls (280,66.6) and (276.6,70) .. (272.4,70) -- (227.6,70) .. controls (223.4,70) and (220,66.6) .. (220,62.4) -- cycle ;

\draw (100,67) node  [color={rgb, 255:red, 255; green, 255; blue, 255 }  ,opacity=1 ] [align=left] {\begin{minipage}[lt]{70.72pt}\setlength\topsep{0pt}
\begin{center}
\textbf{{\footnotesize Chipwhisperer Pro }}\\\textbf{{\footnotesize Capturing Board}}
\end{center}

\end{minipage}};
\draw (132,221) node  [color={rgb, 255:red, 255; green, 255; blue, 255 }  ,opacity=1 ] [align=left] {\begin{minipage}[lt]{70.72pt}\setlength\topsep{0pt}
\begin{center}
\textbf{{\footnotesize Chipwhisperer Nano with STM32F0 Target}}
\end{center}

\end{minipage}};
\draw (306,193) node  [color={rgb, 255:red, 255; green, 255; blue, 255 }  ,opacity=1 ] [align=left] {\begin{minipage}[lt]{40.8pt}\setlength\topsep{0pt}
\begin{center}
{\footnotesize \textbf{STM32F3 Target Board}}
\end{center}

\end{minipage}};
\draw (250,51) node  [color={rgb, 255:red, 255; green, 255; blue, 255 }  ,opacity=1 ] [align=left] {\begin{minipage}[lt]{40.8pt}\setlength\topsep{0pt}
\begin{center}
{\footnotesize \textbf{XMEGA Target Board}}
\end{center}

\end{minipage}};

\end{tikzpicture}

%% file: figs/comparison_with_sota.tex
\begin{tikzpicture}

\definecolor{darkgray176}{RGB}{176,176,176}
\definecolor{darkorange25512714}{HTML}{4FBD5D}
\definecolor{forestgreen4416044}{RGB}{236,156,70}
\definecolor{lightgray204}{RGB}{204,204,204}
\definecolor{steelblue31119180}{HTML}{46A2C4}

\begin{groupplot}[group style={group size=2 by 2, vertical sep=1.2cm}, width=9.3cm, height=4cm]
\nextgroupplot[
xlabel={\small \# Inputs},
xlabel shift={-2ex},
ylabel={\small \# Branch hits},
ylabel shift={-1ex},
legend cell align={left},
tick label style={font=\small},
legend cell align={left},
legend style={
  fill opacity=0.8,
  draw opacity=1,
  text opacity=1,
  at={(0.97,0.03)},
  anchor=south east,
  draw=lightgray204
},
tick align=outside,
tick pos=left,
title={\textbf{Stepper}},
title style={yshift=-2ex},
x grid style={darkgray176},
xmajorgrids,
xmin=-50.205, xmax=1054.305,
xtick style={color=black},
y grid style={darkgray176},
ymajorgrids,
ymin=-87.7, ymax=1841.7,
ytick style={color=black}
]
\addplot [line width=1.4pt, steelblue31119180]
table {%
59.5 1296.9
69.7 1300.5
78.2 1301.1
88.3 1320.1
96.8 1320.1
107 1320.1
117.2 1320.1
125.7 1320.1
135.9 1320.1
146.1 1320.1
154.6 1320.1
164.8 1320.1
173.3 1320.1
183.5 1320.7
193.7 1321.3
202.2 1321.3
212.4 1321.3
220.9 1321.3
231.1 1321.3
241.3 1321.3
249.8 1321.3
260 1321.3
270.1 1321.3
278.6 1321.3
288.8 1321.3
297.3 1321.3
307.5 1321.3
317.7 1321.3
326.2 1321.3
336.4 1321.3
344.9 1321.3
355.1 1321.3
365.3 1321.3
373.8 1321.3
384 1321.3
394.2 1321.3
402.7 1321.3
412.9 1321.3
421.4 1321.3
431.6 1321.3
441.7 1321.3
450.2 1321.3
460.4 1321.3
468.9 1321.3
479.1 1321.3
489.3 1321.3
497.8 1321.3
508 1321.3
518.2 1321.3
526.7 1321.3
536.9 1321.3
545.4 1321.3
555.6 1321.3
565.8 1321.3
574.3 1321.3
584.5 1321.3
594.7 1321.3
603.2 1321.3
613.3 1321.3
621.8 1321.3
632 1321.3
642.2 1321.3
650.7 1321.3
660.9 1321.3
669.4 1321.3
679.6 1321.3
689.8 1321.3
698.3 1321.3
708.5 1321.3
718.7 1321.3
727.2 1321.3
737.4 1321.3
745.9 1321.3
756.1 1321.3
766.3 1321.3
774.8 1321.3
785 1321.3
793.4 1321.3
803.6 1321.3
813.8 1321.3
822.3 1321.3
832.5 1321.3
842.7 1321.3
851.2 1321.3
861.4 1321.3
869.9 1321.3
880.1 1321.3
890.3 1321.3
898.8 1321.3
909 1321.3
917.5 1321.3
927.7 1321.3
937.9 1321.3
946.4 1321.3
956.6 1321.3
966.7 1321.3
975.2 1321.3
985.4 1321.3
993.9 1321.3
1004.1 1321.3
};
\addlegendentry{\small Random}
\addplot [line width=1.4pt, darkorange25512714]
table {%
0 0
47.6 1114.5
51 1211.2
56.1 1307.3
59.5 1315.8
64.6 1320.1
68 1325
73.1 1325
76.5 1325.6
81.6 1326.2
85 1326.2
90 1326.2
93.4 1326.2
96.8 1326.2
101.9 1326.2
105.3 1326.2
110.4 1326.2
113.8 1326.2
118.9 1334.2
122.3 1341.5
127.4 1341.5
130.8 1341.5
135.9 1341.5
139.3 1341.5
142.7 1341.5
147.8 1341.5
151.2 1341.5
156.3 1341.5
159.7 1341.5
164.8 1341.5
168.2 1341.5
173.3 1341.5
176.7 1341.5
181.8 1341.5
185.2 1341.5
188.6 1341.5
193.7 1341.5
197.1 1341.5
202.2 1341.5
205.6 1341.5
210.7 1341.5
214.1 1341.5
219.2 1341.5
222.6 1341.5
227.7 1341.5
231.1 1341.5
234.5 1341.5
239.6 1341.5
243 1341.5
248.1 1341.5
251.5 1341.5
256.6 1341.5
260 1341.5
265 1342.1
268.4 1342.8
273.5 1342.8
276.9 1342.8
280.3 1342.8
285.4 1342.8
288.8 1342.8
293.9 1342.8
297.3 1342.8
302.4 1342.8
305.8 1343.4
310.9 1343.4
314.3 1343.4
319.4 1343.4
322.8 1343.4
326.2 1343.4
331.3 1343.4
334.7 1343.4
339.8 1343.4
343.2 1343.4
348.3 1343.4
351.7 1343.4
356.8 1343.4
360.2 1343.4
365.3 1343.4
368.7 1343.4
372.1 1343.4
377.2 1343.4
380.6 1343.4
385.7 1343.4
389.1 1343.4
394.2 1343.4
397.6 1343.4
402.7 1343.4
406.1 1343.4
411.2 1343.4
414.6 1343.4
418 1343.4
423.1 1343.4
426.5 1343.4
431.6 1343.4
435 1343.4
440 1343.4
443.4 1343.4
448.5 1343.4
451.9 1343.4
457 1343.4
460.4 1343.4
};
\addlegendentry{\small Fuzz'EMup}
\addplot [line width=1.4pt, forestgreen4416044]
table {%
0 0
1 234
2 268
3 299
4 365
5 398
6 416
7 456
8 467
9 491
10 545
11 589
12 610
13 667
14 701
15 745
16 757
17 787
18 803
19 823
20 843
21 854
22 867
23 898
24 904
25 942
26 978
27 995
28 1034
29 1112
30 1154
31 1200
32 1255
33 1278
34 1371
35 1380
36 1380
37 1380
38 1380
39 1380
40 1391
41 1391
42 1391
43 1391
44 1404
45 1404
46 1404
47 1404
48 1413
49 1413
50 1413
51 1413
52 1424
53 1424
54 1424
55 1424
56 1424
57 1437
58 1437
59 1437
60 1437
61 1437
62 1448
63 1448
64 1448
65 1448
66 1448
67 1455
68 1455
69 1455
70 1455
71 1455
72 1455
73 1455
74 1455
75 1459
76 1459
77 1459
78 1459
79 1459
80 1459
81 1462
82 1462
83 1462
84 1462
85 1463
86 1463
87 1463
88 1463
89 1464
90 1464
91 1464
92 1464
93 1464
94 1464
95 1464
96 1464
97 1464
98 1464
99 1464
100 1464
101 1464
102 1464
103 1464
104 1464
105 1464
106 1464
107 1464
108 1464
109 1464
110 1464
111 1464
112 1464
113 1464
114 1464
115 1464
116 1464
117 1464
118 1464
119 1464
120 1464
121 1464
122 1469
123 1469
124 1469
125 1491
126 1491
127 1491
128 1491
129 1491
130 1491
131 1491
132 1491
133 1504
134 1515
135 1515
136 1515
137 1515
138 1515
139 1515
140 1515
141 1515
142 1515
143 1515
144 1515
145 1515
146 1515
147 1515
148 1515
149 1515
150 1515
151 1515
152 1515
153 1515
154 1515
155 1515
156 1515
157 1515
158 1515
159 1515
160 1515
161 1515
162 1515
163 1515
164 1515
165 1515
166 1515
167 1515
168 1515
169 1520
170 1520
171 1520
172 1520
173 1520
174 1520
175 1520
176 1520
177 1520
178 1520
179 1520
180 1520
181 1520
182 1520
183 1520
184 1520
185 1520
186 1520
187 1520
188 1520
189 1520
190 1520
191 1520
192 1520
193 1520
194 1525
195 1525
196 1525
197 1525
198 1525
199 1525
200 1525
201 1525
202 1525
203 1525
204 1525
205 1525
206 1525
207 1525
208 1525
209 1525
210 1525
211 1525
212 1525
213 1525
214 1525
215 1525
216 1525
217 1525
218 1525
219 1525
220 1525
221 1525
222 1525
223 1525
224 1525
225 1525
226 1525
227 1525
228 1525
229 1525
230 1525
231 1525
232 1525
233 1525
234 1525
235 1525
236 1525
237 1525
238 1525
239 1525
240 1525
241 1525
242 1525
243 1561
244 1561
245 1561
246 1561
247 1561
248 1561
249 1561
250 1561
251 1561
252 1561
253 1561
254 1561
255 1561
256 1561
257 1561
258 1561
259 1561
260 1561
261 1561
262 1561
263 1561
264 1561
265 1561
266 1561
267 1561
268 1561
269 1561
270 1561
271 1561
272 1561
273 1561
274 1561
275 1561
276 1561
277 1561
278 1561
279 1561
280 1561
281 1561
282 1561
283 1561
284 1561
285 1561
286 1561
287 1561
288 1561
289 1561
290 1561
291 1561
292 1561
293 1561
294 1561
295 1561
296 1561
297 1561
298 1561
299 1561
300 1561
301 1561
302 1561
303 1561
304 1561
305 1561
306 1561
307 1561
308 1561
309 1561
310 1561
311 1561
312 1561
313 1561
314 1561
315 1561
316 1561
317 1561
318 1561
319 1561
320 1561
321 1561
322 1561
323 1561
324 1561
325 1561
326 1561
327 1561
328 1561
329 1561
330 1561
331 1561
332 1561
333 1561
334 1561
335 1561
336 1561
337 1561
338 1561
339 1561
340 1561
341 1561
342 1561
343 1568
344 1575
345 1575
346 1575
347 1575
348 1575
349 1575
350 1584
351 1584
352 1584
353 1584
354 1584
355 1615
356 1615
357 1615
358 1615
359 1615
360 1632
361 1632
362 1632
363 1632
364 1632
365 1632
366 1661
367 1661
368 1661
369 1661
370 1661
371 1661
372 1661
373 1661
374 1661
375 1661
376 1674
377 1674
378 1674
379 1674
380 1674
381 1674
382 1674
383 1674
384 1674
385 1674
386 1674
387 1674
388 1674
389 1674
390 1674
391 1674
392 1674
393 1674
394 1674
395 1674
396 1674
397 1674
398 1674
399 1674
400 1674
401 1674
402 1674
403 1674
404 1674
405 1674
406 1674
407 1674
408 1674
409 1674
410 1674
411 1674
412 1674
413 1674
414 1674
415 1674
416 1674
417 1674
418 1674
419 1674
420 1674
421 1674
422 1674
423 1674
424 1674
425 1691
426 1691
427 1691
428 1691
429 1691
430 1691
431 1691
432 1691
433 1691
434 1691
435 1691
436 1691
437 1691
438 1691
439 1691
440 1712
441 1712
442 1712
443 1712
444 1732
445 1743
446 1743
447 1743
448 1743
449 1743
450 1743
451 1743
452 1743
453 1751
454 1751
455 1751
456 1751
457 1751
458 1751
459 1751
460 1751
461 1751
462 1751
463 1751
464 1751
465 1751
466 1751
467 1751
468 1751
469 1751
470 1751
471 1751
472 1751
473 1751
474 1751
475 1751
476 1751
477 1754
478 1754
479 1754
480 1754
481 1754
482 1754
483 1754
484 1754
485 1754
486 1754
487 1754
488 1754
489 1754
490 1754
491 1754
492 1754
493 1754
494 1754
495 1754
496 1754
497 1754
498 1754
499 1754
500 1754
501 1754
502 1754
503 1754
504 1754
505 1754
506 1754
507 1754
508 1754
509 1754
510 1754
511 1754
512 1754
513 1754
514 1754
515 1754
516 1754
517 1754
518 1754
519 1754
520 1754
521 1754
522 1754
523 1754
524 1754
525 1754
526 1754
527 1754
528 1754
529 1754
530 1754
531 1754
532 1754
533 1754
534 1754
535 1754
536 1754
537 1754
538 1754
539 1754
540 1754
541 1754
542 1754
543 1754
544 1754
545 1754
546 1754
547 1754
548 1754
549 1754
550 1754
551 1754
552 1754
553 1754
554 1754
555 1754
556 1754
557 1754
558 1754
559 1754
560 1754
561 1754
562 1754
563 1754
564 1754
565 1754
566 1754
567 1754
568 1754
569 1754
570 1754
571 1754
572 1754
573 1754
574 1754
575 1754
576 1754
577 1754
578 1754
579 1754
580 1754
581 1754
582 1754
583 1754
584 1754
585 1754
586 1754
587 1754
588 1754
589 1754
590 1754
591 1754
592 1754
593 1754
594 1754
595 1754
596 1754
597 1754
598 1754
599 1754
600 1754
601 1754
602 1754
603 1754
604 1754
605 1754
606 1754
607 1754
608 1754
609 1754
610 1754
611 1754
612 1754
613 1754
614 1754
615 1754
616 1754
617 1754
618 1754
619 1754
620 1754
621 1754
622 1754
623 1754
624 1754
625 1754
626 1754
627 1754
628 1754
629 1754
630 1754
631 1754
632 1754
633 1754
634 1754
635 1754
636 1754
637 1754
638 1754
639 1754
640 1754
641 1754
642 1754
643 1754
644 1754
645 1754
646 1754
647 1754
648 1754
649 1754
650 1754
651 1754
652 1754
653 1754
654 1754
655 1754
656 1754
657 1754
658 1754
659 1754
660 1754
661 1754
662 1754
663 1754
664 1754
665 1754
666 1754
667 1754
668 1754
669 1754
670 1754
671 1754
672 1754
673 1754
674 1754
675 1754
676 1754
677 1754
678 1754
679 1754
680 1754
681 1754
682 1754
683 1754
684 1754
685 1754
686 1754
687 1754
688 1754
689 1754
690 1754
691 1754
692 1754
693 1754
694 1754
695 1754
696 1754
697 1754
698 1754
699 1754
700 1754
701 1754
702 1754
703 1754
704 1754
705 1754
706 1754
707 1754
708 1754
709 1754
710 1754
711 1754
712 1754
713 1754
714 1754
715 1754
716 1754
717 1754
718 1754
719 1754
720 1754
721 1754
722 1754
723 1754
724 1754
725 1754
726 1754
727 1754
728 1754
729 1754
730 1754
731 1754
732 1754
733 1754
734 1754
735 1754
736 1754
737 1754
738 1754
739 1754
740 1754
741 1754
742 1754
743 1754
744 1754
745 1754
746 1754
747 1754
748 1754
749 1754
750 1754
751 1754
752 1754
753 1754
754 1754
755 1754
756 1754
757 1754
758 1754
759 1754
760 1754
761 1754
762 1754
763 1754
764 1754
765 1754
766 1754
767 1754
768 1754
769 1754
770 1754
771 1754
772 1754
773 1754
774 1754
775 1754
776 1754
777 1754
778 1754
779 1754
780 1754
781 1754
782 1754
783 1754
784 1754
785 1754
786 1754
787 1754
788 1754
789 1754
790 1754
791 1754
792 1754
793 1754
794 1754
795 1754
796 1754
797 1754
798 1754
799 1754
800 1754
801 1754
802 1754
803 1754
804 1754
805 1754
806 1754
807 1754
808 1754
809 1754
810 1754
811 1754
812 1754
813 1754
814 1754
815 1754
816 1754
817 1754
818 1754
819 1754
820 1754
821 1754
822 1754
823 1754
824 1754
825 1754
826 1754
827 1754
828 1754
829 1754
830 1754
831 1754
832 1754
833 1754
834 1754
835 1754
836 1754
837 1754
838 1754
839 1754
840 1754
841 1754
842 1754
843 1754
844 1754
845 1754
846 1754
847 1754
848 1754
849 1754
850 1754
851 1754
852 1754
853 1754
854 1754
855 1754
856 1754
857 1754
858 1754
859 1754
860 1754
861 1754
862 1754
863 1754
864 1754
865 1754
866 1754
867 1754
868 1754
869 1754
870 1754
871 1754
872 1754
873 1754
874 1754
875 1754
876 1754
877 1754
878 1754
879 1754
880 1754
881 1754
882 1754
883 1754
884 1754
885 1754
886 1754
887 1754
888 1754
889 1754
890 1754
891 1754
892 1754
893 1754
894 1754
895 1754
896 1754
897 1754
898 1754
899 1754
900 1754
901 1754
902 1754
903 1754
904 1754
905 1754
906 1754
907 1754
908 1754
909 1754
910 1754
911 1754
912 1754
913 1754
914 1754
915 1754
916 1754
917 1754
918 1754
919 1754
920 1754
921 1754
922 1754
923 1754
924 1754
925 1754
926 1754
927 1754
928 1754
929 1754
930 1754
931 1754
932 1754
933 1754
934 1754
935 1754
936 1754
937 1754
938 1754
939 1754
940 1754
941 1754
942 1754
943 1754
944 1754
945 1754
946 1754
947 1754
948 1754
949 1754
950 1754
951 1754
952 1754
953 1754
954 1754
955 1754
956 1754
957 1754
958 1754
959 1754
960 1754
961 1754
962 1754
963 1754
964 1754
965 1754
966 1754
967 1754
968 1754
969 1754
970 1754
971 1754
972 1754
973 1754
974 1754
975 1754
976 1754
977 1754
978 1754
979 1754
980 1754
981 1754
982 1754
983 1754
984 1754
985 1754
986 1754
987 1754
988 1754
989 1754
990 1754
991 1754
992 1754
993 1754
994 1754
995 1754
996 1754
997 1754
998 1754
999 1754
1000 1754
};
\addlegendentry{\small PowerFuzz}

\nextgroupplot[
xlabel={\small \# Inputs},
xlabel shift={-2ex},
tick label style={font=\small},
legend cell align={left},
legend style={
  fill opacity=0.8,
  draw opacity=1,
  text opacity=1,
  at={(0.97,0.03)},
  anchor=south east,
  draw=lightgray204
},
tick align=outside,
tick pos=left,
title={\textbf{CNC}},
title style={yshift=-2ex},
x grid style={darkgray176},
xmajorgrids,
xmin=-50, xmax=1050,
xtick style={color=black},
y grid style={darkgray176},
ymajorgrids,
ymin=-73.35, ymax=1540.35,
ytick style={color=black}
]
\addplot [line width=1.4pt, steelblue31119180]
table {%
43.2 667.3
57.6 855.6
67.2 895
76 920.1
84.8 923.6
94.4 928.1
108.8 936.1
117.6 938.4
127.2 941.2
136 941.2
144.8 941.2
153.6 941.2
163.2 947
172 947
187.2 947
196.8 947
205.6 947
214.4 947
224 947
232.8 947
241.6 947
251.2 947
260 947
268.8 947
278.4 965.2
287.2 965.2
296 965.2
304.8 965.2
314.4 965.8
323.2 966.4
332 966.4
341.6 978.3
350.4 990.3
359.2 990.3
368.8 990.3
377.6 990.3
386.4 990.3
396 990.3
404.8 990.3
413.6 990.3
423.2 990.3
432 990.3
440.8 990.3
449.6 990.3
459.2 990.3
468 990.3
476.8 990.3
486.4 990.3
495.2 990.3
504 990.3
513.6 990.3
522.4 990.3
531.2 990.3
540.8 990.3
549.6 990.3
558.4 990.3
568 990.3
576.8 990.3
585.6 990.3
594.4 990.3
604 990.3
612.8 990.3
621.6 990.3
631.2 990.3
640 990.3
648.8 990.3
658.4 990.3
667.2 990.3
676 990.3
685.6 990.3
694.4 990.3
703.2 990.9
712.8 993.2
721.6 993.2
730.4 993.2
739.2 993.2
748.8 993.2
757.6 993.2
766.4 993.2
776 993.2
784.8 993.2
793.6 993.2
803.2 993.2
812 993.2
820.8 993.2
830.4 993.2
839.2 993.2
848 993.2
857.6 993.2
866.4 993.2
875.2 993.2
884 993.2
893.6 993.2
902.4 993.2
911.2 993.2
920.8 993.2
929.6 993.2
938.4 993.2
948 993.2
956.8 993.2
};
\addlegendentry{\small Random}
\addplot [line width=1.4pt, darkorange25512714]
table {%
0 0
43.2 675.3
50.4 841.4
57.6 873.9
64 900.1
71.2 904.7
78.4 908.1
85.6 910.4
92.8 911.6
100 912.7
106.4 924.7
113.6 933.8
120.8 933.8
128 933.8
135.2 933.8
142.4 933.8
148.8 933.8
156 933.8
163.2 939.5
170.4 939.5
177.6 957.2
184.8 1100.5
191.2 1101.1
198.4 1102.8
205.6 1102.8
212.8 1102.8
220 1103.9
227.2 1103.9
233.6 1103.9
240.8 1103.9
248 1104.5
255.2 1105.6
262.4 1109.6
269.6 1109.6
276 1109.6
283.2 1109.6
290.4 1109.6
297.6 1121
304.8 1133.6
311.2 1133.6
318.4 1133.6
325.6 1133.6
332.8 1133.6
340 1133.6
347.2 1133.6
353.6 1133.6
360.8 1139.9
368 1139.9
375.2 1139.9
382.4 1139.9
389.6 1139.9
396 1139.9
403.2 1139.9
410.4 1139.9
417.6 1139.9
424.8 1139.9
432 1139.9
438.4 1139.9
445.6 1139.9
452.8 1139.9
460 1139.9
467.2 1139.9
474.4 1139.9
480.8 1139.9
488 1139.9
495.2 1139.9
502.4 1139.9
509.6 1139.9
516 1139.9
523.2 1139.9
530.4 1139.9
537.6 1139.9
544.8 1139.9
552 1139.9
558.4 1139.9
565.6 1139.9
572.8 1140.5
580 1141.6
587.2 1141.6
594.4 1141.6
600.8 1141.6
608 1141.6
615.2 1141.6
622.4 1141.6
629.6 1141.6
636.8 1141.6
643.2 1141.6
650.4 1141.6
657.6 1141.6
664.8 1141.6
672 1141.6
679.2 1141.6
685.6 1141.6
692.8 1141.6
700 1141.6
707.2 1141.6
714.4 1141.6
721.6 1141.6
728 1141.6
735.2 1141.6
742.4 1141.6
};
\addlegendentry{\small Fuzz'EMup}
\addplot [line width=1.4pt, forestgreen4416044]
table {%
0 0
1 143
2 153
3 219
4 234
5 254
6 284
7 301
8 343
9 394
10 435
11 495
12 532
13 564
14 581
15 598
16 621
17 644
18 667
19 684
20 701
21 718
22 735
23 752
24 769
25 786
26 803
27 820
28 837
29 841
30 864
31 881
32 881
33 881
34 881
35 902
36 942
37 954
38 958
39 963
40 969
41 972
42 972
43 983
44 984
45 993
46 993
47 999
48 1013
49 1021
50 1021
51 1025
52 1033
53 1035
54 1043
55 1046
56 1050
57 1050
58 1052
59 1053
60 1055
61 1057
62 1061
63 1065
64 1067
65 1067
66 1067
67 1067
68 1067
69 1067
70 1067
71 1067
72 1067
73 1067
74 1067
75 1067
76 1067
77 1067
78 1067
79 1067
80 1067
81 1067
82 1067
83 1067
84 1067
85 1067
86 1067
87 1067
88 1081
89 1081
90 1081
91 1081
92 1081
93 1081
94 1081
95 1102
96 1110
97 1114
98 1114
99 1114
100 1114
101 1114
102 1114
103 1114
104 1114
105 1114
106 1114
107 1114
108 1120
109 1120
110 1131
111 1133
112 1143
113 1143
114 1143
115 1154
116 1154
117 1154
118 1154
119 1154
120 1154
121 1154
122 1154
123 1154
124 1154
125 1154
126 1154
127 1154
128 1161
129 1161
130 1161
131 1161
132 1161
133 1161
134 1161
135 1164
136 1164
137 1164
138 1164
139 1164
140 1168
141 1168
142 1168
143 1168
144 1170
145 1170
146 1170
147 1170
148 1170
149 1170
150 1170
151 1170
152 1170
153 1170
154 1170
155 1170
156 1170
157 1170
158 1170
159 1170
160 1170
161 1170
162 1170
163 1170
164 1170
165 1170
166 1170
167 1170
168 1175
169 1175
170 1184
171 1184
172 1184
173 1184
174 1184
175 1184
176 1184
177 1184
178 1184
179 1184
180 1184
181 1184
182 1184
183 1184
184 1184
185 1184
186 1184
187 1184
188 1184
189 1184
190 1184
191 1184
192 1184
193 1184
194 1184
195 1184
196 1184
197 1184
198 1184
199 1184
200 1184
201 1184
202 1184
203 1184
204 1184
205 1184
206 1184
207 1184
208 1184
209 1184
210 1184
211 1184
212 1184
213 1184
214 1184
215 1184
216 1184
217 1184
218 1184
219 1184
220 1184
221 1184
222 1184
223 1184
224 1184
225 1184
226 1184
227 1184
228 1184
229 1184
230 1184
231 1184
232 1184
233 1184
234 1184
235 1184
236 1184
237 1184
238 1184
239 1184
240 1184
241 1184
242 1184
243 1184
244 1184
245 1184
246 1184
247 1184
248 1184
249 1184
250 1184
251 1184
252 1184
253 1184
254 1184
255 1184
256 1184
257 1184
258 1184
259 1184
260 1184
261 1184
262 1184
263 1186
264 1188
265 1190
266 1192
267 1192
268 1192
269 1192
270 1192
271 1192
272 1192
273 1192
274 1192
275 1192
276 1192
277 1192
278 1192
279 1192
280 1192
281 1192
282 1192
283 1192
284 1192
285 1192
286 1192
287 1192
288 1192
289 1192
290 1192
291 1192
292 1192
293 1192
294 1192
295 1192
296 1192
297 1192
298 1192
299 1221
300 1221
301 1221
302 1221
303 1221
304 1221
305 1243
306 1254
307 1263
308 1263
309 1263
310 1263
311 1263
312 1263
313 1263
314 1263
315 1263
316 1263
317 1263
318 1263
319 1263
320 1263
321 1263
322 1263
323 1263
324 1263
325 1263
326 1263
327 1263
328 1263
329 1263
330 1263
331 1263
332 1263
333 1263
334 1263
335 1263
336 1263
337 1263
338 1263
339 1263
340 1263
341 1263
342 1263
343 1263
344 1263
345 1263
346 1263
347 1263
348 1263
349 1263
350 1263
351 1263
352 1263
353 1263
354 1263
355 1263
356 1263
357 1263
358 1263
359 1263
360 1263
361 1263
362 1263
363 1263
364 1263
365 1263
366 1263
367 1263
368 1263
369 1263
370 1263
371 1263
372 1263
373 1263
374 1263
375 1263
376 1263
377 1263
378 1263
379 1269
380 1279
381 1283
382 1283
383 1283
384 1283
385 1283
386 1283
387 1283
388 1283
389 1283
390 1283
391 1283
392 1283
393 1283
394 1283
395 1283
396 1283
397 1283
398 1283
399 1283
400 1283
401 1283
402 1283
403 1283
404 1283
405 1283
406 1283
407 1283
408 1283
409 1283
410 1283
411 1283
412 1283
413 1283
414 1283
415 1283
416 1283
417 1283
418 1283
419 1283
420 1283
421 1283
422 1283
423 1283
424 1283
425 1283
426 1283
427 1283
428 1283
429 1283
430 1283
431 1283
432 1283
433 1283
434 1283
435 1283
436 1283
437 1293
438 1297
439 1304
440 1318
441 1318
442 1318
443 1318
444 1318
445 1318
446 1318
447 1318
448 1318
449 1318
450 1318
451 1318
452 1318
453 1318
454 1318
455 1318
456 1318
457 1318
458 1318
459 1318
460 1318
461 1318
462 1318
463 1318
464 1318
465 1318
466 1318
467 1318
468 1318
469 1318
470 1318
471 1318
472 1325
473 1343
474 1351
475 1351
476 1351
477 1351
478 1363
479 1372
480 1385
481 1385
482 1385
483 1385
484 1395
485 1403
486 1410
487 1410
488 1410
489 1421
490 1445
491 1445
492 1445
493 1453
494 1453
495 1453
496 1467
497 1467
498 1467
499 1467
500 1467
501 1467
502 1467
503 1467
504 1467
505 1467
506 1467
507 1467
508 1467
509 1467
510 1467
511 1467
512 1467
513 1467
514 1467
515 1467
516 1467
517 1467
518 1467
519 1467
520 1467
521 1467
522 1467
523 1467
524 1467
525 1467
526 1467
527 1467
528 1467
529 1467
530 1467
531 1467
532 1467
533 1467
534 1467
535 1467
536 1467
537 1467
538 1467
539 1467
540 1467
541 1467
542 1467
543 1467
544 1467
545 1467
546 1467
547 1467
548 1467
549 1467
550 1467
551 1467
552 1467
553 1467
554 1467
555 1467
556 1467
557 1467
558 1467
559 1467
560 1467
561 1467
562 1467
563 1467
564 1467
565 1467
566 1467
567 1467
568 1467
569 1467
570 1467
571 1467
572 1467
573 1467
574 1467
575 1467
576 1467
577 1467
578 1467
579 1467
580 1467
581 1467
582 1467
583 1467
584 1467
585 1467
586 1467
587 1467
588 1467
589 1467
590 1467
591 1467
592 1467
593 1467
594 1467
595 1467
596 1467
597 1467
598 1467
599 1467
600 1467
601 1467
602 1467
603 1467
604 1467
605 1467
606 1467
607 1467
608 1467
609 1467
610 1467
611 1467
612 1467
613 1467
614 1467
615 1467
616 1467
617 1467
618 1467
619 1467
620 1467
621 1467
622 1467
623 1467
624 1467
625 1467
626 1467
627 1467
628 1467
629 1467
630 1467
631 1467
632 1467
633 1467
634 1467
635 1467
636 1467
637 1467
638 1467
639 1467
640 1467
641 1467
642 1467
643 1467
644 1467
645 1467
646 1467
647 1467
648 1467
649 1467
650 1467
651 1467
652 1467
653 1467
654 1467
655 1467
656 1467
657 1467
658 1467
659 1467
660 1467
661 1467
662 1467
663 1467
664 1467
665 1467
666 1467
667 1467
668 1467
669 1467
670 1467
671 1467
672 1467
673 1467
674 1467
675 1467
676 1467
677 1467
678 1467
679 1467
680 1467
681 1467
682 1467
683 1467
684 1467
685 1467
686 1467
687 1467
688 1467
689 1467
690 1467
691 1467
692 1467
693 1467
694 1467
695 1467
696 1467
697 1467
698 1467
699 1467
700 1467
701 1467
702 1467
703 1467
704 1467
705 1467
706 1467
707 1467
708 1467
709 1467
710 1467
711 1467
712 1467
713 1467
714 1467
715 1467
716 1467
717 1467
718 1467
719 1467
720 1467
721 1467
722 1467
723 1467
724 1467
725 1467
726 1467
727 1467
728 1467
729 1467
730 1467
731 1467
732 1467
733 1467
734 1467
735 1467
736 1467
737 1467
738 1467
739 1467
740 1467
741 1467
742 1467
743 1467
744 1467
745 1467
746 1467
747 1467
748 1467
749 1467
750 1467
751 1467
752 1467
753 1467
754 1467
755 1467
756 1467
757 1467
758 1467
759 1467
760 1467
761 1467
762 1467
763 1467
764 1467
765 1467
766 1467
767 1467
768 1467
769 1467
770 1467
771 1467
772 1467
773 1467
774 1467
775 1467
776 1467
777 1467
778 1467
779 1467
780 1467
781 1467
782 1467
783 1467
784 1467
785 1467
786 1467
787 1467
788 1467
789 1467
790 1467
791 1467
792 1467
793 1467
794 1467
795 1467
796 1467
797 1467
798 1467
799 1467
800 1467
801 1467
802 1467
803 1467
804 1467
805 1467
806 1467
807 1467
808 1467
809 1467
810 1467
811 1467
812 1467
813 1467
814 1467
815 1467
816 1467
817 1467
818 1467
819 1467
820 1467
821 1467
822 1467
823 1467
824 1467
825 1467
826 1467
827 1467
828 1467
829 1467
830 1467
831 1467
832 1467
833 1467
834 1467
835 1467
836 1467
837 1467
838 1467
839 1467
840 1467
841 1467
842 1467
843 1467
844 1467
845 1467
846 1467
847 1467
848 1467
849 1467
850 1467
851 1467
852 1467
853 1467
854 1467
855 1467
856 1467
857 1467
858 1467
859 1467
860 1467
861 1467
862 1467
863 1467
864 1467
865 1467
866 1467
867 1467
868 1467
869 1467
870 1467
871 1467
872 1467
873 1467
874 1467
875 1467
876 1467
877 1467
878 1467
879 1467
880 1467
881 1467
882 1467
883 1467
884 1467
885 1467
886 1467
887 1467
888 1467
889 1467
890 1467
891 1467
892 1467
893 1467
894 1467
895 1467
896 1467
897 1467
898 1467
899 1467
900 1467
901 1467
902 1467
903 1467
904 1467
905 1467
906 1467
907 1467
908 1467
909 1467
910 1467
911 1467
912 1467
913 1467
914 1467
915 1467
916 1467
917 1467
918 1467
919 1467
920 1467
921 1467
922 1467
923 1467
924 1467
925 1467
926 1467
927 1467
928 1467
929 1467
930 1467
931 1467
932 1467
933 1467
934 1467
935 1467
936 1467
937 1467
938 1467
939 1467
940 1467
941 1467
942 1467
943 1467
944 1467
945 1467
946 1467
947 1467
948 1467
949 1467
950 1467
951 1467
952 1467
953 1467
954 1467
955 1467
956 1467
957 1467
958 1467
959 1467
960 1467
961 1467
962 1467
963 1467
964 1467
965 1467
966 1467
967 1467
968 1467
969 1467
970 1467
971 1467
972 1467
973 1467
974 1467
975 1467
976 1467
977 1467
978 1467
979 1467
980 1467
981 1467
982 1467
983 1467
984 1467
985 1467
986 1467
987 1467
988 1467
989 1467
990 1467
991 1467
992 1467
993 1467
994 1467
995 1467
996 1467
997 1467
998 1467
999 1467
1000 1467
};
\addlegendentry{\small PowerFuzz}

\nextgroupplot[
xlabel={\small \# Inputs},
xlabel shift={-2ex},
ylabel={\small \# Branch hits},
ylabel shift={-1ex},
tick label style={font=\small},
legend cell align={left},
legend style={
  fill opacity=0.8,
  draw opacity=1,
  text opacity=1,
  at={(0.97,0.03)},
  anchor=south east,
  draw=lightgray204
},
tick align=outside,
tick pos=left,
title={\textbf{GPS}},
title style={yshift=-2ex},
x grid style={darkgray176},
xmajorgrids,
xmin=-50, xmax=1050,
xtick style={color=black},
y grid style={darkgray176},
ymajorgrids,
ymin=-52.25, ymax=1097.25,
ytick style={color=black}
]
\addplot [line width=1.4pt, steelblue31119180]
table {%
47.7 753.8
57 816.3
63.6 829.4
71.5 837.5
83.5 840
91.4 847.5
104.7 849.4
112.6 849.4
119.2 850.7
127.2 853.2
135.1 855
141.8 856.3
149.7 856.3
157.7 860
164.3 860
172.2 863.1
180.2 863.8
186.8 865
194.8 865
202.7 865
209.3 865
217.3 865
225.2 865.6
231.9 865.6
239.8 866.3
247.8 866.9
254.4 866.9
262.3 866.9
270.3 867.5
276.9 867.5
284.9 867.5
292.8 871.9
299.4 873.1
307.4 873.1
315.3 873.1
322 873.1
329.9 873.1
337.9 873.1
344.5 873.1
352.4 873.1
360.4 873.1
367 873.1
375 873.8
382.9 874.4
389.5 875.6
397.5 875.6
405.4 875.6
412.1 875.6
420 875.6
427.9 875.6
434.6 875.6
442.5 875.6
450.5 875.6
457.1 875.6
465 875.6
473 875.6
479.6 875.6
487.6 875.6
495.5 876.3
502.1 877.5
510.1 877.5
518 877.5
524.7 877.5
532.6 877.5
540.6 877.5
547.2 877.5
555.1 877.5
563.1 880
569.7 882.5
577.7 882.5
585.6 882.5
592.2 882.5
600.2 882.5
608.1 882.5
614.8 882.5
622.7 883.1
630.7 883.8
637.3 883.8
645.2 883.8
653.2 883.8
659.8 883.8
667.8 883.8
675.7 883.8
682.3 883.8
690.3 883.8
698.2 883.8
704.9 883.8
712.8 883.8
720.8 883.8
727.4 883.8
735.3 883.8
743.3 883.8
749.9 883.8
757.9 883.8
765.8 883.8
772.4 885
780.4 885
788.3 885
795 885
802.9 885
};
\addlegendentry{\small Random}
\addplot [line width=1.4pt, darkorange25512714]
table {%
37.1 649.5
42.4 705.1
47.7 789.4
53 818.2
58.3 828.8
63.6 839.4
68.9 839.4
74.2 839.4
79.5 843.2
84.8 844.4
90.1 845
95.4 848.8
100.7 851.9
104.7 851.9
110 856.9
115.3 857.5
120.6 857.5
125.9 857.5
131.2 857.5
136.5 863.1
141.8 863.1
147.1 863.1
152.4 863.1
157.7 865.6
163 865.6
168.3 868.1
173.6 868.8
178.9 868.8
184.2 869.4
189.5 870.6
194.8 878.1
200.1 884.4
205.4 884.4
210.7 884.4
216 884.4
221.3 884.4
226.6 885
231.9 885
235.8 885
241.1 885
246.4 885
251.7 890
257 893.1
262.3 893.1
267.6 893.1
272.9 893.1
278.2 893.1
283.5 893.1
288.8 893.1
294.1 893.1
299.4 893.1
304.7 895
310 895
315.3 895
320.6 896.3
325.9 896.9
331.2 896.9
336.5 896.9
341.8 898.1
347.1 898.1
352.4 898.1
357.7 898.1
361.7 898.1
367 898.1
372.3 898.1
377.6 898.1
382.9 898.1
388.2 898.1
393.5 898.1
398.8 898.1
404.1 898.1
409.4 898.1
414.7 898.1
420 898.1
425.3 898.1
430.6 898.1
435.9 898.1
441.2 898.1
446.5 898.1
451.8 898.1
457.1 898.1
462.4 898.1
467.7 898.1
473 898.1
478.3 898.1
483.6 898.1
488.9 898.1
492.9 898.1
498.2 898.1
503.5 898.1
508.8 898.1
514.1 898.1
519.4 898.1
524.7 898.1
530 898.1
535.3 898.1
540.6 898.1
545.9 898.1
551.2 898.1
556.5 898.1
};
\addlegendentry{\small Fuzz'EMup}
\addplot [line width=1.4pt, forestgreen4416044]
table {%
0 0
1 50
2 100
3 150
4 200
5 250
6 300
7 350
8 400
9 450
10 475
11 500
12 525
13 550
14 575
15 600
16 612
17 625
18 637
19 650
20 662
21 675
22 687
23 700
24 712
25 725
26 737
27 750
28 750
29 787
30 787
31 787
32 812
33 812
34 812
35 812
36 824
37 824
38 824
39 824
40 834
41 834
42 834
43 834
44 834
45 834
46 834
47 834
48 834
49 834
50 834
51 834
52 834
53 834
54 834
55 834
56 834
57 834
58 834
59 838
60 838
61 838
62 838
63 838
64 838
65 838
66 838
67 838
68 838
69 838
70 838
71 838
72 838
73 838
74 838
75 838
76 843
77 843
78 843
79 851
80 851
81 851
82 851
83 863
84 863
85 863
86 863
87 863
88 863
89 863
90 863
91 876
92 876
93 876
94 876
95 876
96 885
97 885
98 885
99 885
100 885
101 885
102 885
103 885
104 885
105 885
106 885
107 885
108 885
109 885
110 885
111 885
112 885
113 885
114 885
115 885
116 885
117 885
118 885
119 885
120 885
121 885
122 885
123 885
124 885
125 885
126 885
127 885
128 885
129 885
130 885
131 885
132 885
133 885
134 885
135 885
136 885
137 885
138 885
139 885
140 885
141 885
142 885
143 885
144 885
145 885
146 885
147 885
148 885
149 885
150 885
151 885
152 885
153 885
154 885
155 885
156 885
157 885
158 885
159 885
160 885
161 885
162 885
163 885
164 885
165 885
166 885
167 885
168 885
169 885
170 885
171 885
172 885
173 885
174 885
175 885
176 885
177 885
178 885
179 898
180 898
181 898
182 898
183 898
184 898
185 898
186 898
187 898
188 898
189 898
190 898
191 898
192 898
193 898
194 898
195 898
196 912
197 912
198 912
199 912
200 912
201 912
202 921
203 921
204 921
205 921
206 921
207 921
208 921
209 921
210 921
211 921
212 921
213 921
214 921
215 921
216 921
217 921
218 921
219 921
220 921
221 921
222 921
223 921
224 921
225 921
226 921
227 921
228 921
229 921
230 921
231 921
232 921
233 921
234 921
235 921
236 921
237 921
238 921
239 921
240 921
241 921
242 921
243 921
244 921
245 921
246 921
247 921
248 921
249 921
250 921
251 921
252 921
253 921
254 921
255 921
256 921
257 921
258 921
259 925
260 925
261 925
262 925
263 925
264 925
265 925
266 925
267 925
268 925
269 925
270 925
271 925
272 925
273 925
274 925
275 925
276 925
277 926
278 926
279 926
280 926
281 926
282 926
283 926
284 926
285 926
286 926
287 926
288 926
289 926
290 926
291 926
292 926
293 926
294 928
295 928
296 928
297 928
298 928
299 928
300 928
301 928
302 928
303 928
304 928
305 928
306 928
307 928
308 928
309 928
310 928
311 928
312 928
313 928
314 928
315 928
316 928
317 928
318 928
319 928
320 928
321 928
322 928
323 928
324 928
325 941
326 941
327 946
328 946
329 946
330 946
331 952
332 952
333 952
334 952
335 952
336 952
337 952
338 952
339 952
340 952
341 952
342 964
343 964
344 964
345 964
346 964
347 964
348 964
349 964
350 964
351 964
352 964
353 964
354 964
355 964
356 964
357 964
358 964
359 971
360 971
361 971
362 971
363 971
364 971
365 971
366 971
367 971
368 971
369 971
370 971
371 971
372 971
373 971
374 971
375 971
376 971
377 971
378 971
379 971
380 984
381 984
382 984
383 984
384 984
385 984
386 984
387 984
388 984
389 984
390 984
391 984
392 984
393 984
394 984
395 984
396 984
397 984
398 984
399 984
400 984
401 984
402 984
403 984
404 984
405 984
406 984
407 984
408 984
409 984
410 984
411 984
412 984
413 984
414 984
415 984
416 984
417 984
418 984
419 984
420 984
421 984
422 984
423 984
424 984
425 984
426 984
427 984
428 984
429 984
430 984
431 984
432 984
433 984
434 984
435 984
436 984
437 984
438 984
439 984
440 984
441 984
442 984
443 984
444 984
445 984
446 984
447 984
448 984
449 984
450 984
451 984
452 984
453 984
454 984
455 984
456 984
457 993
458 993
459 993
460 993
461 993
462 1004
463 1004
464 1004
465 1004
466 1004
467 1004
468 1004
469 1004
470 1004
471 1004
472 1004
473 1004
474 1004
475 1004
476 1004
477 1004
478 1004
479 1004
480 1004
481 1012
482 1012
483 1012
484 1017
485 1017
486 1017
487 1017
488 1017
489 1017
490 1017
491 1017
492 1017
493 1017
494 1017
495 1017
496 1017
497 1017
498 1017
499 1017
500 1017
501 1017
502 1017
503 1017
504 1017
505 1017
506 1017
507 1017
508 1017
509 1017
510 1017
511 1017
512 1017
513 1017
514 1017
515 1017
516 1017
517 1017
518 1017
519 1017
520 1017
521 1017
522 1017
523 1017
524 1017
525 1017
526 1017
527 1017
528 1017
529 1017
530 1017
531 1017
532 1017
533 1017
534 1023
535 1023
536 1026
537 1026
538 1026
539 1026
540 1026
541 1026
542 1026
543 1026
544 1026
545 1026
546 1026
547 1026
548 1026
549 1026
550 1026
551 1026
552 1026
553 1026
554 1026
555 1026
556 1026
557 1026
558 1026
559 1026
560 1026
561 1026
562 1026
563 1026
564 1026
565 1026
566 1026
567 1026
568 1026
569 1026
570 1026
571 1026
572 1026
573 1026
574 1026
575 1026
576 1026
577 1026
578 1026
579 1026
580 1026
581 1026
582 1026
583 1026
584 1026
585 1026
586 1026
587 1026
588 1026
589 1026
590 1026
591 1026
592 1026
593 1026
594 1026
595 1026
596 1026
597 1026
598 1026
599 1026
600 1026
601 1026
602 1026
603 1026
604 1026
605 1026
606 1026
607 1026
608 1026
609 1026
610 1026
611 1026
612 1026
613 1026
614 1026
615 1026
616 1026
617 1026
618 1026
619 1026
620 1026
621 1026
622 1026
623 1026
624 1026
625 1026
626 1026
627 1026
628 1026
629 1026
630 1026
631 1026
632 1026
633 1026
634 1035
635 1035
636 1035
637 1035
638 1035
639 1035
640 1035
641 1035
642 1035
643 1035
644 1035
645 1035
646 1035
647 1035
648 1035
649 1035
650 1035
651 1035
652 1035
653 1035
654 1035
655 1035
656 1035
657 1035
658 1035
659 1035
660 1045
661 1045
662 1045
663 1045
664 1045
665 1045
666 1045
667 1045
668 1045
669 1045
670 1045
671 1045
672 1045
673 1045
674 1045
675 1045
676 1045
677 1045
678 1045
679 1045
680 1045
681 1045
682 1045
683 1045
684 1045
685 1045
686 1045
687 1045
688 1045
689 1045
690 1045
691 1045
692 1045
693 1045
694 1045
695 1045
696 1045
697 1045
698 1045
699 1045
700 1045
701 1045
702 1045
703 1045
704 1045
705 1045
706 1045
707 1045
708 1045
709 1045
710 1045
711 1045
712 1045
713 1045
714 1045
715 1045
716 1045
717 1045
718 1045
719 1045
720 1045
721 1045
722 1045
723 1045
724 1045
725 1045
726 1045
727 1045
728 1045
729 1045
730 1045
731 1045
732 1045
733 1045
734 1045
735 1045
736 1045
737 1045
738 1045
739 1045
740 1045
741 1045
742 1045
743 1045
744 1045
745 1045
746 1045
747 1045
748 1045
749 1045
750 1045
751 1045
752 1045
753 1045
754 1045
755 1045
756 1045
757 1045
758 1045
759 1045
760 1045
761 1045
762 1045
763 1045
764 1045
765 1045
766 1045
767 1045
768 1045
769 1045
770 1045
771 1045
772 1045
773 1045
774 1045
775 1045
776 1045
777 1045
778 1045
779 1045
780 1045
781 1045
782 1045
783 1045
784 1045
785 1045
786 1045
787 1045
788 1045
789 1045
790 1045
791 1045
792 1045
793 1045
794 1045
795 1045
796 1045
797 1045
798 1045
799 1045
800 1045
801 1045
802 1045
803 1045
804 1045
805 1045
806 1045
807 1045
808 1045
809 1045
810 1045
811 1045
812 1045
813 1045
814 1045
815 1045
816 1045
817 1045
818 1045
819 1045
820 1045
821 1045
822 1045
823 1045
824 1045
825 1045
826 1045
827 1045
828 1045
829 1045
830 1045
831 1045
832 1045
833 1045
834 1045
835 1045
836 1045
837 1045
838 1045
839 1045
840 1045
841 1045
842 1045
843 1045
844 1045
845 1045
846 1045
847 1045
848 1045
849 1045
850 1045
851 1045
852 1045
853 1045
854 1045
855 1045
856 1045
857 1045
858 1045
859 1045
860 1045
861 1045
862 1045
863 1045
864 1045
865 1045
866 1045
867 1045
868 1045
869 1045
870 1045
871 1045
872 1045
873 1045
874 1045
875 1045
876 1045
877 1045
878 1045
879 1045
880 1045
881 1045
882 1045
883 1045
884 1045
885 1045
886 1045
887 1045
888 1045
889 1045
890 1045
891 1045
892 1045
893 1045
894 1045
895 1045
896 1045
897 1045
898 1045
899 1045
900 1045
901 1045
902 1045
903 1045
904 1045
905 1045
906 1045
907 1045
908 1045
909 1045
910 1045
911 1045
912 1045
913 1045
914 1045
915 1045
916 1045
917 1045
918 1045
919 1045
920 1045
921 1045
922 1045
923 1045
924 1045
925 1045
926 1045
927 1045
928 1045
929 1045
930 1045
931 1045
932 1045
933 1045
934 1045
935 1045
936 1045
937 1045
938 1045
939 1045
940 1045
941 1045
942 1045
943 1045
944 1045
945 1045
946 1045
947 1045
948 1045
949 1045
950 1045
951 1045
952 1045
953 1045
954 1045
955 1045
956 1045
957 1045
958 1045
959 1045
960 1045
961 1045
962 1045
963 1045
964 1045
965 1045
966 1045
967 1045
968 1045
969 1045
970 1045
971 1045
972 1045
973 1045
974 1045
975 1045
976 1045
977 1045
978 1045
979 1045
980 1045
981 1045
982 1045
983 1045
984 1045
985 1045
986 1045
987 1045
988 1045
989 1045
990 1045
991 1045
992 1045
993 1045
994 1045
995 1045
996 1045
997 1045
998 1045
999 1045
1000 1045
};
\addlegendentry{\small PowerFuzz}

\nextgroupplot[
xlabel={\small \# Inputs},
xlabel shift={-2ex},
tick label style={font=\small},
legend cell align={left},
legend style={
  fill opacity=0.8,
  draw opacity=1,
  text opacity=1,
  at={(0.97,0.03)},
  anchor=south east,
  draw=lightgray204
},
tick align=outside,
tick pos=left,
title={\textbf{Soldering Station}},
title style={yshift=-2ex},
x grid style={darkgray176},
xmajorgrids,
xmin=-50, xmax=1050,
xtick style={color=black},
y grid style={darkgray176},
ymajorgrids,
ymin=-71.9, ymax=1509.9,
ytick style={color=black}
]
\addplot [line width=1.4pt, steelblue31119180]
table {%
52.1 957.8
66.9 1024.3
75.4 1026.2
84.8 1027.5
94.1 1027.5
102.6 1029.4
112 1030.4
120.5 1032.5
129.9 1034.4
139.2 1037.4
151.6 1040
161 1040
170.3 1040
178.8 1041.5
188.2 1043
197.5 1044.9
206.1 1044.9
215.4 1044.9
224 1044.9
233.3 1045.9
242.6 1045.9
251.2 1045.9
260.5 1045.9
269.8 1045.9
278.4 1045.9
287.7 1045.9
297 1045.9
305.6 1045.9
314.9 1045.9
323.5 1045.9
332.8 1045.9
342.1 1045.9
350.7 1045.9
360 1045.9
369.4 1045.9
377.9 1045.9
387.2 1045.9
396.6 1045.9
405.1 1045.9
414.5 1045.9
423 1045.9
432.3 1045.9
441.7 1045.9
450.2 1045.9
459.6 1045.9
468.9 1045.9
477.4 1045.9
486.8 1045.9
496.1 1045.9
504.7 1045.9
514 1045.9
522.6 1045.9
531.9 1045.9
541.2 1045.9
549.8 1045.9
559.1 1045.9
568.4 1045.9
577 1045.9
586.3 1045.9
595.6 1045.9
604.2 1045.9
613.5 1045.9
622.1 1045.9
631.4 1045.9
640.7 1045.9
649.3 1045.9
658.6 1045.9
668 1045.9
676.5 1045.9
685.8 1045.9
695.2 1045.9
703.7 1045.9
713.1 1045.9
721.6 1045.9
730.9 1045.9
740.3 1045.9
748.8 1045.9
758.2 1045.9
767.5 1045.9
776 1045.9
785.4 1045.9
794.7 1045.9
803.3 1045.9
812.6 1045.9
821.2 1045.9
830.5 1045.9
839.8 1045.9
848.4 1045.9
857.7 1045.9
867 1045.9
875.6 1045.9
884.9 1045.9
894.2 1045.9
902.8 1045.9
912.1 1045.9
920.7 1045.9
930 1045.9
939.3 1045.9
947.9 1045.9
957.2 1045.9
};
\addlegendentry{\small Random}
\addplot [line width=1.4pt, darkorange25512714]
table {%
43.5 927.9
50.5 978.4
57.5 1016.8
63.8 1026.6
70.8 1032.3
77.8 1034.4
84.8 1035.3
91 1036.2
98 1036.2
105 1036.2
112 1036.2
119 1036.2
125.2 1036.2
132.2 1036.2
139.2 1036.2
146.2 1067.1
152.4 1068.8
159.4 1069.1
166.4 1069.1
173.4 1069.1
180.4 1069.1
186.6 1069.1
193.6 1070.1
200.6 1071.6
207.6 1072
213.8 1072
220.8 1072
227.8 1072
234.8 1072
241.8 1072
248.1 1072
255.1 1072
262.1 1072
269.1 1072
275.3 1072
282.3 1072
289.3 1072.5
296.3 1073.1
303.3 1073.1
309.5 1073.1
316.5 1073.1
323.5 1073.1
330.5 1073.1
336.7 1073.1
343.7 1073.1
350.7 1073.1
357.7 1073.1
364.7 1073.1
370.9 1073.1
377.9 1083.7
384.9 1083.7
391.9 1083.7
398.1 1083.7
405.1 1083.7
412.1 1083.7
419.1 1083.7
426.1 1083.7
432.3 1083.7
439.3 1083.7
446.3 1083.7
453.3 1084.3
459.6 1084.6
466.6 1084.6
473.6 1084.6
480.6 1084.6
487.6 1084.6
493.8 1084.6
500.8 1084.6
507.8 1084.6
514.8 1084.6
521 1084.6
528 1084.6
535 1084.6
542 1084.6
549 1084.6
555.2 1084.6
562.2 1084.6
569.2 1084.6
576.2 1084.6
582.4 1084.6
589.4 1084.6
596.4 1084.6
603.4 1084.6
610.4 1084.6
616.6 1084.6
623.6 1084.6
630.6 1084.6
637.6 1084.6
643.9 1084.6
650.9 1084.6
657.9 1084.6
664.9 1084.6
671.9 1084.6
678.1 1084.6
685.1 1084.6
692.1 1084.6
699.1 1084.6
705.3 1084.6
712.3 1084.6
719.3 1084.6
};
\addlegendentry{\small Fuzz'EMup}
\addplot [line width=1.4pt, forestgreen4416044]
table {%
0 0
1 143
2 153
3 219
4 234
5 254
6 284
7 301
8 343
9 394
10 435
11 495
12 532
13 564
14 581
15 598
16 621
17 644
18 667
19 684
20 701
21 718
22 735
23 752
24 769
25 786
26 803
27 820
28 837
29 841
30 864
31 881
32 881
33 881
34 881
35 902
36 942
37 954
38 958
39 963
40 969
41 972
42 972
43 983
44 984
45 993
46 993
47 993
48 993
49 993
50 993
51 993
52 993
53 993
54 993
55 993
56 993
57 993
58 993
59 993
60 993
61 993
62 993
63 993
64 993
65 993
66 993
67 993
68 993
69 993
70 993
71 993
72 993
73 993
74 993
75 993
76 999
77 1013
78 1021
79 1021
80 1021
81 1021
82 1021
83 1021
84 1021
85 1021
86 1021
87 1021
88 1021
89 1021
90 1021
91 1021
92 1021
93 1021
94 1021
95 1021
96 1025
97 1025
98 1025
99 1025
100 1025
101 1025
102 1025
103 1025
104 1025
105 1025
106 1025
107 1025
108 1025
109 1025
110 1033
111 1033
112 1033
113 1033
114 1033
115 1033
116 1033
117 1033
118 1033
119 1033
120 1033
121 1033
122 1033
123 1033
124 1033
125 1033
126 1033
127 1033
128 1035
129 1043
130 1046
131 1050
132 1050
133 1052
134 1053
135 1055
136 1057
137 1061
138 1065
139 1067
140 1067
141 1067
142 1067
143 1067
144 1067
145 1067
146 1067
147 1067
148 1067
149 1067
150 1067
151 1067
152 1067
153 1067
154 1067
155 1067
156 1067
157 1067
158 1067
159 1067
160 1067
161 1067
162 1067
163 1081
164 1081
165 1102
166 1110
167 1114
168 1114
169 1114
170 1120
171 1120
172 1154
173 1154
174 1161
175 1164
176 1168
177 1170
178 1170
179 1170
180 1170
181 1170
182 1170
183 1170
184 1170
185 1170
186 1170
187 1170
188 1170
189 1170
190 1170
191 1170
192 1170
193 1170
194 1170
195 1170
196 1170
197 1170
198 1170
199 1170
200 1170
201 1175
202 1175
203 1184
204 1184
205 1184
206 1184
207 1184
208 1184
209 1184
210 1184
211 1184
212 1184
213 1184
214 1184
215 1184
216 1184
217 1184
218 1184
219 1184
220 1184
221 1184
222 1184
223 1184
224 1184
225 1184
226 1184
227 1184
228 1184
229 1184
230 1184
231 1184
232 1184
233 1184
234 1184
235 1184
236 1184
237 1184
238 1184
239 1184
240 1184
241 1184
242 1184
243 1184
244 1184
245 1184
246 1184
247 1184
248 1184
249 1184
250 1184
251 1184
252 1184
253 1184
254 1184
255 1184
256 1184
257 1184
258 1184
259 1184
260 1184
261 1184
262 1191
263 1191
264 1191
265 1198
266 1198
267 1198
268 1209
269 1209
270 1209
271 1209
272 1209
273 1219
274 1219
275 1219
276 1219
277 1219
278 1219
279 1219
280 1219
281 1219
282 1219
283 1219
284 1219
285 1219
286 1219
287 1219
288 1219
289 1219
290 1219
291 1219
292 1219
293 1219
294 1219
295 1219
296 1219
297 1219
298 1219
299 1219
300 1219
301 1219
302 1219
303 1219
304 1219
305 1219
306 1219
307 1219
308 1219
309 1219
310 1219
311 1219
312 1219
313 1219
314 1219
315 1219
316 1219
317 1219
318 1219
319 1219
320 1219
321 1219
322 1219
323 1219
324 1219
325 1219
326 1219
327 1219
328 1219
329 1219
330 1219
331 1219
332 1219
333 1240
334 1262
335 1278
336 1297
337 1320
338 1340
339 1358
340 1380
341 1380
342 1380
343 1380
344 1380
345 1380
346 1380
347 1380
348 1380
349 1380
350 1380
351 1380
352 1380
353 1380
354 1380
355 1380
356 1380
357 1380
358 1385
359 1393
360 1393
361 1399
362 1399
363 1399
364 1412
365 1412
366 1412
367 1412
368 1412
369 1412
370 1412
371 1412
372 1412
373 1412
374 1412
375 1421
376 1421
377 1421
378 1421
379 1421
380 1421
381 1421
382 1421
383 1421
384 1421
385 1421
386 1421
387 1421
388 1421
389 1421
390 1428
391 1428
392 1428
393 1428
394 1428
395 1428
396 1431
397 1431
398 1431
399 1431
400 1431
401 1431
402 1431
403 1438
404 1438
405 1438
406 1438
407 1438
408 1438
409 1438
410 1438
411 1438
412 1438
413 1438
414 1438
415 1438
416 1438
417 1438
418 1438
419 1438
420 1438
421 1438
422 1438
423 1438
424 1438
425 1438
426 1438
427 1438
428 1438
429 1438
430 1438
431 1438
432 1438
433 1438
434 1438
435 1438
436 1438
437 1438
438 1438
439 1438
440 1438
441 1438
442 1438
443 1438
444 1438
445 1438
446 1438
447 1438
448 1438
449 1438
450 1438
451 1438
452 1438
453 1438
454 1438
455 1438
456 1438
457 1438
458 1438
459 1438
460 1438
461 1438
462 1438
463 1438
464 1438
465 1438
466 1438
467 1438
468 1438
469 1438
470 1438
471 1438
472 1438
473 1438
474 1438
475 1438
476 1438
477 1438
478 1438
479 1438
480 1438
481 1438
482 1438
483 1438
484 1438
485 1438
486 1438
487 1438
488 1438
489 1438
490 1438
491 1438
492 1438
493 1438
494 1438
495 1438
496 1438
497 1438
498 1438
499 1438
500 1438
501 1438
502 1438
503 1438
504 1438
505 1438
506 1438
507 1438
508 1438
509 1438
510 1438
511 1438
512 1438
513 1438
514 1438
515 1438
516 1438
517 1438
518 1438
519 1438
520 1438
521 1438
522 1438
523 1438
524 1438
525 1438
526 1438
527 1438
528 1438
529 1438
530 1438
531 1438
532 1438
533 1438
534 1438
535 1438
536 1438
537 1438
538 1438
539 1438
540 1438
541 1438
542 1438
543 1438
544 1438
545 1438
546 1438
547 1438
548 1438
549 1438
550 1438
551 1438
552 1438
553 1438
554 1438
555 1438
556 1438
557 1438
558 1438
559 1438
560 1438
561 1438
562 1438
563 1438
564 1438
565 1438
566 1438
567 1438
568 1438
569 1438
570 1438
571 1438
572 1438
573 1438
574 1438
575 1438
576 1438
577 1438
578 1438
579 1438
580 1438
581 1438
582 1438
583 1438
584 1438
585 1438
586 1438
587 1438
588 1438
589 1438
590 1438
591 1438
592 1438
593 1438
594 1438
595 1438
596 1438
597 1438
598 1438
599 1438
600 1438
601 1438
602 1438
603 1438
604 1438
605 1438
606 1438
607 1438
608 1438
609 1438
610 1438
611 1438
612 1438
613 1438
614 1438
615 1438
616 1438
617 1438
618 1438
619 1438
620 1438
621 1438
622 1438
623 1438
624 1438
625 1438
626 1438
627 1438
628 1438
629 1438
630 1438
631 1438
632 1438
633 1438
634 1438
635 1438
636 1438
637 1438
638 1438
639 1438
640 1438
641 1438
642 1438
643 1438
644 1438
645 1438
646 1438
647 1438
648 1438
649 1438
650 1438
651 1438
652 1438
653 1438
654 1438
655 1438
656 1438
657 1438
658 1438
659 1438
660 1438
661 1438
662 1438
663 1438
664 1438
665 1438
666 1438
667 1438
668 1438
669 1438
670 1438
671 1438
672 1438
673 1438
674 1438
675 1438
676 1438
677 1438
678 1438
679 1438
680 1438
681 1438
682 1438
683 1438
684 1438
685 1438
686 1438
687 1438
688 1438
689 1438
690 1438
691 1438
692 1438
693 1438
694 1438
695 1438
696 1438
697 1438
698 1438
699 1438
700 1438
701 1438
702 1438
703 1438
704 1438
705 1438
706 1438
707 1438
708 1438
709 1438
710 1438
711 1438
712 1438
713 1438
714 1438
715 1438
716 1438
717 1438
718 1438
719 1438
720 1438
721 1438
722 1438
723 1438
724 1438
725 1438
726 1438
727 1438
728 1438
729 1438
730 1438
731 1438
732 1438
733 1438
734 1438
735 1438
736 1438
737 1438
738 1438
739 1438
740 1438
741 1438
742 1438
743 1438
744 1438
745 1438
746 1438
747 1438
748 1438
749 1438
750 1438
751 1438
752 1438
753 1438
754 1438
755 1438
756 1438
757 1438
758 1438
759 1438
760 1438
761 1438
762 1438
763 1438
764 1438
765 1438
766 1438
767 1438
768 1438
769 1438
770 1438
771 1438
772 1438
773 1438
774 1438
775 1438
776 1438
777 1438
778 1438
779 1438
780 1438
781 1438
782 1438
783 1438
784 1438
785 1438
786 1438
787 1438
788 1438
789 1438
790 1438
791 1438
792 1438
793 1438
794 1438
795 1438
796 1438
797 1438
798 1438
799 1438
800 1438
801 1438
802 1438
803 1438
804 1438
805 1438
806 1438
807 1438
808 1438
809 1438
810 1438
811 1438
812 1438
813 1438
814 1438
815 1438
816 1438
817 1438
818 1438
819 1438
820 1438
821 1438
822 1438
823 1438
824 1438
825 1438
826 1438
827 1438
828 1438
829 1438
830 1438
831 1438
832 1438
833 1438
834 1438
835 1438
836 1438
837 1438
838 1438
839 1438
840 1438
841 1438
842 1438
843 1438
844 1438
845 1438
846 1438
847 1438
848 1438
849 1438
850 1438
851 1438
852 1438
853 1438
854 1438
855 1438
856 1438
857 1438
858 1438
859 1438
860 1438
861 1438
862 1438
863 1438
864 1438
865 1438
866 1438
867 1438
868 1438
869 1438
870 1438
871 1438
872 1438
873 1438
874 1438
875 1438
876 1438
877 1438
878 1438
879 1438
880 1438
881 1438
882 1438
883 1438
884 1438
885 1438
886 1438
887 1438
888 1438
889 1438
890 1438
891 1438
892 1438
893 1438
894 1438
895 1438
896 1438
897 1438
898 1438
899 1438
900 1438
901 1438
902 1438
903 1438
904 1438
905 1438
906 1438
907 1438
908 1438
909 1438
910 1438
911 1438
912 1438
913 1438
914 1438
915 1438
916 1438
917 1438
918 1438
919 1438
920 1438
921 1438
922 1438
923 1438
924 1438
925 1438
926 1438
927 1438
928 1438
929 1438
930 1438
931 1438
932 1438
933 1438
934 1438
935 1438
936 1438
937 1438
938 1438
939 1438
940 1438
941 1438
942 1438
943 1438
944 1438
945 1438
946 1438
947 1438
948 1438
949 1438
950 1438
951 1438
952 1438
953 1438
954 1438
955 1438
956 1438
957 1438
958 1438
959 1438
960 1438
961 1438
962 1438
963 1438
964 1438
965 1438
966 1438
967 1438
968 1438
969 1438
970 1438
971 1438
972 1438
973 1438
974 1438
975 1438
976 1438
977 1438
978 1438
979 1438
980 1438
981 1438
982 1438
983 1438
984 1438
985 1438
986 1438
987 1438
988 1438
989 1438
990 1438
991 1438
992 1438
993 1438
994 1438
995 1438
996 1438
997 1438
998 1438
999 1438
1000 1438
};
\addlegendentry{\small PowerFuzz}
\end{groupplot}

\end{tikzpicture}

%% file: sections/5_conclusion.tex
\section{Conclusion}
\label{sec:conclusion}

Gray-box firmware fuzzing assumes the full visibility of firmware and therefore not applicable when verifying embedded systems with proprietary, encrypted or obfuscated firmware. In this paper, we presented \textbf{PowerFuzz}, the first statistical black-box firmware fuzzing framework that leverages physical power side-channel measurements. PowerFuzz captures the MCU's power consumption as a time-series signal in response to each fuzzer-generated input and applies a two-stage similarity analysis, combining dynamic time warping for coarse deviation window identification and a growing-window Pearson correlation analysis for precise branch localization to identify newly executed branches from power traces. These identified branches drive the dynamic construction of a high-level control flow graph of the black-box firmware, which in turn guides a prioritized branch selection mechanism that directs the fuzzing engine's mutation effort toward deeper, unexplored regions of the firmware's execution space. Experimental evaluation across three embedded hardware platforms using ten firmware benchmarks demonstrated that PowerFuzz can achieve branch coverage comparable (within 13.5\%) to gray-box fuzzing. The results also highlight that PowerFuzz can significantly outperform (up to 22\%) the state-of-the-art black-box fuzzer based on electromagnetic emanations. 